\documentclass[useAMS,usenatbib]{mn2e}
\usepackage{graphicx,epsfig}
\usepackage[usenames]{color}
\usepackage{url}

\newcommand{\M}[1]{\mathbf{#1}}         
\newcommand{\T}[1]{#1^{\top}}           
\newcommand{\nicefrac}[2]{\leavevmode\kern.1em
            \raise.5ex\hbox{\the\scriptfont0 #1}\kern-.1em
      /\kern-.15em\lower.25ex\hbox{\the\scriptfont0 #2}}

\title[Stellar populations in galactic discs]{Star formation history of barred disc galaxies}
\author[P. S\'anchez-Bl\'azquez et al.]{P. S\'anchez-Bl\'azquez$^{1,2,3,4}$
\thanks{E-mail: psanchezb@iac.es}  
P. Ocvirk$^{5,6}$,  B. K. Gibson$^{4}$, I. P\'erez$^{7}$, R.F. Peletier$^{8}$\\
$^{1}$Departamento de F\'{\i}sica Te\'orica, Universidad Aut\'onoma de Madrid, E-28049, Cantoblanco, Madrid, Spain\\
$^{2}$Instituto de Astrof\'{\i}sica de Canarias, E-38200 La Laguna, Tenerife, Spain\\
$^{3}$Departamento de Astrof\'{\i}sica, Universidad de La Laguna, E-38205 La Laguna, Tenerife, Spain\\
$^{4}$Jeremiah Horrocks Institute, University of Central Lancashire, Preston, PR1~2HE, UK\\
$^{5}$Astrophysikalisches Institut Potsdam, an der Sternwarte 16, D-14482 Postdam, Germany\\
$^{6}$Observatoire Astronomique de Strasbourg, 11 rue de l"universit\'e, 67000 Strasbourg, France\\
$^{7}$Departamento de F\'{\i}sica Te\'orica y del Cosmos Universidad de Granada, E-18071, Granada, Spain\\
$^{8}$Kapteyn Astronomical Institute, Rijksuniversiteit Groningen, 9700 AV Groningen, The Netherlands}
\begin{document}

\date{ }

\pagerange{\pageref{firstpage}--\pageref{lastpage}} \pubyear{2002}

\maketitle

\label{firstpage}

\begin{abstract}
We present the first results of a pilot study aimed at understanding 
the influence of bars on the evolution of galaxy discs through the study 
of their stellar content.  We examine here the kinematics, star 
formation history, mass-weighted, luminosity-weighted, and single 
stellar population (SSP) equivalent ages and metallicities for four
galaxies ranging from lenticulars to late-type spirals.
The data employed extends to 2-3 disc scalelengths,
with S/N(\AA)$>$50.  Several techniques are explored to 
derive star formation histories and SSP-equivalent 
parameters, each of which are shown to be compatible.
We demostrate that the age-metallicity degeneracy is highly reduced by using 
spectral fitting techniques --instead of indices-- to derive these parameters.
Our results are robust 
to the choices of stellar population models.  We found that the majority 
of the stellar mass in our sample is composed of old ($\sim$10~Gyr) 
stars. This is true in the bulge and the disc region, even beyond
two disc scalelengths.
In the bulge region, we find that the 
young, dynamically cold, structures produced by the presence of the bar
(e.g., nuclear discs or rings) are responsible for shaping the bulges' age 
and metallicity gradients, as suggested by Peletier et al. 
(2007)\nocite{Pel07}.  In the disc region, a larger fraction of young 
stars is present in the external parts of the disc compared with the 
inner disc.  The disc growth is, therefore, compatible with a moderate 
inside-out formation scenario, where the luminosity-weighted age changes from
$\sim$10~Gyrs in the centre, to $\sim$4~Gyrs at two disc scalelengths,
depending upon the galaxy.
However, the presence of substructure, like star forming 
rings, can produce stellar population trends that are not directly 
related with the growing of the disc but to the bar potential. For two 
galaxies, we compare the metallicity and age gradients of the disc major
axis with 
that of the bar, finding very important differences. 
In particular, the stellar 
population of the bar is more similar to the bulge than to the disc, 
indicating that, at least in  those two galaxies, bars formed long ago and have survived 
to the present day.
\end{abstract}


\section{Introduction}

During disc assembly, secular evolution must have played a role in
shaping the structure of disc galaxies as we see them at
z=0. Non-axisymmetric instabilities, particularly bars, drive a
substantial redistribution of mass and angular momentum in the disc
(Sellwood \& Wilkinson 1993\nocite{SW93}; Sellwood 1981\nocite{Sell81}; Pfenniger \& Friedli 1991\nocite{PF91};
Athanassoula 2003\nocite{Ath03}; Debattista et al. 2006\nocite{Deb06}). Since 
bars appear to affect strongly the
overall dynamics and evolution of galaxies, it is reasonable to
presume that they also play a significant role in shaping galactic
chemical evolution, 
since mixing by global gas flows will clearly
change abundance profiles in the disc (see Friedli
1998\nocite{Fried98}; Roskar et al. 2008b\nocite{Rosk08b};
S\'anchez-Bl\'azquez et al. 2009a\nocite{SB09a}); for example, numerical
simulations (Friedli, Benz \& Kennicutt 1994\nocite{Fried94}) predict
that the stellar abundance gradient in disc galaxies is flattened by
macroscopic mixing induced by the bar. The slope of the abundance
gradients in both the stellar and gaseous component are reduced in a
few dynamical timescales by more than 50\%.

Several of the relationships observed in our own Galaxy 
used to constrain chemical evolution models, such as the
age-metallicity relation (Edvardsson et al. 1993\nocite{Ed93}; Carraro
et al. 1998\nocite{Carr98}; Feltzing \& Gonzalez 2001\nocite{FG01};
Rocha-Pinto et al. 2000\nocite{RP00}; Twarog
1980ab\nocite{Twa80a}\nocite{Twa80b}; Feltzing et
al. 2001\nocite{Felt01}), the metallicity distribution of G- and
K-dwarf
stars (e.g., Rocha-Pinto \& Maciel 1996\nocite{RPM96}; Wyse \& Gilmore
1995\nocite{WG95}; Flynn \& Morell 1997\nocite{FM97}; Favata et
al. 1997\nocite{Fav97}; Haywood 2001\nocite{Hay01}) or the temporal
evolution of metallicity gradients (Friel et
al. 2002\nocite{Friel02}; Maciel et al. 2003\nocite{Mac03};
Stanghellini et al. 2006\nocite{Stan06}; Magrini et
al. 2009\nocite{Mag09}) could be  strongly influenced by the presence of
radial migrations.  

The mechanisms proposed for being
primary drivers of radial migration are numerous
(Spitzer \& Schwarzschild 1953\nocite{SS53}; Barbanis \&
Woltjer 1967\nocite{BW67}; Friedli 1998; Binney \& Lacey
1988\nocite{BL88}; Fuchs 2001\nocite{Fuchs01}; Sellwood \& Binney
2002\nocite{SB02}; Roskar et al. 2008a; S\'anchez-Bl\'azquez et
al. 2009; Minchev \& Falmaey 2009). 
However, in the present work we are interested in constraining the influence of a
bar upon this migration. This  is important, not only because the Milky Way has
a bar, but because
recent simulations suggest that the 
interaction between the bar and the spiral structure can result in 
significantly faster angular momentum changes, 
making mixing mechanisms due to spiral arms much more efficient in 
barred than in non-barred galaxies (Minchev \& Falmaey 2009).

Several studies have found that barred galaxies have a shallower
gas-phase metallicity gradient than non-barred ones (e.g., Pagel \&
Edmunds 1981\nocite{PE81}; Vila-Costas \& Edmunds 1992\nocite{VCE92};
Edmunds \& Roy 1993; Martin \& Roy 1994\nocite{MR94}; Zaritsky et
al. 1994; Ryder 1995). However, an equivalent study for the stellar
abundances remains to be done.
Studies of the gas-phase abundances provide with
a present-day snapshot of the interstellar medium abundances. On the other
hand, study of the stellar ages and metallicities give us 
`archaeological' clues as to the chemical and dynamical 
evolution of the galactic structures.
It is very important to study both because gas and stars suffer from very
different evolutionary processes; the gas is mainly dominated by the 
gravitational torque of the non-axisymmetric mass
component, while the evolution of the stellar component is mainly affected by
different orbital mixing.

In general, stellar population studies in the disc region of spiral
galaxies are sparse, with exception of our own galaxy (see Freeman \&
Bland-Hawthorn 2002\nocite{FB02} for a review; Friel et al. 2002\nocite{Friel02}; Yong
et al. 2006\nocite{Yong06}; Carraro et al.  2007\nocite{Carr07}) and for 
some nearby galaxies, such as M33 (Monteverde et
al. 1997\nocite{Mon97}; Barker et al. 2008\nocite{Bar08}; Williams et al.\ 2009\nocite{Will09b};
Cioni et al. 2009\nocite{Cio09}), M100 (Beauchamp \& Hardy 1997\nocite{BH97}); M81
(Hughes et al. 1994; Tikhonov et al. 2005; Williams et al. 2008;
Davdige 2006a, 2008); NGC2403 (Davidge 2007); NGC 300 (Vlajic et
al. 2008; Kudritzki et al 2008; Urbaneja et al. 2005; Gogarten et
al. 2010) or M31 (Worthey et al. 2005).
For external galaxies outside the Local Group, stellar population
gradients have been mainly investigated using (mostly optical) colours
(e.g., de Jong 1996\nocite{dJ96}; Peletier \& Balcells
1996\nocite{PB96}; Jansen et al\ 2000; Bell \& de Jong
2000\nocite{BdJ00}; MacArthur et al. 2004\nocite{Mac04}; Taylor et
al. 2005;  Mu\~noz-Mateos et al. 2007, 2009 (using UV-GALEX colours); 
Prochaska 
et al. 2010).  These studies found that
disc galaxies tend to be bluer in the external parts.  This trend has
been interpreted as the consequence of stellar population gradients,
in the sense that galaxies are older and more metal rich in their
centers than in the external parts.  However, there are large discrepancies
in the  magnitude of the stellar population gradients 
derived by different authors.
This is because it is extremely difficult to disentangle the
effects of age, metallicity, and dust extinction (on average, $\sim$1~mag
in the central regions, see Ganda et al. 2009) using only colours.
Spectroscopic studies may help to alleviate the associated degeneracies, but
the low-surface brightness of the
disc region and the nebular emission lines filling some of the most important
age-diagnostic absorption lines make these studies very difficult.
Some pioneering works tried to overcome the difficulty of measuring
low-surface brightness absorption lines in the disc using narrow-band
imaging, or performing Fabry-P\'erot interferometry with Tunable Filters
(Beauchamp \& Hardy 1997\nocite{BH97}; Molla, Hardy \& Beauchamp
1999\nocite{Moll99}; Ryder, Fenner \& Gibson 2005\nocite{Ryd05}).
Unfortunately, the low spectral resolution and poor S/N ratio of the
data compromised the results in each case.  Furthermore,
these works were restricted to the study of just a few indices 
(Mgb, Fe5270, Fe5335)
which limited their ability to break the age-metallicity degeneracy.
More recently, Yoachim \& Dalcanton (2008)\nocite{YD08}, and
MacArthur, Gonz\'alez \& Courteau (2009)\nocite{Mac09} have measured
line-strength indices and star formation histories, repectively,  as a function of  radius
using long-slit spectroscopy.  While the former found very strong
radial age gradients and flat metallicity gradients 
for the only galaxy for which they could derive stellar population gradients,
(FGC 1440, and  Scd galaxy),  the
latter reported very mild gradients in both age and metallicity.

These differences maybe the consequence of the  very reduced samples (9 and 1 galaxy),
although they can also be 
due to the different techniques used to derive 
stellar parameters in both studies --while the former obtained single
stellar population (SSP hereafter)-equivalent parameters, the latter
derived luminosity and mass-weighted values.
The difference could also be due to the uncertainties of the emission-line
corrections that affect more strongly the ages derived with 
single indices than those obtained with full spectral fitting. 
It is one of the motivation of this work to explore some of these issues.
Despite heroic efforts, we really only have information pertaining
to the inner disc for
most galaxies ($\sim$1$-$2 scalelengths, in the MacArthur et al.\ 2009 data),
and only for 9 galaxies.

This is part of a long-term project aimed at studying the star
formation history of the different components of a sample of face-on
disc galaxies with deep long-slit observations taken at the GEMINI
telescope.  One of our main goals is to understand the influence of
bars  on  the evolution of these systems, for which, in a forthcoming
paper we will compare with analogous non-barred galaxies.
This study is complemented by that of the stellar populations  in
galaxy bars and their bulges (P\'erez, S\'anchez-Bl\'azquez \& Zurita
2007; P\'erez, S\'anchez-Bl\'azquez \& Zurita 2009; P\'erez \&
S\'anchez-Bl\'azquez 2010, in preparation).  In this paper, we present
a pilot study using data from 4 galaxies. The aim of our work here 
is to present
the different analysis techniques, their limitations and robustness,
and the dependence of the results on the choice of different
parameters.

\section{Observations and Data Reduction}
\label{sec:obs}

The galaxies were selected from the Third Reference Catalogue (RC3: de Vaucouleurs et al. 1991) to
cover a range in morphological types from S0 to Sc. They were imposed
to be nearby ($cz$ $<$ 3000 kms$^{-1}$) and barred.  For the earlier
types, we selected the sample to overlap with that of P\'erez,
S\'anchez-Bl\'azquez \& Zurita (2007)\nocite{PSZ07} as, for these
galaxies, spectra along the bar have already been obtained at Siding
Spring Observatory and the Isaac Newton Telescope (P\'erez et
al. 2007, 2009\nocite{PSZ09}). This will allow us to compare the gradients along the 
bar with those along the disc (see \S~\ref{sec:bars}).

\begin{table*}
\begin{tabular}{rrrrrrr}
\hline
\hline
\multicolumn{1}{c}{gal} & \multicolumn{1}{c}{PA}  & \multicolumn{1}{c}{i}     &   \multicolumn{1}{c}{Morph}    &  \multicolumn{1}{c}{r$_e$(kpc)}  & \multicolumn{1}{c}{r$_s$ (kpc)} & 
 \\
\multicolumn{1}{c}{(1)} & \multicolumn{1}{c}{(2)}  & \multicolumn{1}{c}{(3)}     &   \multicolumn{1}{c}{(4)}    &  \multicolumn{1}{c}{(5)}  & \multicolumn{1}{c}{(6)} 
\\
\hline
NGC 1358   &180  & 38.8 (iii)   &  SAB(R)0     & 1.17(a) & 5.5(a)        \\ 
NGC 1365   &225  & 40.0(i)      &  (R')SBb(s)b & 0.96(c) & 6.5(b)         \\ 
NGC 1433   &210  & 33.0(ii)     &  (R)SB(rs)ab& 0.26(c)  & 1.8(c)   \\
NGC 1672   &140  & 36.0 (iv)    & (R')SB(r)bc &  0.40(c) & 1.6 (c) &\\ 
\hline
\end{tabular}
\caption{Sample of galaxies observed; columns: (1) galaxy name; (2) PA: Position angle of the slit in degrees from N through E;
(3) inclination  in degrees (obtained from different references) (i) Roy \& Walsh (1997)
(ii) Ryder et al. (1996) (iii) Dumas, Emsellem \& Ferruit  (2007) (iv) Baumgart \& Peterson (1986); 
(4) Morphological type (from Hyperleda); (5) effective radius
of the bulge; (6) scalelength of the disc. The references for the effective radius and the scale-length 
of the disc are the following: 
(a) Dong \& de Robertis (2006); (b) Z\'anmar-S\'anchez et al.\ (2008); (c) Baumgart \& Peterson (1986).
\label{table1}}
\end{table*}

\begin{figure*}
\begin{tabular}{cc}
\resizebox{0.32\textwidth}{!}{\includegraphics[angle=0,bb=99 226 493 617]{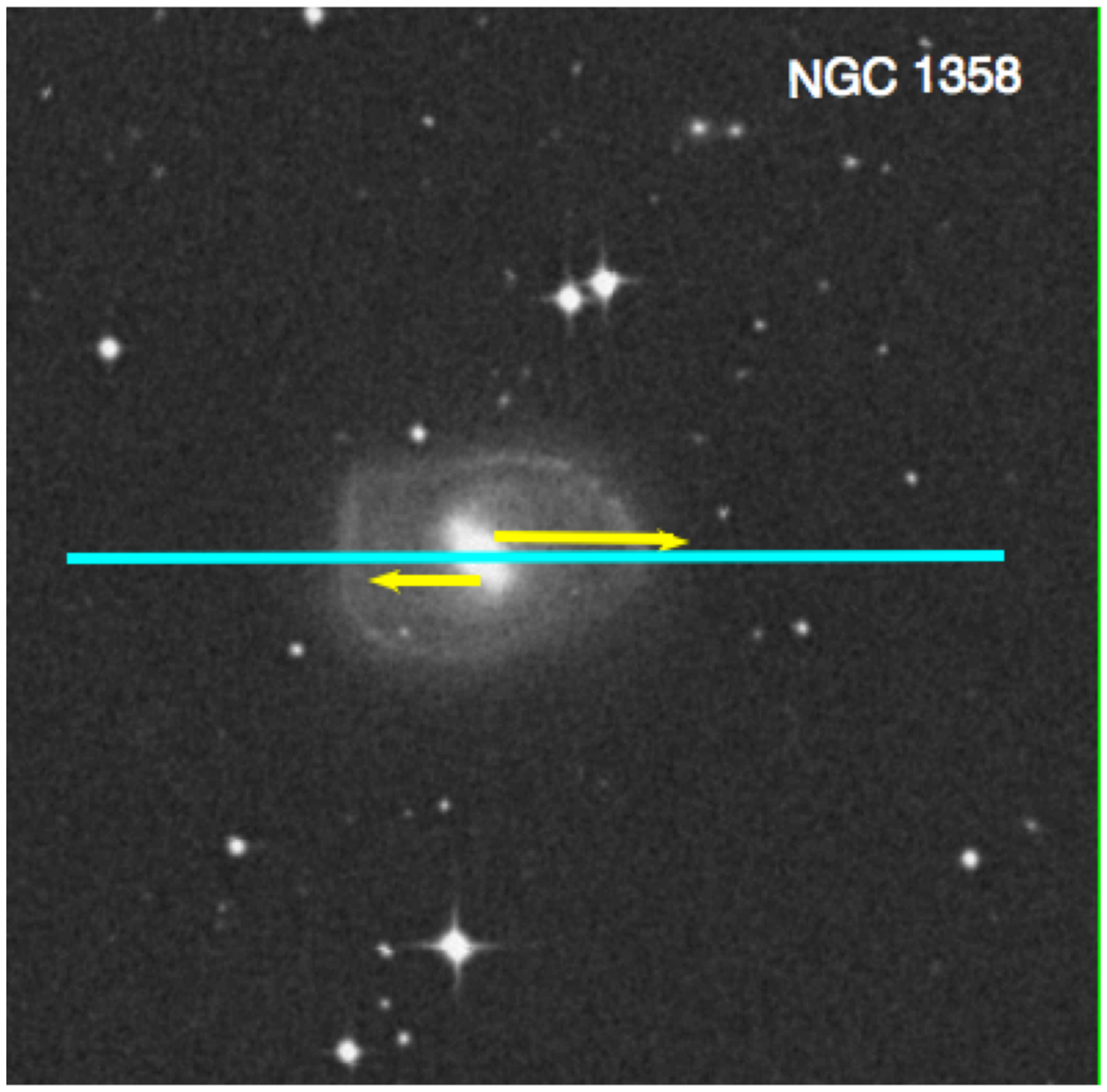}}&
\resizebox{0.32\textwidth}{!}{\includegraphics[angle=0,bb=189 314 405 530]{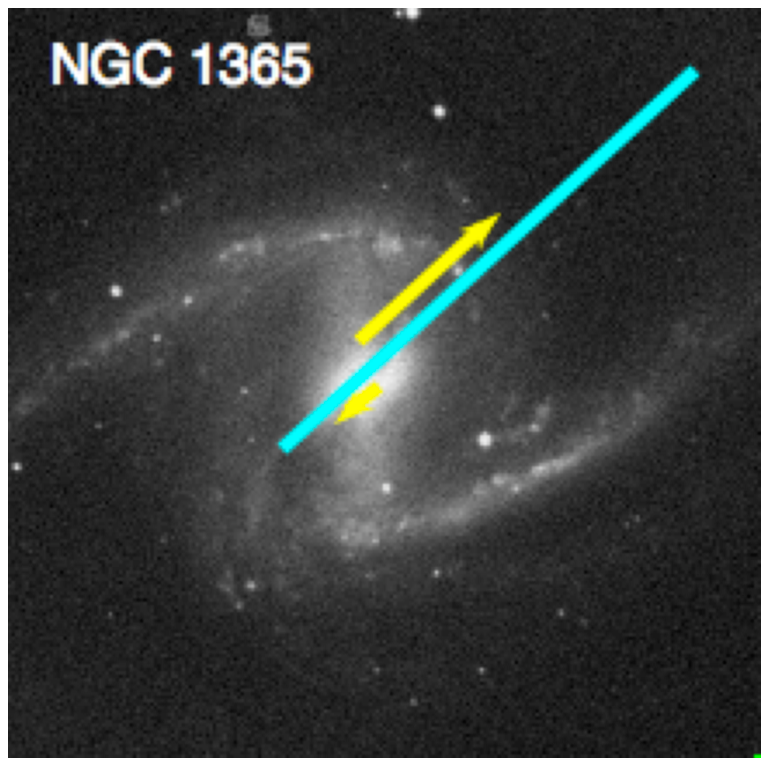}}\\
\resizebox{0.31\textwidth}{!}{\includegraphics[angle=0,bb= 189 314 405 530]{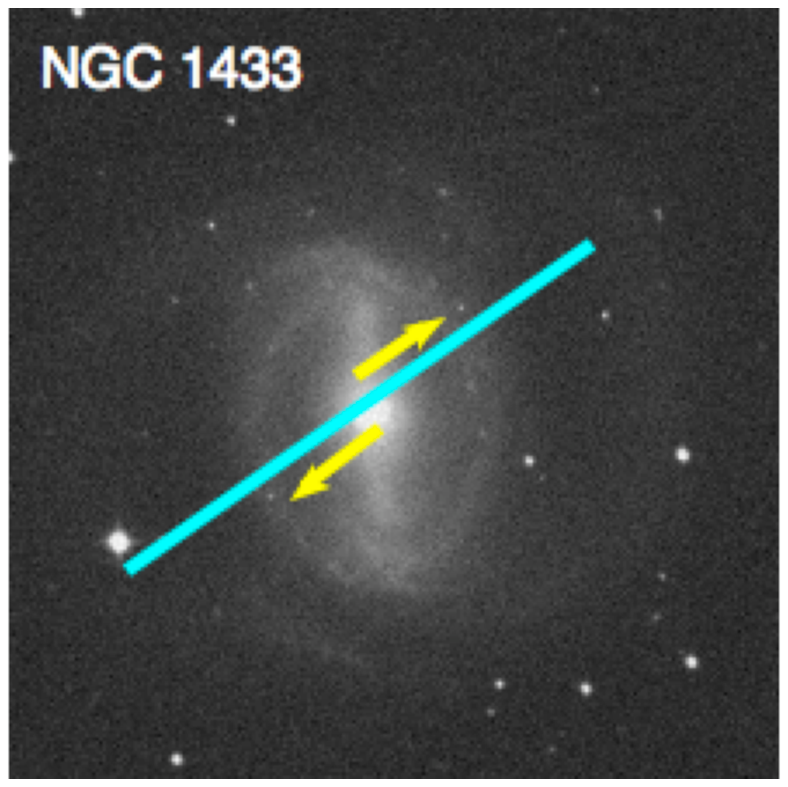}}&
\resizebox{0.32\textwidth}{!}{\includegraphics[angle=0,bb= 180 303 413 538]{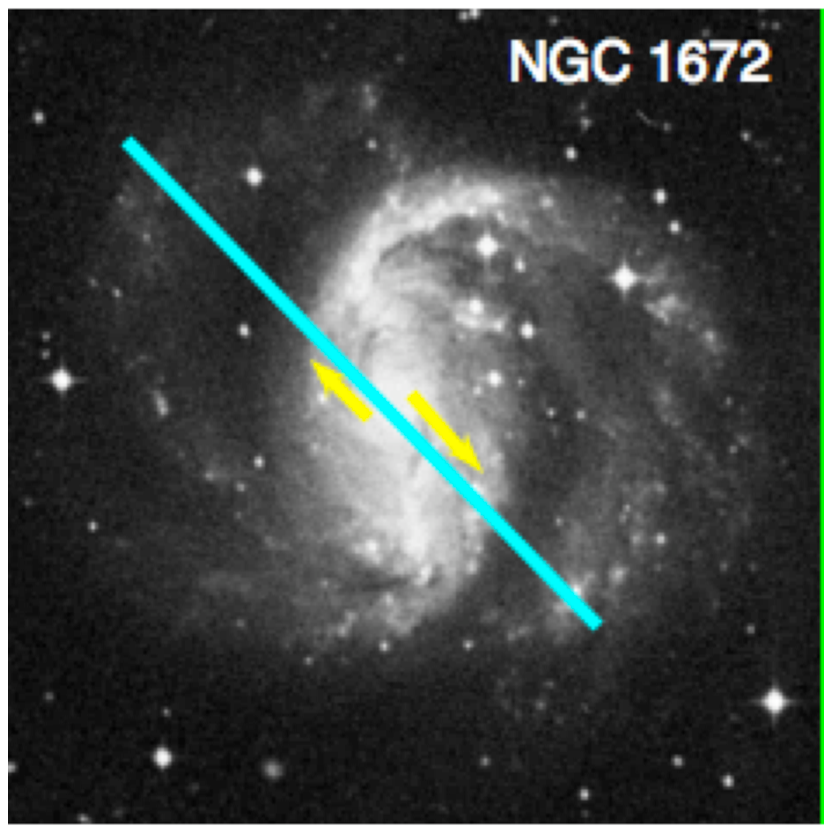}}
\end{tabular}
\caption{Observational setup for our sample of galaxies.
The images all cover an area of 450$\times$450$\arcsec$. The yellow arrows indicate
the extent of the stellar population gradients presented here.}
\end{figure*}
All observations were taken with the Gemini South Multi-Object Spectrograph
(GMOS) (Hook et al. 2004\nocite{H04}) as part of Gemini Programs GS-2008B-73
and GS-2008B-85.
Long-slit spectroscopy along the major axes of the galaxies was obtained
with a slit width of 0.75$"$ with the B600-G5303 grating.
We performed a 4$\times$2 binning giving a spectral resolution 
of FWHM=3.5~\AA~  (measured on the sky lines of the images) and a spatial scale of 0.146 arcsec/pixel.
Our spectra  cover the wavelength range 4050-6750~\AA, with  gaps
between $\sim$4920-5025 and 5865-5930\AA.

The major obstacle to measuring reliable 
stellar populations at large radii  is the sky
subtraction, as the light of the galaxy is only a few 
percent of the background. 
Nod-and-Shuffle spectroscopy allows very accurate sky subtraction
and is available on GMOS. However, the length of the slit in this mode is only
1.83 arcmin and our objects span between 2 and 5 arcmin on the sky.
In order to obtain a good sky subtraction after every target exposure 
of 1800s, a sky exposure of 400s was taken  slightly
offset ($\sim$200 arcsec) from the target. Total exposure
times for each galaxy were 1800$\times$6 (3h), while a 400s exposure on 
the sky was taken before and after each target exposure.
These long exposures on sky minimise 
the amount of noise introduced in this step.

Standard data reduction procedures (flat-fielding, cosmic ray removal,
wavelength calibration, sky subtraction and fluxing) were performed
with REDUCEME (Cardiel
1999\nocite{Car99})\footnote{\tt www.ucm.es/info/Astrof/software/reduceme/reduceme.html}. This
reduction package allows a parallel treatment of data and error frames
from the initial steps of the reduction process and, therefore,
generates an associated error spectrum of each individual data spectrum
which is quite realistic and takes  into account the propagation of the
errors in each reduction step.  Initial reduction of the CCD frames
involved bias and dark current subtraction, removal of pixel-to-pixel
sensitivity variations (using flat-field exposures of a tungsten
calibration lamp) and correction for two-dimensional low frequency
scale sensitivity variations (using twilight sky exposures).  Prior to
the wavelength calibration, arc frames were used to correct for 
C-distortion in the images. Spectra were converted to a linear
wavelength scale using typically 80 arc lines fitted by 3rd order
polynomials, with RMS errors of $\sim$0.05\AA. The images were not
distorted in the spatial direction and, therefore, S-correction was
not applied. 
Before performing any distortion
correction or wavelength calibration, the sky images were multiplied by a
factor corresponding to the exposure time of the science frames and were
subtracted. 

From each fully reduced galaxy frame, a final frame was created by extracting
spectra along the slit, binning in the spatial direction to guarantee a minimum
signal-to-noise ratio per \AA~(S/N) of 50 in a 50\AA-wide region centered on 5100~\AA.
The spectroscopic data extend to $\sim$2 scalelengths for most of the galaxies, except
for NGC~1358, where $\sim$3 scalelengths are reached. Only for two of the
four galaxies (NGC~1358 and NGC~1365) do
our data extend beyond corotation.

\subsection{Notes on the Sample}
\begin{itemize}
\item NGC~1358:
This galaxy is a Seyfert, S0/a galaxy with a bar.
For this galaxy, long-slit spectroscopy along the bar was presented 
in P\'erez, S\'anchez-Bl\'azquez \& Zurita (2009).
The galaxy  has  an inner gaseous spiral structure (with size $<$1 kpc; Dumas,  Emsellem \& Ferruit 2007\nocite{Dum07}).
The outer bar has a corotation radius of 5.1 kpc.

\item NGC~1365:  This galaxy is a barred SB(s)b galaxy. It has a nuclear 
ring with radius $\sim$ 7 arcsec.
Several studies have reported super-solar metallicities in the gaseous components, 
measured in its HII regions (Roy \& Walsh 1997).
 
\item NGC~1433: A prototypical ringed barred galaxy (Treuthardt et al. 2008\nocite{Treut08}), 
with a morphological type (R')SB(r)ab.
It has  a circumnuclear ring with a diameter about 0.1 times the size of the bar
(at r$\sim$ 19$"$, Crocker et al. 1996\nocite{Crock96};
Hameed \& Devereuz 1999\nocite{HD99}; Buta et al. 2001; Comeron et al.\ 2007), and an outer 
pseudo-ring about twice the diameter of the bar.
Spectra along the bar were presented by P\'erez et al. (2009).

\item NGC~1672:
An SB(s)b, barred spiral galaxy, with a circumnuclear ring of star
formation at r$\sim$0.350 kpc (S\'ersic \& Pastoriza 1965\nocite{SP65}; Comeron et al. 2007) surrounding a nucleus with low-level
activity (Veron-Cetty \& Veron 1986\nocite{VV86}), also containing 
a star forming region.
The bar size is 5.1 kpc (Buta 1987) and the inner Lindblad resonance (ILR) 
is calculated to be 
0.49 kpc from the galaxy center, coincident with the edge of the circumnuclear ring.

\end{itemize}

\section{Kinematics, Dynamics and Emission Line Removal}
\begin{figure}
\centering
\resizebox{0.45\textwidth}{!}{\includegraphics[angle=-90]{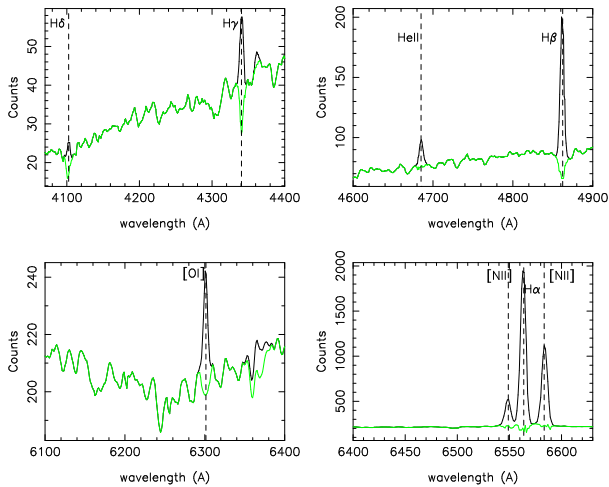}}
\caption{Central spectrum of NGC~1365 before (black line) and after
  (green line) subtracting the emission spectra using {\tt
    GANDALF}.\label{gandalf_spectra}}
\end{figure}

One of the most important drawbacks to studying the stellar population of disc
galaxies has been that  the stellar absorption line spectra are highly contaminated by 
nebular emission lines. 
However, this situation has changed with the development of new
specialist software that allows the simultaneous fitting of absorption
and emission lines.  We used {\tt GANDALF} (Sarzi et
al. 2006\nocite{Sar06}) to perform this task. {\tt GANDALF} fits
simultaneously the absorption and emission lines treating the latter
as additional Gaussians. In a first step, emission lines are masked
and the absorption line spectrum is fit with the penalised
pixel-fitting {\tt pPXF} (Cappellari \& Emsellem 2004\nocite{CE04}),
using as templates the new stellar population models of Vazdekis et
al. (2010) based on the MILES library (S\'anchez-Bl\'azquez et
al. 2006\nocite{SB06}; Cenarro et al. 2007\nocite{Cen07})\footnote{The models
are available at {\tt miles.iac.es}}.  In this step, 
radial velocities and 
velocity dispersions ($\sigma$, hereafter) for the stellar component
are derived.
The best values of velocity and $\sigma$ and the best template mix are then
used as initial values for the derivation of emission lines using {\tt 
GANDALF}.
Emission-line equivalent widths, radial velocities, and $\sigma$
for the gaseous component are derived in this second step.
The fit allows for a low-order Legendre polynomial in order to account for
small differences in the continuum shape between the pixel spectra and
the templates. The best fitting template mix is determined by a
$\chi^2$ minimization in pixel space.
 Emission line spectra at each radius 
were substracted from the observed spectra for the subsequent analysis.
Figure~\ref{gandalf_spectra} shows the central spectrum of NGC~1365
before and after subtracting the emission spectra.  Figure~\ref{gandalf_curves} shows the
line-of-sight velocity and the $\sigma$ of the stellar and
gaseous components as a function of radius for our sample of galaxies.
We fit the position and the $\sigma$ of each atomic species
independently. Sarzi et al. (2006) caution {\tt GANDALF} users
about this procedure when emission lines cannot be measured
confidently, as contamination due to template mistmatch can
still be important and result in spurious detections. This problem
affects the H$\beta$ line, which we
cannot measure in all spectra because it does fall in the gap of the
detector. Furthermore, spurious detections would result in velocities
that differ from those measured with other lines and, in our case, 
the velocities measured with all the emission lines are very consistent 
(see \S~\ref{gas_velocity}), giving support to our approach of 
fitting all the lines independently.

The errors in the radial velocity and velocity dispersion were
computed through numerical simulation. In each simulation, a
bootstrapped galaxy spectrum, obtained using the error spectrum 
(see \S~\ref{sec:obs}), was
fed into the algorithm (a different optimal template is computed in
each simulation). Errors in the final parameters were computed as the
unbiased standard deviation of the different solutions. The final
errors are expected to be quite realistic, as they incorporate all the
uncertainties of the entire reduction process, from the first steps
(e.g., flat fielding) to the final measurements of the parameters.
We did not attempt to correct the velocities for inclination,
as this correction is very uncertain and very large
for nearly face-on galaxies.

\begin{figure*}
\resizebox{0.4\textwidth}{!}{\includegraphics[angle=-90]{n1358.halpha.new.gas.ps}}
\resizebox{0.4\textwidth}{!}{\includegraphics[angle=-90]{n1365.halpha.new2.gas.ps}}
\resizebox{0.4\textwidth}{!}{\includegraphics[angle=-90]{n1433.halpha.new.gas.ps}}
\resizebox{0.4\textwidth}{!}{\includegraphics[angle=-90]{n1672.halpha.new2.gas.ps}}
\caption{Line-of-sight velocity as a function of radius and velocity dispersion 
for the stellar component (black points) and the gaseous components (colored points).
The kinematics for the gas component has been measured using different emission 
lines as indicated in the insets.\label{gandalf_curves}}
\end{figure*}
\begin{figure}
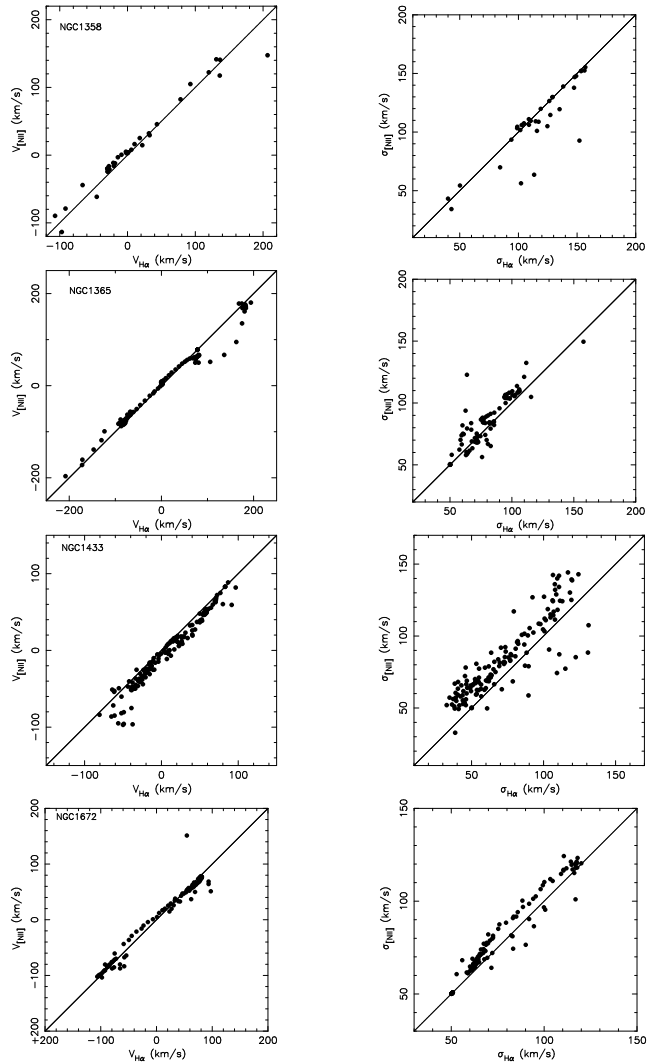

\resizebox{0.2\textwidth}{!}{\includegraphics[angle=-90]{vel_comparacion.n1358.ps}}
\resizebox{0.2\textwidth}{!}{\includegraphics[angle=-90]{sigma_comparacion.n1358.ps}}
\resizebox{0.2\textwidth}{!}{\includegraphics[angle=-90]{vel_comparacion.n1365.ps}}
\resizebox{0.2\textwidth}{!}{\includegraphics[angle=-90]{sigma_comparacion.n1365.ps}}
\resizebox{0.2\textwidth}{!}{\includegraphics[angle=-90]{vel_comparacion.n1433.ps}}
\resizebox{0.2\textwidth}{!}{\includegraphics[angle=-90]{sigma_comparacion.n1433.ps}}
\resizebox{0.2\textwidth}{!}{\includegraphics[angle=-90]{vel_comparacion.n1672.ps}}\hspace{1.3cm}
\resizebox{0.2\textwidth}{!}{\includegraphics[angle=-90]{sigma_comparacion.n1672.ps}}
\caption{Comparison of the radial velocity and velocity dispersion of the emission 
lines H$\alpha$ and [NII] measured with {\tt GANDALF}.\label{fig:comparison}}
\end{figure}

\subsection{Reliability of the derived $\sigma$}
\begin{figure}
\resizebox{0.4\textwidth}{!}{\includegraphics[angle=-90]{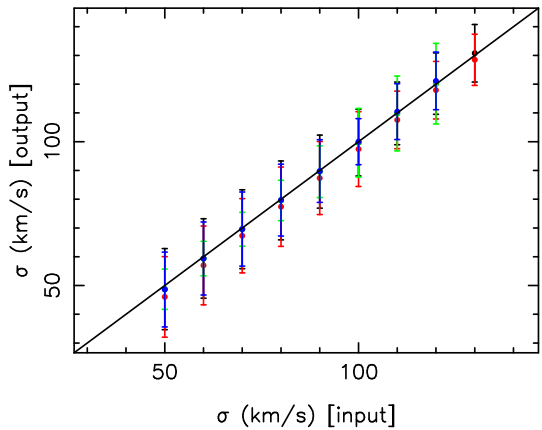}}
\caption{Comparison of the input vs. calculated velocity dispersion in 
a series of simulations performed in synthetic spectra with different
values of metallicities and corresponding to different star formation 
histories (see text for details). 
The solid lines indicate the 1:1 relation. \label{test_sigmas}}
\end{figure}

Given the resolution of our data $\sigma_{\rm inst}\sim 90$ kms$^{-1}$, 
it is worth testing if the derived velocity dispersions (in some cases 
around $\sim$50 kms$^{-1}$ are reliable. For this reason we 
perform a series of simulations using synthetic spectra corresponding
to  different  star formation histories,
trying to mimic as far as possible the real data.
In particular, we 
took the parametrized star formation histories given as examples in the MILES  webpage 
({\tt http://www.iac.es})
that include exponential, constant and composition of different burst with 
metallicities ranging from $[Z/H]=-2.32$ to $0.22$. 
We first broadened the synthetic spectra to  the nominal instrumental 
resolution of our data by convolving with a Gaussian broadening function 
and then we broadened again those spectra with velocity dispersions 
from 50 to 120 kms$^{-1}$ in steps
of 10 kms$^{-1}$. Afterwards,  we
added noise to simulate the signal-to-noise of the observed spectra 
(S/N(\AA)$\sim$50 which is the minimum we used to bin our data).
The velocity dispersion was then measured using {\tt ppxf} using the same 
parameters as we used for the real data. The results are plotted in Fig.~\ref{test_sigmas}.
As can be seen, we do not detect any significant systematic effect between the input and 
output velocity dispersions.
A similar result was found by Toloba et al.\ (2011)\nocite{T11} using a different code.
In particular, these authors found that for S/N(\AA)$>$50, the relative error in the derived
$\sigma$ was always lower than 10\% for $\sigma$ down to half the instrumental resolution.
\subsection{Stellar Radial Velocities and Velocity Dispersion}

All our galaxies show clear rotation, while the $\sigma$ profiles behave 
quite differently. 
The $\sigma$ variations with radius are very small for NGC~1672 and NGC~1365, as
found in previous studies (Ganda et al. 2006\nocite{Gan06}; P\'erez et
al. 2009) for late-type spirals.  For the early-type spirals NGC~1358
(S0) and NGC~1433 (Sab), the velocity dispersion profiles decline with
increasing radius, 
as has been found in many elliptical
galaxies (e.g., D'Onofrio et al. 1995\nocite{DO95};
S\'anchez-Bl\'azquez et al. 2006\nocite{SB06a}).  The galaxy NGC~1433
has a very strong local minimum in the velocity dispersion at the
position of the ring (Buta et al. 2007), indicating the colder nature
of this structure, but then the trend of $\sigma$ is 
to decrease towards the external
regions, as in NGC~1358.

In P\'erez et al. (2009), we found that {\it all} early-type spirals with
bars show disc-like structures in their centres, inferred from their 
kinematics. These
structures are characterised by a central velocity dispersion minimum 
($\sigma$-drops), central $\sigma$ plateaus, and/or a clear rotating
inner disc or ring, as derived by the line-of-sight position-velocity
diagram. For the present sample, a $\sigma$-drop is present in the
early-type spiral NGC 1365. 

\subsection{Gas Velocities and Velocity Dispersion}
\label{gas_velocity}
The  kinematics of the gaseous component have been derived using different emission lines.  
In principle,  the  H$\alpha$ emission could be dominated
by gas with different kinematics than that contributing to the [NII]
emission. However, we do not find any significant difference,
neither in the line-of-sight velocity nor in the velocity dispersion
measured for different emission lines, whether it is  H$\alpha$ or
[NII] (see Fig.~\ref{fig:comparison}). 
Gas always rotates faster and shows structures like 
wiggles (see also Ganda et al. 2006).
Deviations from circular motions (most likely caused by the bar)
are clearly evident in all the curves, while the radial velocities of the stellar components
are much smoother.

\section{Star Formation Histories}
\label{sec:sfh}
Disc galaxies are obviously composite systems in which  several
generations of stars contribute to the integrated
light spectrum. While for spheroidal systems, single stellar populations are, in
many cases, not a bad approximation of the star formation history, for
disc galaxies this is certainly not true.  In the last few years,
new models of stellar populations have been developed which
predict not only the strength of individual characteristics, but
the entire spectral energy distribution for a population of a
given age and metallicity (e.g. Vazdekis 1999\nocite{Vaz99}; Bruzual
\& Charlot 2003, BC03 hereafter\nocite{BC03}; PEGASE-HR (LeBorgne et al. 2004)\nocite{PEG-HR}, Coelho et
al. 2007\nocite{Coel07}; Vazdekis et al. 2010). This  has led to
the parallel development of numerical techniques to derive the star
formation history of galaxies using as much information as possible
from the spectra (Heavens et al. 2000; Panter et al. 2003;
Cid Fernandes et al.\ 2005; Ocvirk et al. 2006ab; Tojeiro et al. 2007; Koleva et. 2009).

We use here {\tt STECKMAP} (STEllar Content and Kinematics via Maximum a Posteriori, Ocvirk et al. 2006a,b
\nocite{Ocv06a,Ocv06b}), along with 2 sequences of stellar population models taken from Bruzual \& Charlot (2003) and Vazdekis et al. (2010), 
spanning an age range $10^8 - 1.7.10^{10}$ yr divided in 30 logarithmic age bins, and a metallicity range [Z/H]=[+0.2, -1.3].
{\tt STECKMAP} is a Bayesian method that simultaneously
recovers the kinematic and stellar population properties via a maximum {\it a posteriori}
algorithm.
It has been extensively tested and used in a variety of applications, ranging from globular clusters 
(Sharina \&  Davoust 2008; Ocvirk 2010)\nocite{sharina09,ocvirk2010} to stripped galaxies
 (Pappalardo et al. 2010)\nocite{pappalardo2010}. It is a public tool and can
be 
obtained at \url{http://astro.u-strasbg.fr/~ocvirk/}. We note also that an online service 
has been recently launched, allowing the user to submit their spectrum, run the code on the remote server, 
and collect the results. The latter can be accessed at \url{http://astar.aip.de:20202/steckmap}.
The method is not parametric and does not make any {\it a priori} assumption 
regarding the star formation history. The only condition that {\tt STECKMAP} imposes is that the 
different unknowns,  namely the stellar age distribution (hereafter, LIGHT),  the  age-metallicity 
relation (hereafter, AMR), 
 the  line-of-sight velocity distribution (LOSVD) or the broadening 
function (hereafter, BF), have to be smooth in order to avoid extreme oscillating solutions 
that are not robust and most likely unphysical. 
The objective function, i.e., the function to minimise, is defined as:
\begin{equation}
Q\mu=\chi^2(\M{s}(\M{x,Z,g})) + P_\mu(\M{x,Z,g}),
\end{equation}
which is a penalised $\chi^2$, where $\M{s}$ is the model spectrum resulting from the 
LIGHT $\M{x}$, the AMR $\M{Z}$, and the BF $\M{g}$. The penalisation $P_\mu$ can be written as: 
$P_\mu(\M{x,Z,g})=\mu_{\M{x}}P(\M{x})+\mu_{\M{Z}}P(\M{Z})+\mu_{{\rm{v}}}P(\M{g})$, where the function $P$ gives 
high values for solutions with strong oscillations (ie. LIGHT and AMR changing rapidly with time or noisy BF)
and small values for smoothly varying solutions.
As explained in Ocvirk et al. (2006a), adding the penalisation $P$
to the objective function is exactly equivalent to injecting {\it a priori} 
information  into the problem.
In practice, this is like imposing an {\it a priori} probability density to the solution as 
$f_{\rm prior}(\M{x}) = \exp(-\mu_{\M{x}}P(\M{x}))$ (see Ocvirk et al.\ 2006 for more details).

For this work, we define $P$ as a quadratic function of the unknown $\M{x}$, involving a 
kernel $\M{L}$, as in Eq. (29) of Ocvirk et al. (2006a):

\begin{equation}
P(\M{x})=\T{\M{x}}\cdot \T{\M{L}} \cdot \M{L} \cdot \M{x}.
\end{equation}

We use a Laplacian smooting kernel for the LIGHT and BF (i.e. $\M{L}=\M{D_2}$), and a gradient kernel for the 
AMR (i.e. $\M{L}=\M{D_1}$), as in Ocvirk (2010)\nocite{ocvirk2010}, using the definitions of Ocvirk et al. (2006a).

Choosing the right values of the smoothing parameters $\mu_{\M{x},\M{Z},\rm{v}}$ is not a trivial problem. 
In principle, one could choose the values giving the smaller $\chi^2$ in
the fit, but this usually yields a wide range of smoothing parameters, spanning typically 3-4 decades, in which the
fit is acceptable. Although this affects the detailed shape of the LIGHT, AMR and BF, it does not impact the overall interpretation of the fit, and we checked via Monte Carlo simulations that the global shape of the solutions, and in particular the integrated quantities such as luminosity and mass-weighted ages and metallicities are stable for this range of smoothing parameters.
The final chosen values were $\mu_{\M{z}}=10^2$ and
$\mu_{\M{x}}=1$.

For this work, we have not fit
simultaneously the star formation histories and the kinematics,
and we use the values obtained with {\tt ppxf}. The reasons
are explained in detail in Appendix~\ref{a:sigma_z}. Basically, the 
existing degeneracy between the metallicity and the velocity 
dispersion (Koleva et al. 2007)  biases the mean-weighted metallicites
if both parameters are fitted at the same time. 
We found that this degeneracy affects mostly the mass-weighted values 
of the metallicity (see Appendix), in the sense that  the derived mass-weighted 
metallicity values are higher if the kinematics are not fixed. 

Error bars were calculated by means of Monte Carlo simulations
where each pixel is perturbed randomly according to the error
spectrum, and assuming a Gaussian distribution. 
However, the errors calculated in this manner are very
small, due to the high signal-to-noise ratio of our data.
For this high S/N data, the errors in the theoretical models
have usually a stronger impact on the solution (Ocvirk et al. 2006a).
Several studies have now compared the results in the star formation
histories recovered using different models (Koleva et
al.\ 2008\nocite{Kol08}; Coelho, Mendes de Oliveira \& Cid Fernandes
2009\nocite{Coel09}; Gonz\'alez 2009 \& Cid Fernandes 2010). In
particular, Koleva et al. (2008) found that the success in the
recovery of the age and metallicity of a single synthetic population depends critically on the atmospheric
parameter coverage of the stellar library used in the SSP models. They
found that results using MILES and PEGASE-HR are very
consistent for ages larger than 1 Gyr and $[{\rm Z}/{\rm H}]>-1$, while the results using BC03 differ at metallicities far
from solar.

For this reason, we make use of  several different stellar population models and
check the consistency of our results to this choice.
We employ 
both the new stellar population models by Vazdekis et al. (2010, V10
hereafter) that include  the  new and improved stellar library MILES
(S\'anchez-Bl\'azquez et al. 2006\nocite{SB06}; Cenarro et
al. 2007\nocite{Cen07} and Bruzual \& Charlot (2003, BC03 hereafter).


In order to deal with possible flux calibration errors, we multiply 
the model by a
smooth non-parametric transmission curve, representing 
the instrumental response 
multiplied by the interstellar extinction. This curve has 30 nodes spread uniformally along the 
wavelength range, and the transmission curve is obtained by spline-interpolating between the nodes. 
The latter are treated as additional parameters and adjusted during the minimisation procedure. This
 continuum
matching technique is similar in essence to the multiplicative polynomial used 
by {\tt NBursts}
(Chillingarian et al. 2007)\nocite{NBURSTS} or {\tt ULySS}
 (Koleva et al.\ 2009b). It has also been successfully used by Ocvirk (2010) and will be detailed in a 
forthcoming paper.

\begin{figure}
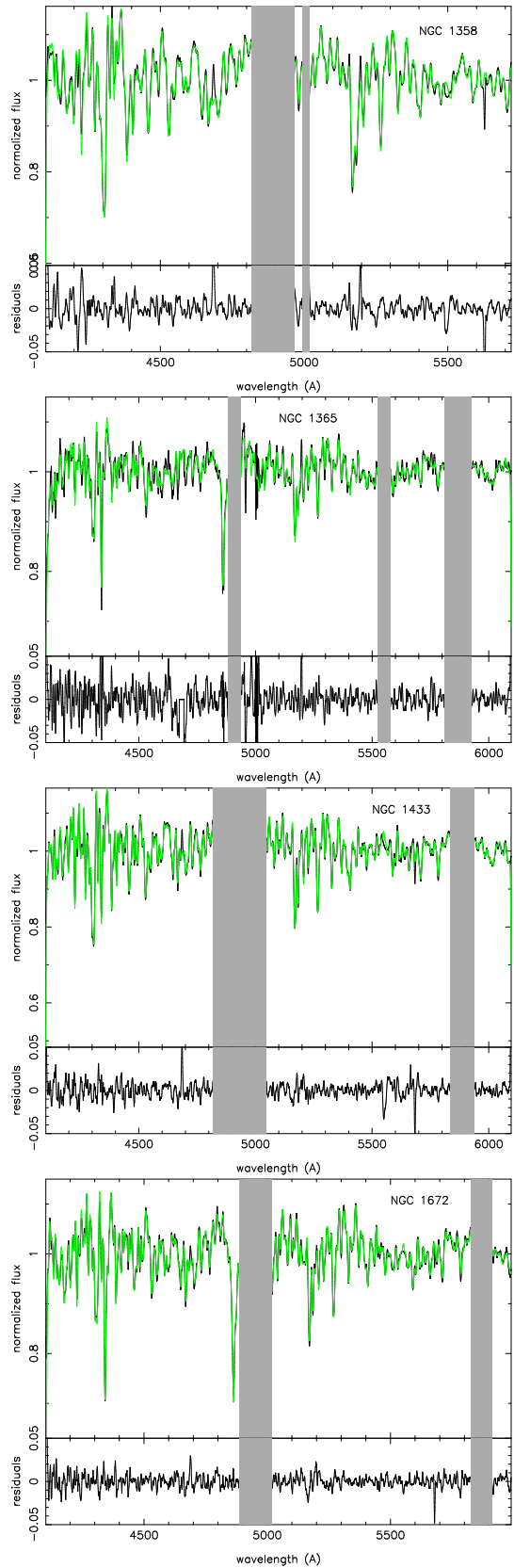

\centering
\resizebox{0.4\textwidth}{!}{\includegraphics[angle=-90]{n1358.bestfit2.ps}} 
\resizebox{0.4\textwidth}{!}{\includegraphics[angle=-90]{n1365.bestfit2.ps}} 
\resizebox{0.4\textwidth}{!}{\includegraphics[angle=-90]{n1433.bestfit2.ps}} 
\resizebox{0.4\textwidth}{!}{\includegraphics[angle=-90]{n1672.bestfit2.ps}} 
\caption{Central spectrum of the galaxies (black line) with the 
best fit  obtained
by {\tt STECKMAP} (green line). The grey areas  show  the masked parts of the spectra: two 
of them correspond to the gap of the detectors while others mask areas of the spectrum
affected by sky residuals. \label{spectra.fit}}
\end{figure}

We fit the wavelength range $\lambda\lambda$4150-6100~\AA~ masking the regions
corresponding to the detector gaps and masking other regions where the
spectrum was affected by sky residuals. In some cases, the TiO
molecular band was also  masked,  as the models did not seem 
to fit it properly.  Figure~\ref{spectra.fit} shows the central
spectrum of the four galaxies analysed in this study, together with the
best fit obtained by {\tt STECKMAP} using the V10 stellar population
models. The underlying continuum shape has been removed from  
both the model and the data.

\begin{figure*}
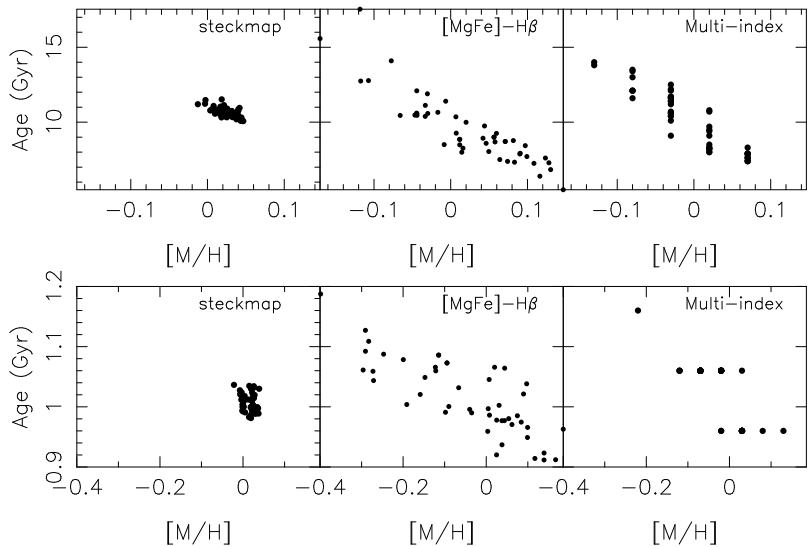

\resizebox{0.6\textwidth}{!}{\includegraphics[angle=-90]{agez_degeneracy.ps}} 
\resizebox{0.6\textwidth}{!}{\includegraphics[angle=-90]{agez_degeneracy2.ps}}
\caption{Measured age-metallicity using three different techniques (as indicated in the labels),
on two synthetic spectra with ages 1 (bottom panel) and 10 Gyr (upper panel) and solar metallicity, 
perturbed with Gaussian 
noise expected
for a spectrum with a S/N(\AA)~50.\label{age-meta.degeneracy}}
\end{figure*}

One of the most important problems affecting the stellar population studies is the 
well-known age-metallicity degeneracy, i.e., the colours and spectral characteristics
of a given population can be mimic with another population younger but more metal rich 
or older but more metal poor. 
To see the extent by which 
the results obtained with  {\tt steckmap} are  affected by this degeneracy we use a 
10~Gyr and a 1~Gyr solar metallicity SSP from 
Vazdekis et al. (2010) and added noise to obtain a spectrum with a signal-to-noise per \AA~ of 50. Then
we performed 50 Montecarlo simulations in which each pixel was perturbed with the associated 
noise error, following a Gaussian distribution. For each simulation a new value of the age and metallicity 
was derived using  three different techniques: (1) spectral fitting with {\tt steckmap}; (2) classical index-index
diagram fit (using [MgbFe]-H$\beta$ diagram) and (3) multi-index fitting technique (following Proctor \& Sansom 2002).
The techniques (2) and (3) have been extensively used by several authors to derive their stellar population 
SSP-equivalent parameters (e.g., Trager et al. 2000; Kuntschner et al. 2006; S\'anchez-Bl\'azquez et al. 2006, 2007;
Paudel et al. 2010, among many others).
The results are displayed in Fig~\ref{age-meta.degeneracy}. 
It can be seen that the age-metallicity degeneracy is 
very much reduced using {\tt steckmap} compared with the other two 
widely-used techniques. 
\section{Results}

\subsection{Stellar Population Gradients}
\label{sec:results:grad}
\begin{figure*}
\resizebox{0.3\textwidth}{!}{\includegraphics[angle=-90]{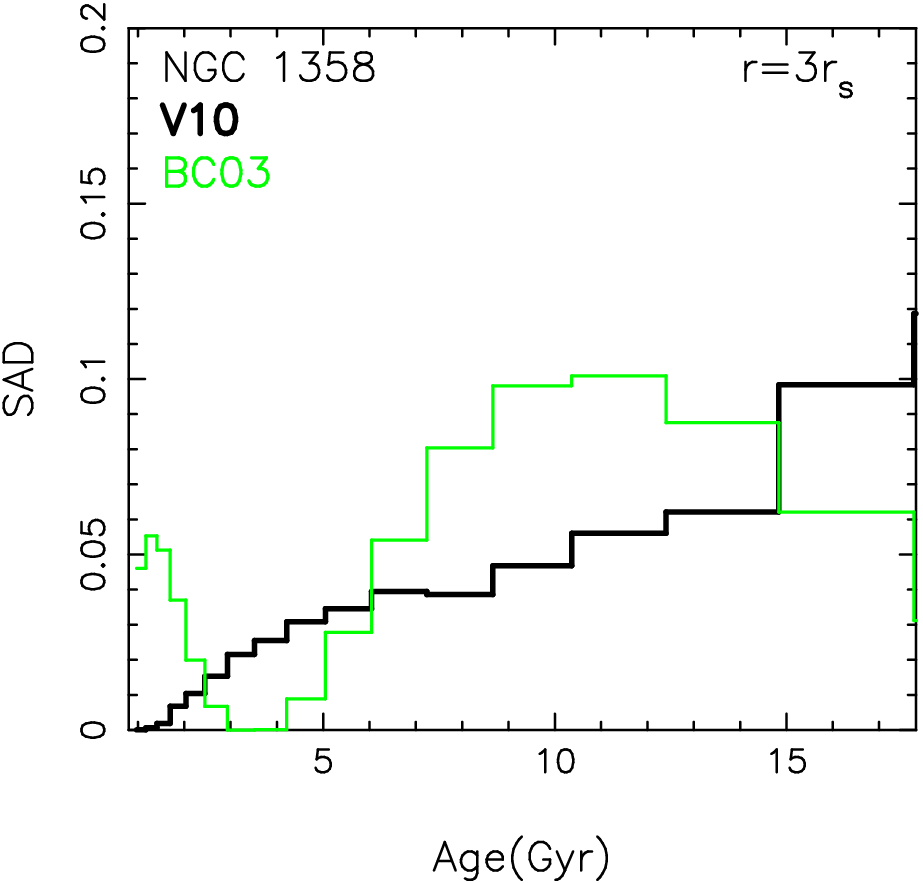}} 
\resizebox{0.3\textwidth}{!}{\includegraphics[angle=-90]{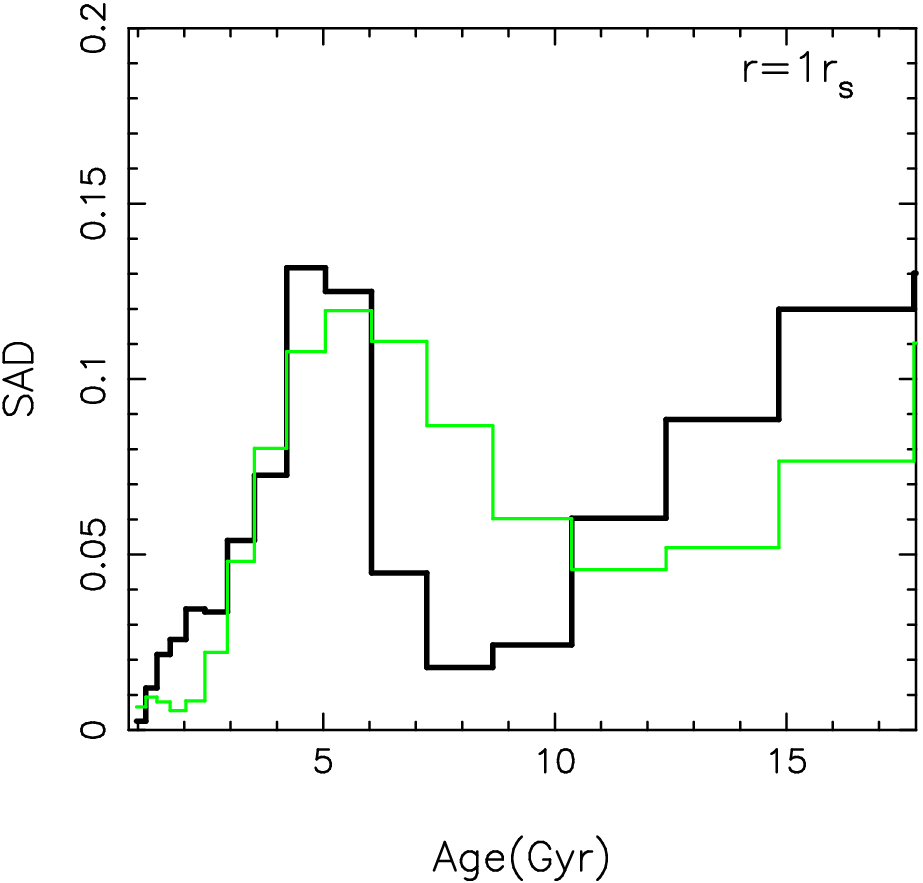}}
\resizebox{0.3\textwidth}{!}{\includegraphics[angle=-90]{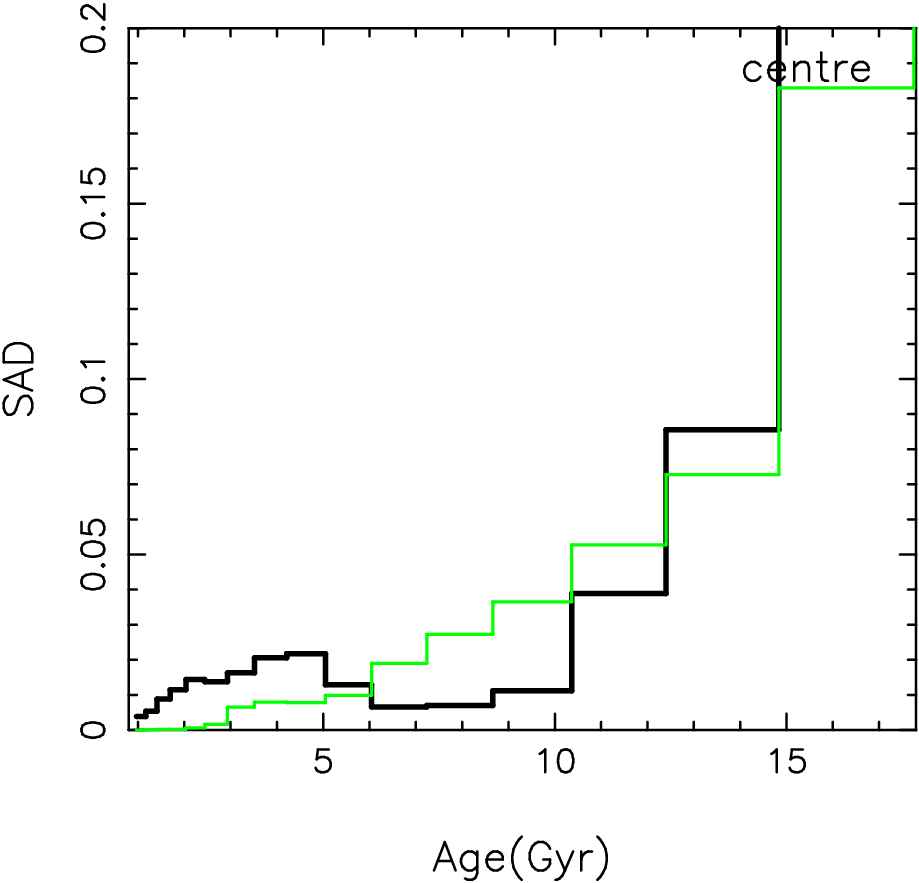}}

\resizebox{0.3\textwidth}{!}{\includegraphics[angle=-90]{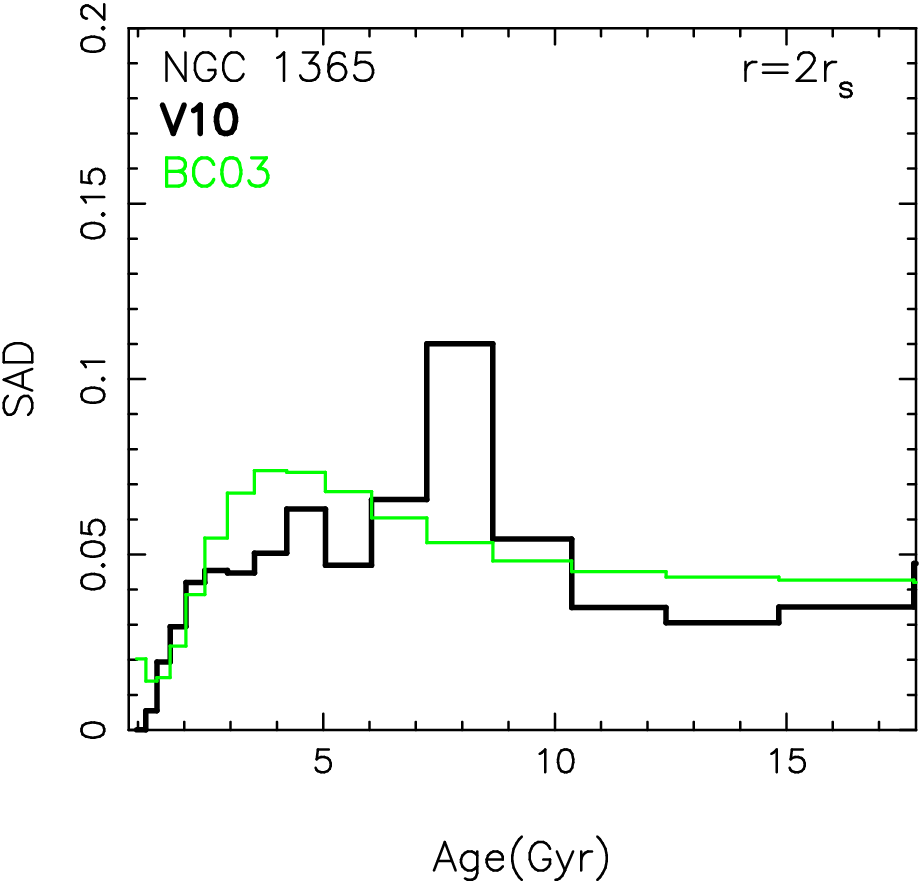}}
\resizebox{0.3\textwidth}{!}{\includegraphics[angle=-90]{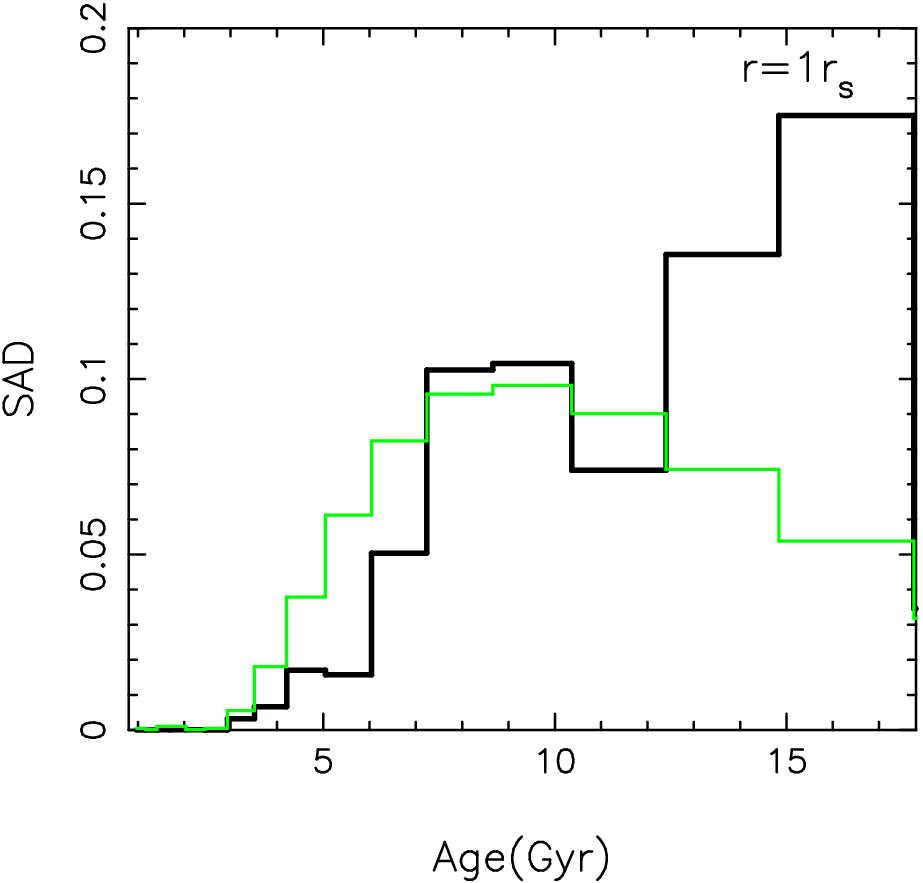}}
\resizebox{0.3\textwidth}{!}{\includegraphics[angle=-90]{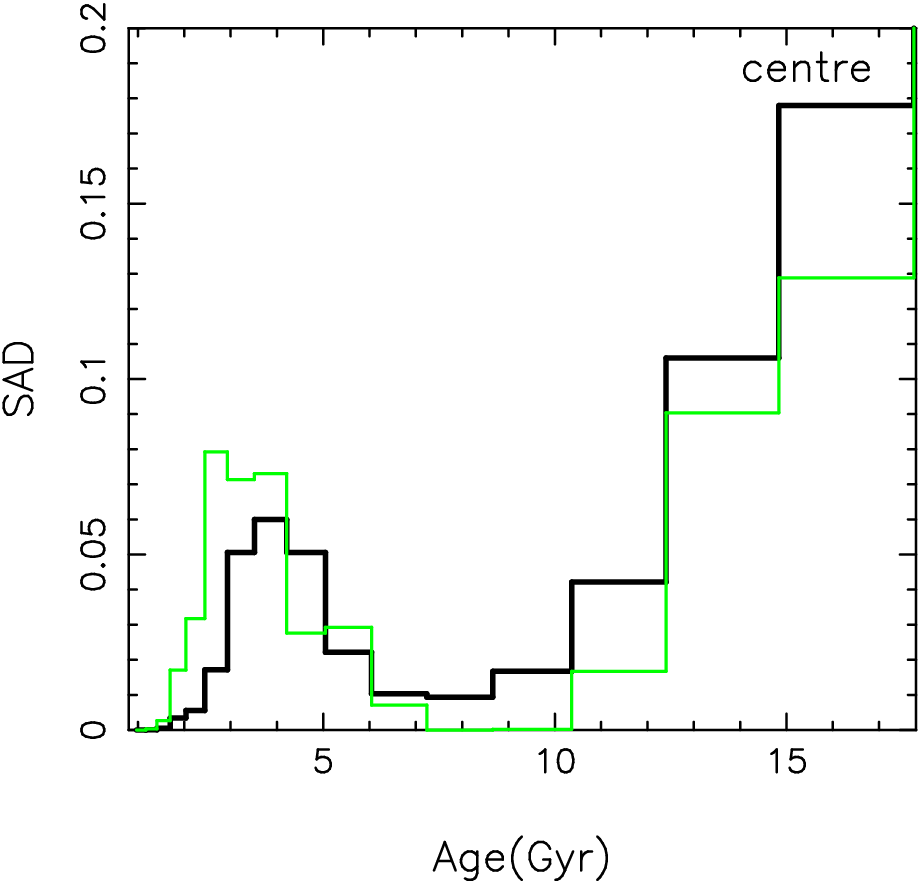}}

\resizebox{0.3\textwidth}{!}{\includegraphics[angle=-90]{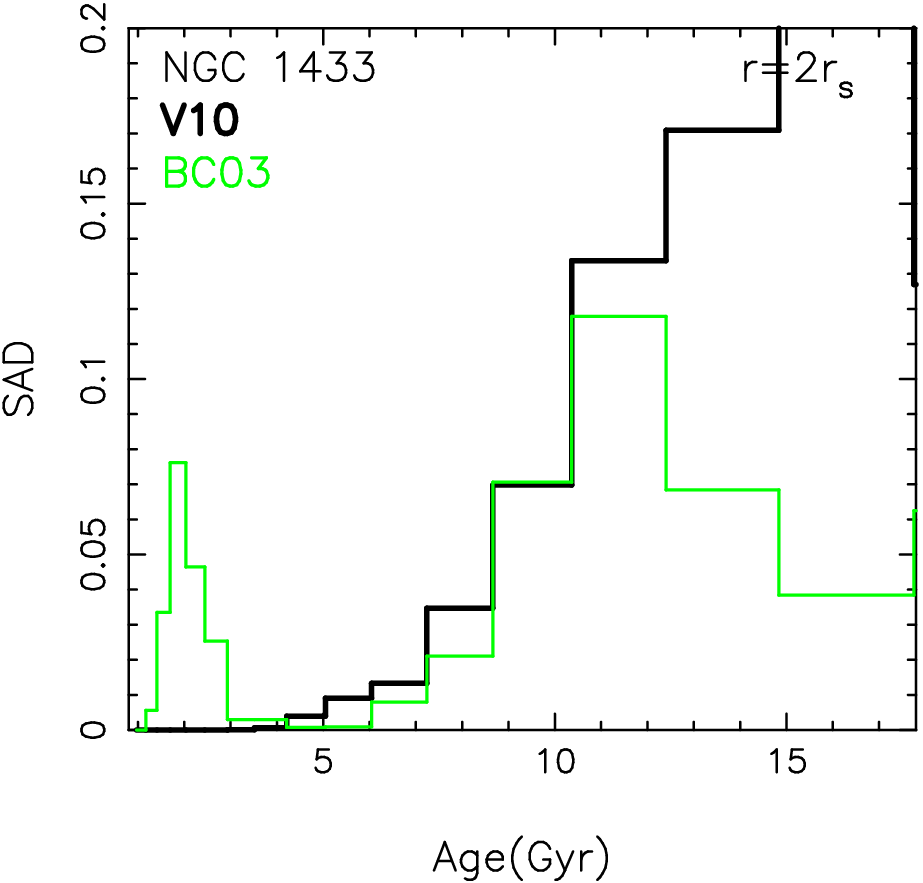}}
\resizebox{0.3\textwidth}{!}{\includegraphics[angle=-90]{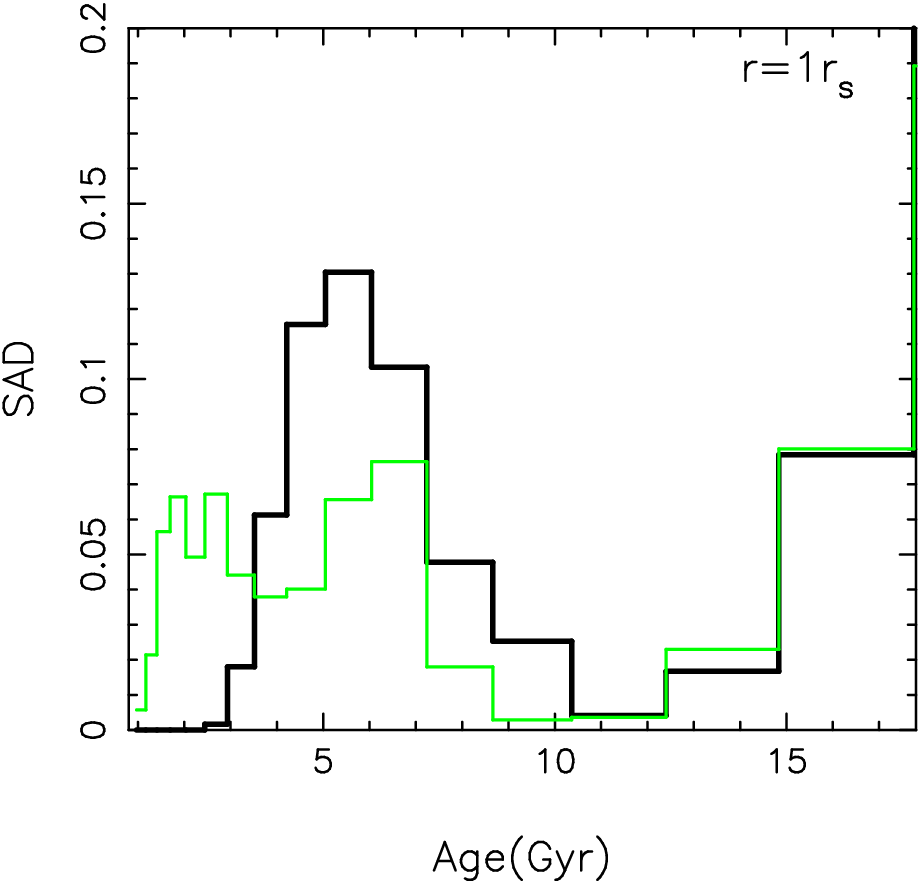}}
\resizebox{0.3\textwidth}{!}{\includegraphics[angle=-90]{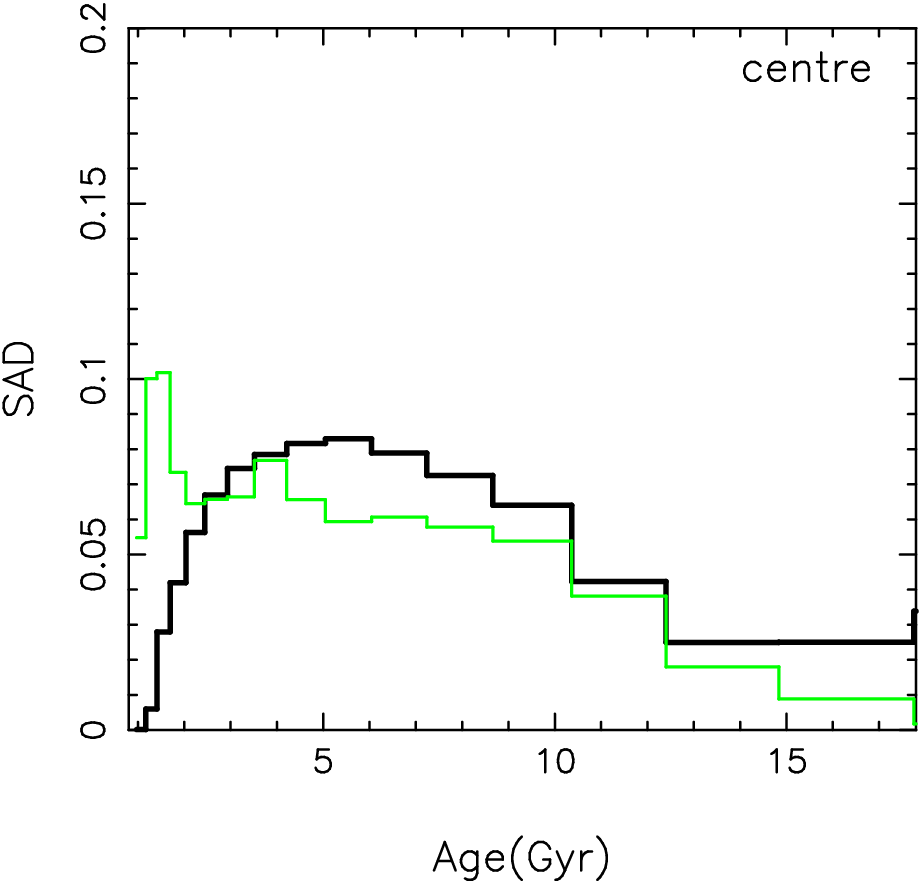}}

\resizebox{0.3\textwidth}{!}{\includegraphics[angle=-90]{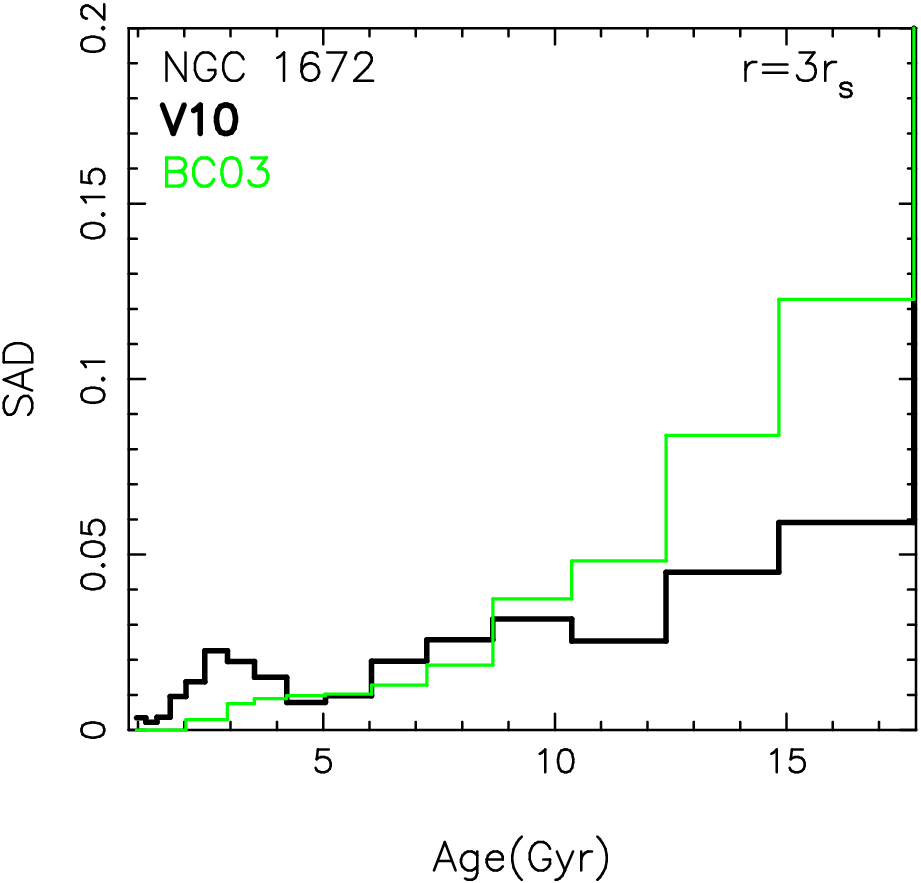}}
\resizebox{0.3\textwidth}{!}{\includegraphics[angle=-90]{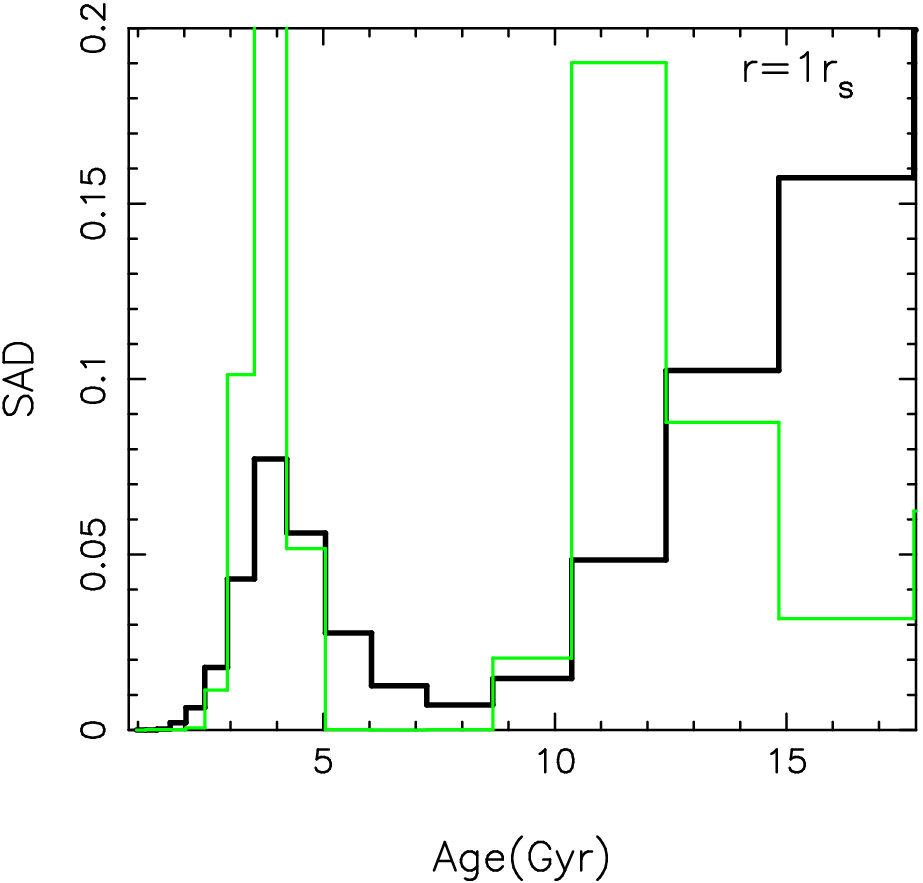}} 
\resizebox{0.3\textwidth}{!}{\includegraphics[angle=-90]{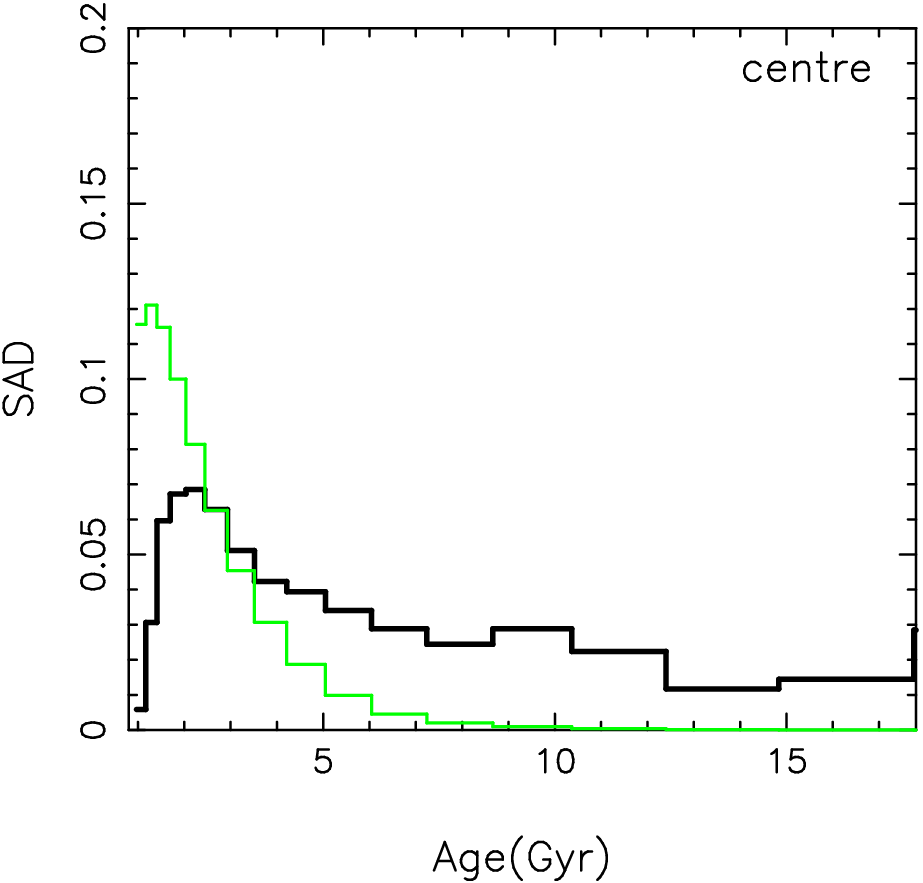}}
\caption{Stellar flux fractions at different ages calculated at different 
scalelengths of the disc.
Black lines indicate the results obtained when  the V10 models are used while in green we  have plotted
the results obtained with the B03 models. \label{sfh_examples}}
\end{figure*}

Figure~\ref{sfh_examples} shows some examples of the derived flux
fraction contributions of stars in the various age bins at different
radii for our sample of galaxies using both  sets  of models, V10 and
BC03.  In general, the star formation histories that we obtain are
compatible with being exponentially declining with timescales $\tau$ between 2
and 7 Gyr.  Secondary bursts of star formation are seen at several radii
within the sample.

\begin{figure*}
\resizebox{0.4\textwidth}{!}{\includegraphics[angle=-90]{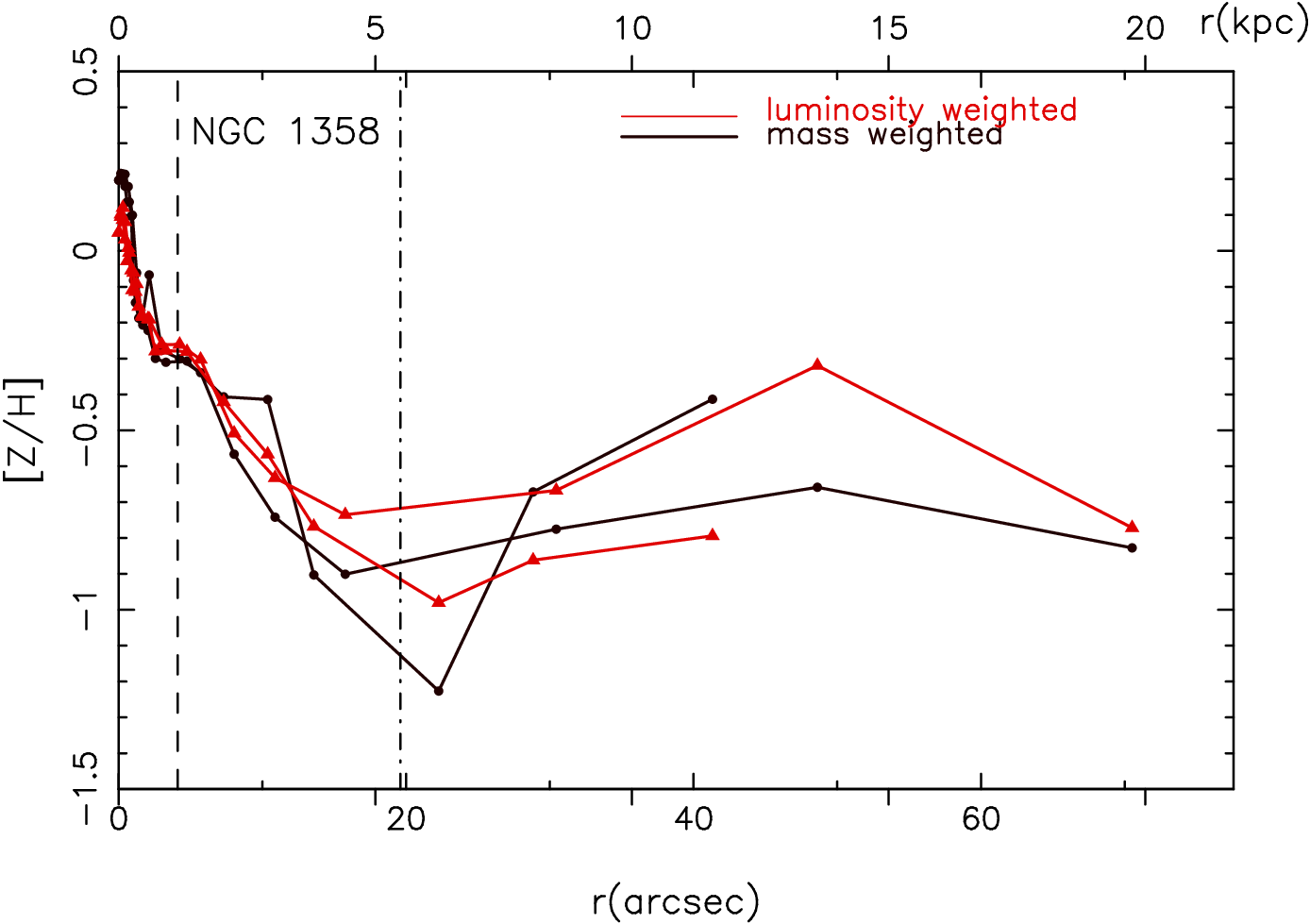}}
\resizebox{0.4\textwidth}{!}{\includegraphics[angle=-90]{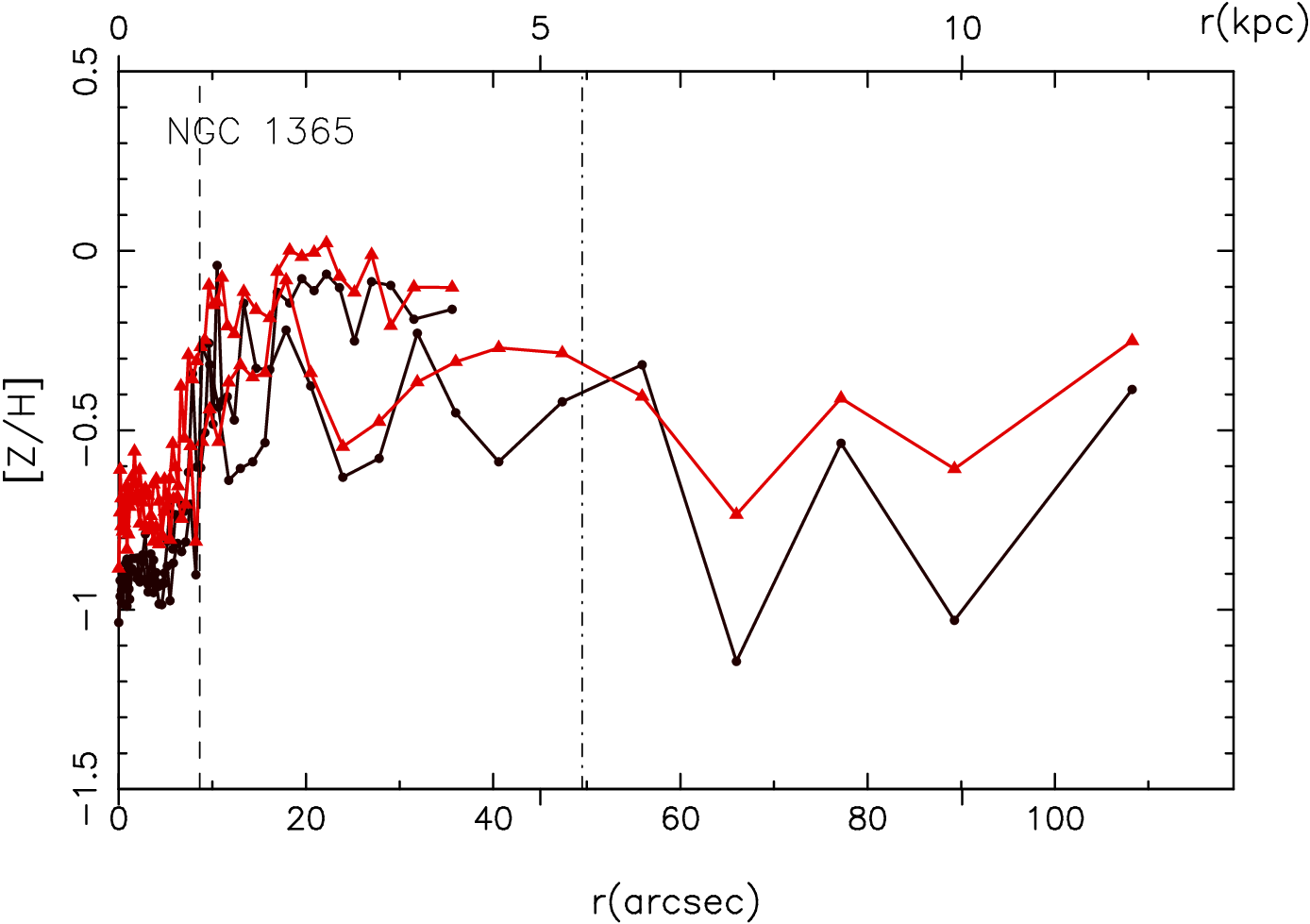}}
\resizebox{0.4\textwidth}{!}{\includegraphics[angle=-90]{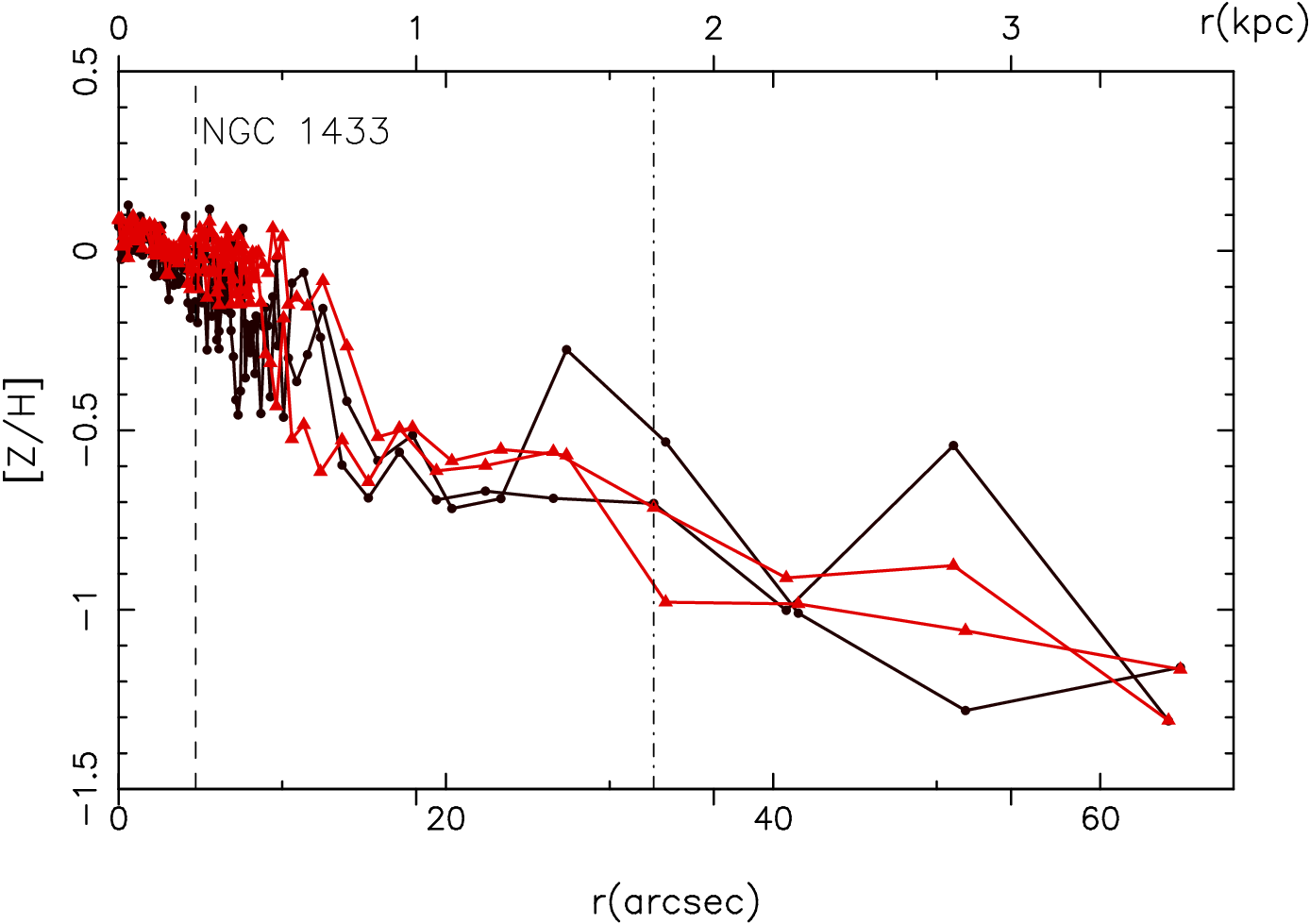}}
\resizebox{0.4\textwidth}{!}{\includegraphics[angle=-90]{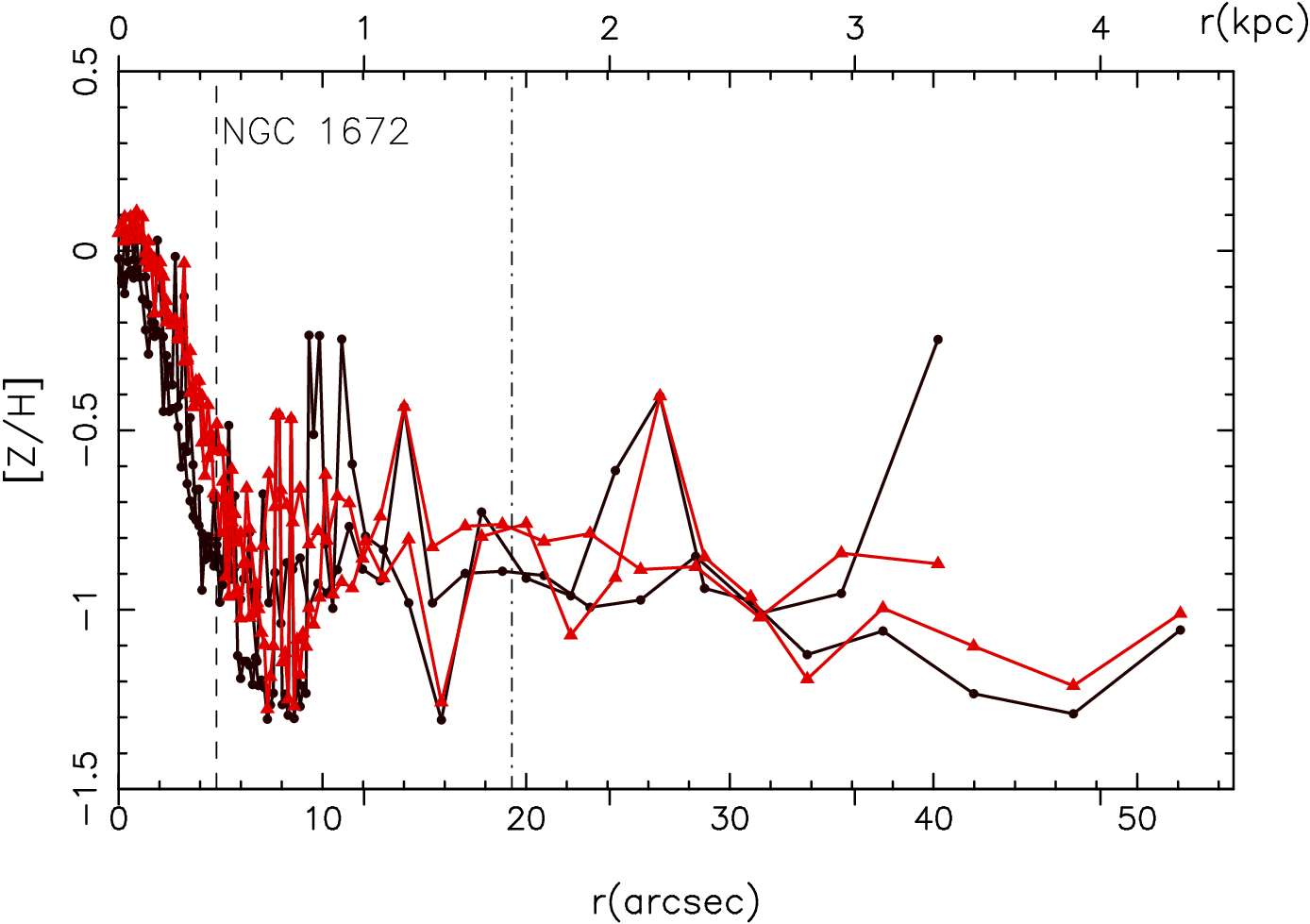}}
\caption{Mass- and luminosity-weighted  metallicity gradients derived from the 
recovered star formation history using V10 models.
Vertical dashed  lines indicate  the position of the effective radius of the bulge  
while dot-dashed lines show  the disc scale-length, obtained from the references indicated in Table~\ref{table1}.\label{age_z.weighted}}
\end{figure*}

\begin{figure}
\resizebox{0.5\textwidth}{!}{\includegraphics[angle=-90]{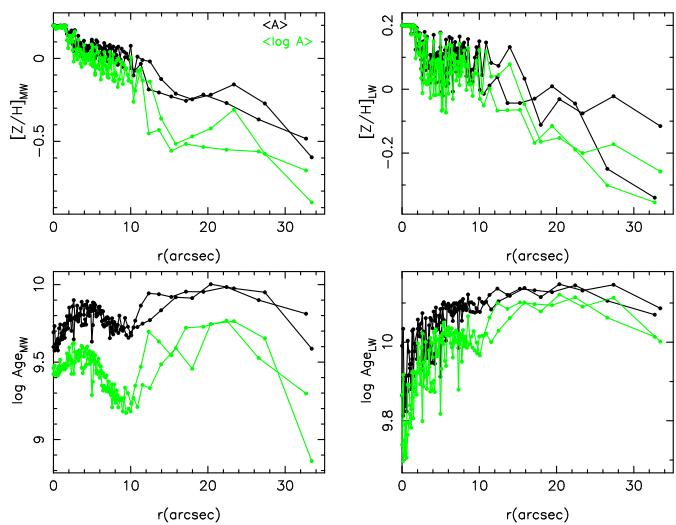}}
\caption{Comparison of the mass- and luminosity-weighted age
  and metallicity for NGC1433 when the means are performed  using a logarithmic
  scale (green line) and  using  a linear scale (i.e. $\log$ removed in Eq. \ref{e:LWMWdef}, black
  line).\label{comparison_lin_log}}
\end{figure}

To quantify the gradients in a way that can be compared with theoretical predictions, 
Figs~\ref{age_z.weighted} and \ref{age_z.weighted2}  show the mass- and luminosity-weighted  metallicities and
ages   as a function of  radius for our sample of galaxies obtained 
with V10 stellar population models.
The mass- and luminosity-weighted quantities are obtained respectively as:
\begin{equation}
\begin{array}{rcl}
\left\langle q \right\rangle_{\rm MW} &=& \frac{\sum_{i} {\rm{MASS}}(i) \log q_i}{\sum_{i} {\rm{MASS}}(i)} \, , \\
&& \\
\left\langle q \right\rangle_{\rm LW} &=& \frac{\sum_{i} {\rm{LIGHT}}(i) \log q_i}{\sum_{i} {\rm LIGHT}(i)} \, ,
\end{array}
\label{e:LWMWdef}
\end{equation}
where $q$ is the physical parameter we want to estimate, i.e., age
or metallicity, and ${\rm MASS}(i)$ and ${\rm LIGHT}(i)$ are respectively 
the reconstructed mass and flux contributions of the stars in the $i$-th age bin, 
as returned by {\tt STECKMAP}.
We have chosen here to weight both the ages and metallicities in a base-10 logarithmic scale,
as these values are more similar to the SSP-equivalent parameters (see Sec~/ref{sec:ssp}). 
The mean values would have changed if, instead, linear averages were
computed. As an example we  plot in  Fig.~\ref{comparison_lin_log} the
mass- and luminosity-weighted age and metallicities for NGC~1433
comparing the values when instead of log(age) and [Z/H], Z and age are
averaged.  As can be seen, there can be important differences in the
weighted values of age and metallicity depending on how the mean is
computed.  It is important then, to specify the way in which the average is 
calculated.

\begin{figure*}
\resizebox{0.4\textwidth}{!}{\includegraphics[angle=-90]{n1358.age.sfr.weighted6.ps}}
\resizebox{0.4\textwidth}{!}{\includegraphics[angle=-90]{n1365.age.sfr.weighted6.ps}}
\resizebox{0.4\textwidth}{!}{\includegraphics[angle=-90]{n1433.age.sfr.weighted6.ps}}
\resizebox{0.4\textwidth}{!}{\includegraphics[angle=-90]{n1672.age.sfr.weighted6.ps}}
\caption{ Mass- and luminosity-weighted age  gradients derived from the 
recovered star formation history. The meaning of the colors and symbols are the same 
as in Fig.~\ref{age_z.weighted}\label{age_z.weighted2}}
\end{figure*}

We explore, in appendix~\ref{a:comparisonmodels} if these gradients are model dependent. We found that for populations 
younger than $\sim$3~Gyr\footnote{We have not explored the whole parameter space because we are limited
by the values measured in our sample and, therefore, we cannot provide with a range of ages and metallicites
where the models agree or disagree as done, e.g., in Koleva et al. (2008)}, 
metallicities with BC03 tend to be lower and the ages
higher than those derived 
with V10. For the rest of the populations, the mean age and metallicity values are not dependent
of the model used. In our sample of galaxies, significant differences are only visible in the bulge
region of NGC~1433 and the ring region in NGC~1672.
In the figures, we indicate the position of the effective radius of the bulge and the scale-length 
of the discs (see Table~\ref{table1}). We now analyse the populations in these two regions separately.
\subsection{Bulge Region}
The stellar population parameter profiles in the bulge regions show a
variety of behaviour. The luminosity-weighted metallicity profile decreases in 3 of the 4
galaxies but increases in NGC~1365. The luminosity-weighted 
age profile in the bulge region also behaves differently depending on the galaxy, with profiles that do not 
always increase  or decrease monotonically.  The luminosity-weighted age
reflects  the presence of dynamically cold structures, where there has
been (or where there is currently) recent star formation,  manifesting itself as central
discs (e.g. in NGC~1433) or rings (NGC~1365 and NGC~1433 at r$\sim$7
and 10 arcsec respectively, and NGC~1672 at r$\sim$ 9 arcsec).  The
mass-weighted values are much flatter, indicating that the bulge
regions  are  dominated in mass by an old stellar population (10-15 Gyr),
although the central discs in NGC~1672 and NGC~1433 show mass-weighted
ages slightly younger ($\sim$6 Gyr).  Very similar conclusions were
reached by MacArthur et al. (2009) in their larger sample of galaxies,
where also a large variety of age and metallicity gradients were
found while the mass-weighted ages were always old.

Curiously, NGC~1365 shows a much lower metallicity in the bulge region
than the other galaxies of our sample.  NGC~1365 is an Sb galaxy with a star formation
history that can be reproduced with an exponentially declining
function with $\tau \sim 6$ Gyr at all  radii within the bulge region except in the nuclear ring (radius $\sim$ 7 arcsec)
where there is current star formation. 
The most puzzling aspect is that 
the  gas phase metallicity of this galaxy  is quite
high in this region, $\sim 12+\log O/H=9.15$
(Roy \& Wash 1997; Pilyugin, Contini \& Vilchez 2004).
Furthermore, contrasting the low stellar stellar metallicity values in the bulge region, 
this galaxy  has the largest metallicity of the sample in the disc region
(see next Section and Fig.~\ref{age_z.weighted2}). 
One possibility is that the metal poor stellar population in the center formed 
from low-metallicity gas that could have been both, accreted or driven from the external
parts of the galaxy to the circumnuclear region by the bar. Subsequently, the newly formed stars 
enriched the interestellar medium to the levels we observe now. To reach the 
level of enrichment that is observed, the star formation had to be rather intense.
It has been suggested that NGC~1365 has recently ($\sim$ 1 Gyr ago) experienced an interaction 
that made the disc unstable, producing the development of a  bar (Zanmar-S\'anchez et al. 2007). 
Inverse metallicity gradients  in the central 
parts of merger remnants has been observed in the other galaxies in the local Universe
(Rupke et al. 2010; Kewley et al. 2010).

\subsection{Disc Region}
\label{sec:test2}

In general, the inner disc region of all the galaxies (as in the bulge 
region) is dominated by an old
population (see Fig.~\ref{age_z.weighted2}). The mass-weighted age gradients in the 
disc region are always quite flat with a mean age around 10-12 Gyr.  The
luminosity-weighted age gradient is either flat or slightly negative
(at 2 scalelengths, the age decrease is 
4 Gyr for NGC~1672 and $\sim$4.7 Gyr for NGC~1358) consistent with an 
inside-out formation of the
disc.  

The mass- and luminosity-weighted metallicity
gradients show very similar slopes (see Fig.~\ref{age_z.weighted}), suggesting very little
evolution with time in the gradients of our galaxies.

An aside needs to be made here: as we mention in Sec.~\ref{sec:sfh}, the stellar population 
parameters have been calculated fixing the  kinematics  to the values 
obtained with {\tt GANDALF}.
We have noticed that when this is not done, a degeneracy between the $\sigma$ and metallicity 
appears, producing a  systematic effect for which luminosity weighted values  are  lower
than the mass-weighted ones (see Appendix~\ref{a:sigma_z}). This is true using
both, {\tt STARLIGHT} and 
{\tt STECKMAP}. This effect is also visible in some of the gradients shown by MacArthur et al. (2009).
We speculate that their mass-weighted metallicities  can be also biased due to this degeneracy.

\subsection{Comparison Between the Disk and the Bar Regions}
\label{sec:bars}
Bars are very important structures that can influence the evolution of
disc galaxies enormously. However, little is known about their 
own destiny. In particular, for our study, it is important to know if they can
survive for a  long time or  whether  they are easily destroyed. There is still no 
consensus in the literature about this point (Bournaud \& Combes 
2005; Shen \& Sellwood 2004; Berentzen et al. 2007). If bars are not
long-lasting structures, the fact that a galaxy does not have a bar
now, does not mean that it did not have it in the recent past. In this
case, the lack of differences in the stellar gradients would not mean
that the bar did not have an influence on  producing radial mixing.

\begin{figure*}
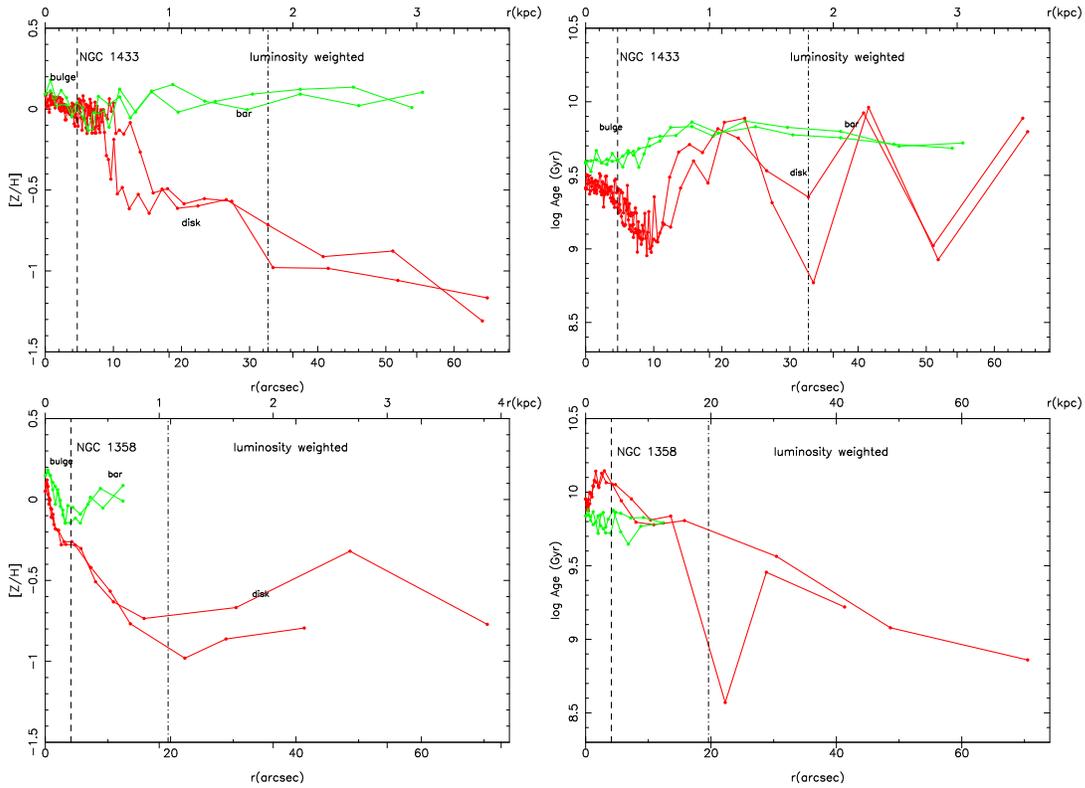

\resizebox{0.4\textwidth}{!}{\includegraphics[angle=-90]{n1433.z.lw.components.new.ps}}
\resizebox{0.4\textwidth}{!}{\includegraphics[angle=-90]{n1433.age.components.new.ps}}
\resizebox{0.4\textwidth}{!}{\includegraphics[angle=-90]{n1358.z.components.new.ps}}
\resizebox{0.4\textwidth}{!}{\includegraphics[angle=-90]{n1358.age.components.new.ps}}
\caption{Luminosity-weighted ages and metallicities  as a function of  radius measured 
along the bar (green line)
and along the major axis of the disc (red line). \label{ii.components}}
\end{figure*}

Two of the galaxies of the sample presented here, NGC~1358 and
NGC~1433 were studied by P\'erez et al. (2009)\nocite{PSZ09}. There,
we presented long slit spectroscopy along the bar of the
galaxies, finding three different types of bars according to their 
age and metallicity distribution as a function of
radius: (1) Bars with 
young/intermediate age and negative metallicity gradients. These
bars are present in galaxies with the lowest velocity dispersion 
in our sample; (2) Bars with zero metallicity gradient and younger
stellar populations at the end of the bars; and (3) Bars with old 
stellar populations and positive metallicity gradients. This type 
was found predominantly in those galaxies with a higher central velocity 
dispersion. NGC~1358 belongs to the third group while 
NGC~1433  is part of the second. 
One of the results from that study was that the majority  of the bars
in our sample (biased towards morphologically classified early-type spirals and 
S0) were composed of old stellar populations. 
However, the fact that the
bars  have old ages does not mean, necessarily, that the structure is
old, as the bar might have been formed, recently, out of old stars in
the disc. One way to disentangle these two options is comparing the
stellar population of the disc and the bar.
In P\'erez et al. (2009) we performed the analysis by comparing line-strength
indices with SSP models. 
There, we used the data to perform
an analysis similar to the one presented here and ran {\tt STECKMAP} on the 
spectra as a function of  bar radius.
Figure~\ref{ii.components} shows the comparison of the 
luminosity-weighted age and metallicity along the disc and the bar.  
As can be seen in 
the two studied galaxies, the stars in the bar are older and more
metal rich than those of the disc. Furthermore, the gradient in both
parameters is much flatter  in the bar. In general, the stellar content of the bar
is more similar to that of the bulge than to the disc.
This  confirms that, at least for some galaxies, the bars are, in fact, long-lasting structures.
The two galaxies for which we have been able to perform the comparison are very early-types.
It will be interesting to perform this comparison for more late-type spirals, as there must 
be a correlation between the age of the bars and the morphological type of the galaxy
(e.g., Gadotti 2008). 


\section{Line-strength Indices and Single Stellar Populations}
\label{sec:ssp}
Most of the studies of unresolved stellar populations have  not derived  
whole star formation histories, but SSP-equivalent ages and
metallicities using combinations of line-strength indices, most
commonly Lick/IDS indices (Trager et al. 2000a\nocite{T00a};
S\'anchez-Bl\'azquez et al. 2006b\nocite{SB06b}; Ganda et
al. 2007\nocite{Gan07}).  This approach has the advantage that  it allows one
to derive not only average ages and metallicities, but also to obtain
an estimation of the relative chemical abundance ratios, as the
sensitivity of individual Lick indices to the abundance of different
chemical species is different and has been calibrated by different
authors (Tripicco \& Bell 1995\nocite{TB95}; Korn, Maraston \& Thomas
2005\nocite{Korn05}; Houdashelt, Trager \& Worthey
2005\nocite{Houd05}). New models that predict the whole spectra 
for populations with different chemical composition are being built now
(e.g., Coehlo et al.\ 2009; Lee et al. 2009) and, in the future, it will be posible
to derive this information also from  full spectral fitting.
The index analysis  has the disadventage
that higher signal-to-noise ratio is  needed  
and, also, that if the sensitivity of the indices to changes
in the chemical abundances of different elements is not properly 
calibrated, the results might be biased depending on the indicator
used to perform the analysis (more than with the full spectral fitting). 
Furthermore, the comparison of indices
with SSP-models can only give us SSP-equivalent parameters and those
are degenerate to different star formation histories.
In this section, we derive SSP-equivalent ages, metallicities
and $\alpha$/Fe ratios by comparing Lick/IDS indices to the new models
of V10 in order to compare these values with the
luminosity- and mass-weighted values derived in
Sec.~\ref{sec:results:grad}.

The  canonical models of V10  were built using an empirical
stellar library (MILES, S\'anchez-Bl\'azquez et al. 2006) with
chemical abundance ratios matching those of the solar neighbourhood
and with isochrones assuming a solar partition for the different
elements. Because we want to   derive not  only global metallicities, but 
also chemical abundance  ratios, we modified the
predictions for individual indices for different patterns of chemical
abundances.  Following the approach of several previous authors (e.g.,
Tantalo, Chiosi \& Bressan 1998\nocite{Tan98}; Trager et
al. 2000a\nocite{T00a}; Thomas, Maraston \& Bender 2003\nocite{TMB03};
S\'anchez-Bl\'azquez et al. 2009b), we did not change the different
elements individually but assumed that some of them are linked
nucleosynthetically and, therefore, vary in lock-step.  We group the
different elements in enhanced (E hereafter), including N, Ne, Na, Mg, C,
Si and S (i.e., most $\alpha$-elements released in Type II SN plus N and C); 
depressed, including Ca and Fe, and unchanged (the rest), in
such a way that the global metallicity of the models remains
unchanged.  We developed models with enhancements from [E/Fe]=$-0.1$
to 0.5 in steps of 0.05.  In all cases, C is enhanced only half the
amount of the other enhanced elements. The reason for that is that
enhancements of the order of $\sim$0.3 dex bring the C/O ratio very
close to the values at which a carbon star is formed (see Houdsashelt
et al. 2002; Korn et al.\ 2005).  The response functions by Korn et
al. (2005) are calculated for variations in the abundances of
different elements of 0.3 dex.  The responses for different
 enhancements different from 0.3 dex were obtained using the procedures described in
S\'anchez-Bl\'azquez et al. (2009b) (originally described by Trager et
al. 2000).

We derived SSP-equivalent parameters age, [Fe/H] and
$\alpha$/Fe-abundance ratios for all the galaxies of the sample using the
$\chi^2$ minimization technique detailed in Proctor \& Sansom (2002)
and Proctor et al. (2004ab). The technique involves the comparison of
as many indices as possible to simple stellar population models.  We
find that it is extremely difficult to fit simultaneously all the
indices with a single stellar population, most likely because the star
formation histories of our sample cannot be approximated by an
SSP. 
After several tests we decided to avoid the bluest region of the
spectra (the H$\delta$ indices and the CN) because they deviated the
most in all the fitted spectra.  
These indices are more sensitive to the presence of young populations
than the rest of the Lick indices and, therefore, removing them from 
the fit we are biasing our equivalent ages toward older values.
Contrary to what was done in
Proctor et al. (2004ab)\nocite{Pro04b}\nocite{Pro04a} we did use the
same indices to fit all the spectra. This is because different indices
are sensitive to different populations and different chemical species
so if the galaxies do not share the same chemical abundance patterns as 
the models we are using, the results can be biased by using different
sets of indices at different radius.  Furthermore, the use of
different indicators to obtain SSP-equivalent parameters will give
different results in case the star formation history is  not 
a single burst (S\'anchez-Bl\'azquez et al. 2006b; Schiavon et
al. 2007\nocite{Schi07}).  For these reasons, all the indices that in
some of the galaxies of our sample fell in the gaps of the detectors
were eliminated from the fit.  In total, we use 10 Lick/IDS indices
(ie., all the Lick/IDS indices that our spectral range allowed us to
measure excluding those that fell in the gaps of the detector).

\begin{figure*}
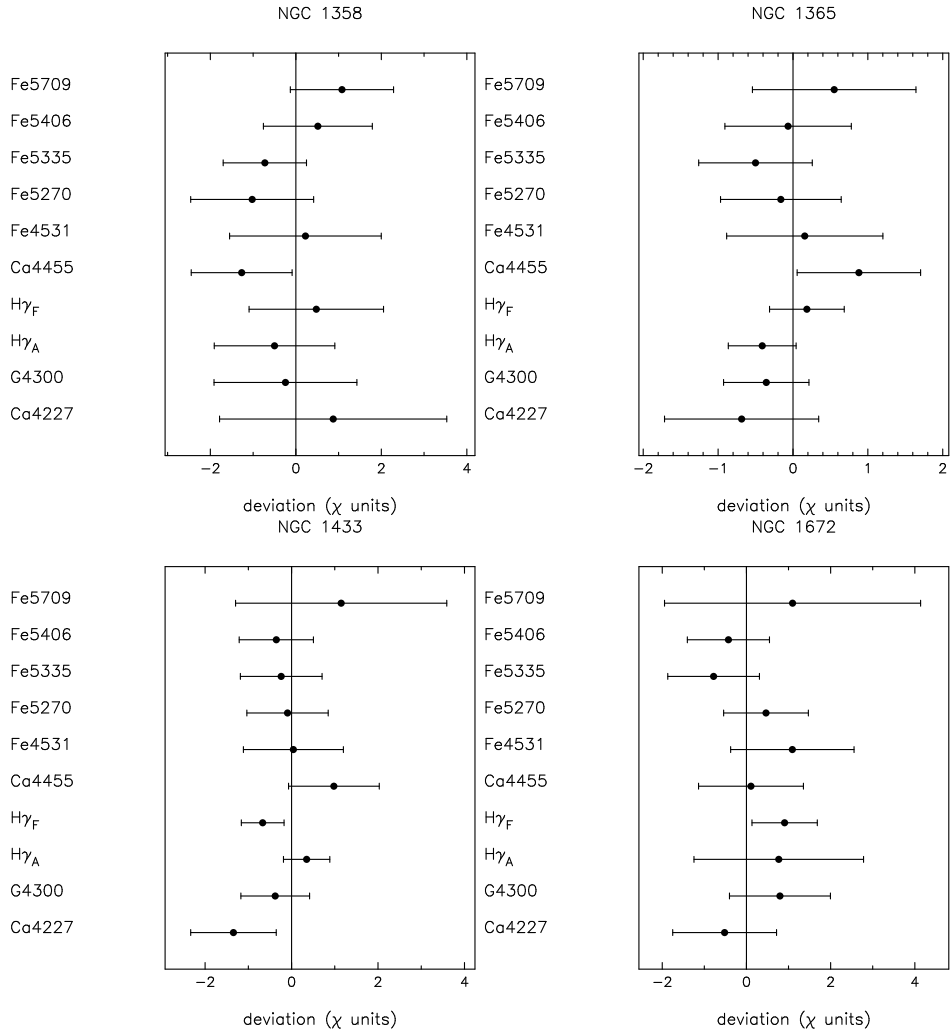

\resizebox{0.35\textwidth}{!}{\includegraphics[angle=-90]{n1358.chi.ps}}
\resizebox{0.35\textwidth}{!}{\includegraphics[angle=-90]{n1365.chi.ps}}
\resizebox{0.35\textwidth}{!}{\includegraphics[angle=-90]{n1433.chi.ps}}
\resizebox{0.35\textwidth}{!}{\includegraphics[angle=-90]{n1672.chi.ps}}
\caption{Average deviation in units of error (i.e., $\chi$) of the measured
indices from the indices corresponding to the best-fitting. Error bars represent
the scatter in the deviation.\label{ssp_residuals}}
\end{figure*}

Residuals from the best-fitting (observed value minus best fitting
value expressed in terms of the index errors) indices
for all the galaxies
are summarised in Fig.~\ref{ssp_residuals}.  The points represent the
average deviation from the best-fitting values while the error bars
show the RMS for all the points as a function of radius.
As can be seen, for the set of the indices that we have used, the 
 mean differences between the fitted and the measured indices are very
small. This indicates that despite that fact that the stellar populations of disc 
galaxies cannot be reproduced by  a single burst, most of the 
 Lick indices are sensitive to the same stellar populations  once the bluest indices
are excluded.

\begin{figure*}
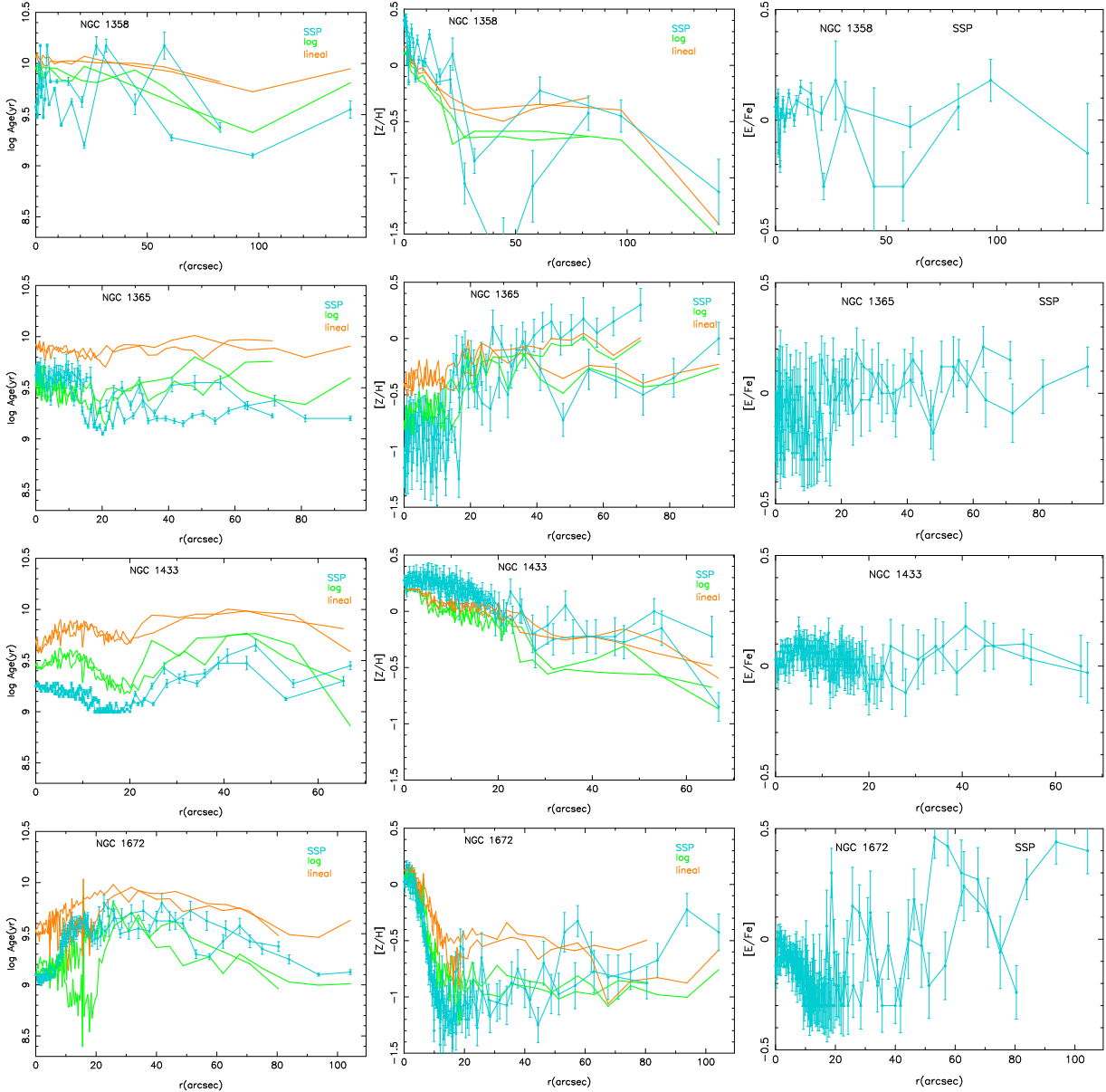

\resizebox{0.3\textwidth}{!}{\includegraphics[angle=-90]{n1358.age.ssp.comp.new.ps}}
\resizebox{0.3\textwidth}{!}{\includegraphics[angle=-90]{n1358.z.ssp.comp.new.ps}}
\resizebox{0.3\textwidth}{!}{\includegraphics[angle=-90]{n1358.en.ssp.new.ps}}

\resizebox{0.3\textwidth}{!}{\includegraphics[angle=-90]{n1365.age.ssp.comp.new.ps}}
\resizebox{0.3\textwidth}{!}{\includegraphics[angle=-90]{n1365.z.ssp.comp.new.ps}}
\resizebox{0.3\textwidth}{!}{\includegraphics[angle=-90]{n1365.en.ssp.new.ps}}

\resizebox{0.3\textwidth}{!}{\includegraphics[angle=-90]{n1433.age.ssp.comp.new.ps}}
\resizebox{0.3\textwidth}{!}{\includegraphics[angle=-90]{n1433.z.ssp.comp.new.ps}}
\resizebox{0.3\textwidth}{!}{\includegraphics[angle=-90]{n1433.en.ssp.new.ps}}

\resizebox{0.3\textwidth}{!}{\includegraphics[angle=-90]{n1672.age.ssp.comp.new.ps}}
\resizebox{0.3\textwidth}{!}{\includegraphics[angle=-90]{n1672.z.ssp.comp.new.ps}}
\resizebox{0.3\textwidth}{!}{\includegraphics[angle=-90]{n1672.en.ssp.new.ps}}

\caption{Single stellar population equivalent age, metallicity, and [E/Fe],
as a function of  radius for the galaxies of the sample.\label{ssp_gradients}}
\end{figure*}

Figure~\ref{ssp_gradients} shows the derived age, [Z/H] and [E/Fe]
profiles for our sample of galaxies. The shape of the profiles is
more similar to the luminosity-weighted than to the mass-weighted
values derived in the Sec.~\ref{sec:sfh}, as expected. However, the 
SSP-equivalent  ages  are, in general, younger. This is in
agreement with the results of Serra \& Trager (2007)\nocite{ST07} who
compared the SSP-equivalent ages and metallicities with the V-band
luminosity weighted ones (not exactly the same as the one derive
here) using artificial star formation histories of two instantaneous  bursts.
 Figure~\ref{ssp_gradients} also shows the mean linear-weighted values
compared with the log-weighted values. It can be seen that the log-weighted
values are more similar to the SSP-values, with a trend for the linear
weighted ages to be larger, as expected.

\begin{figure*}
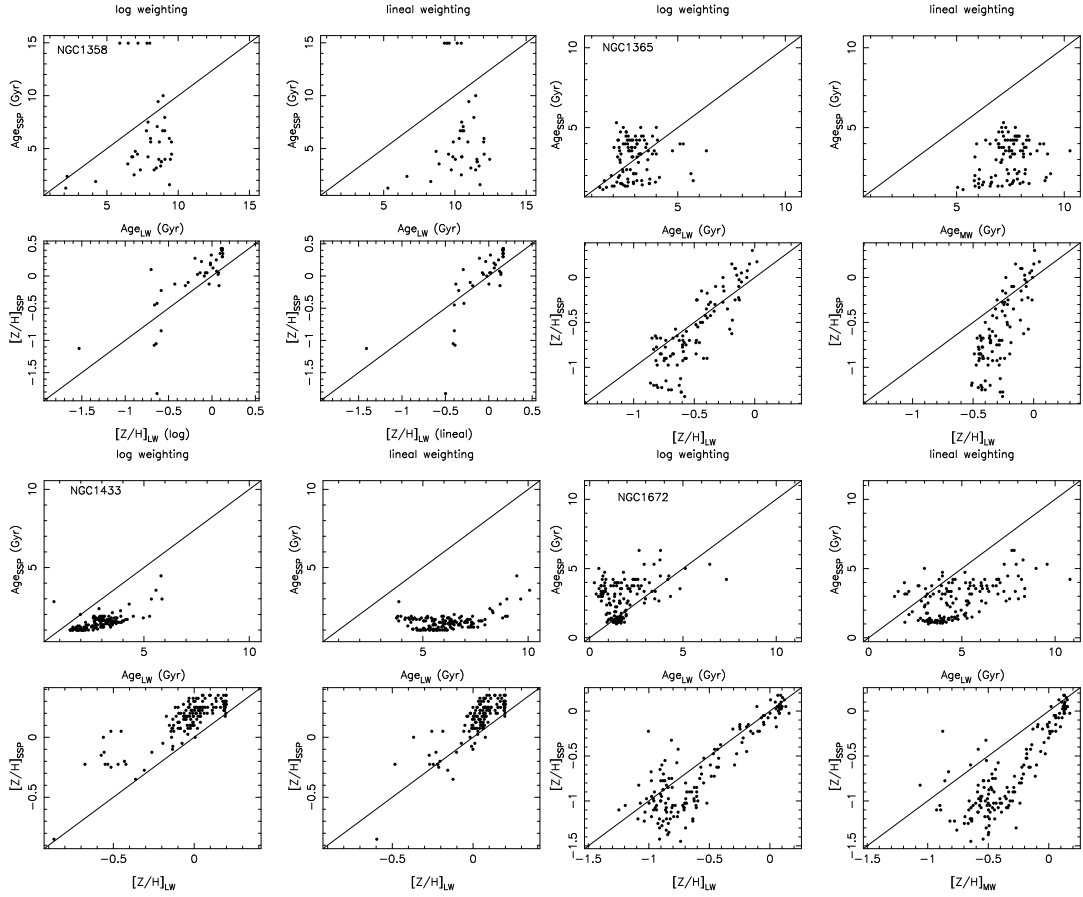

\resizebox{0.4\textwidth}{!}{\includegraphics[angle=-90]{n1358.compara_new.ps}}
\resizebox{0.4\textwidth}{!}{\includegraphics[angle=-90]{n1365.compara_new.ps}}
\resizebox{0.4\textwidth}{!}{\includegraphics[angle=-90]{n1433.compara_new.ps}}
\resizebox{0.4\textwidth}{!}{\includegraphics[angle=-90]{n1672.compara_new.ps}}
\caption{Comparison of the SSP-equivalent parameters and the luminosity- and mass-weighted values
for the 4 galaxies of this study.\label{compara_ssp.lw}}
\end{figure*}

Figure~\ref{compara_ssp.lw} shows a 1:1 comparison between the
SSP-equivalent ages and metallicities and the luminosity and
weighted- values for the different galaxies.  As can be seen,
luminosity-weighted values and SSP-equivalent values tend to agree
for old ages and high metallicities. As the ages get lower,
SSP-equivalent values are systematically smaller than the
luminosity-weighted and the same trend is observed for metallicities.
Mass-weighted ages, on the contrary  are  always higher than the
SSP-equivalent ones and much more constant, reflecting what we
mentioned before, that the galaxies are dominated in mass by an old
stellar population, independently of radius.  The mass-weighted
metallicity, however, follows a similar trend than the
luminosity-weighted one and, in the same way, agrees with the
SSP-equivalent one for high metallicity-values.

Several authors have analysed stellar populations of bulges using
line-strength indices and derived SSP-equivalent stellar population
parameters (e.g., Trager, Dalcanton \& Weiner 1999\nocite{TDW99};
Goudfrooij, Gorgas \& Jablonka 1999, Proctor \& Sansom 2002,
Falc\'on-Barroso, Peletier \& Balcells 2002; Morelli et al., Thomas \&
Davies 2006, Moorthy \& Holztman 2006; 
Peletier et al. 2007). Some of them have presented gradients as a function of  
radius. In general, it is found that bulges of galaxies have negative
metallicity gradients ($\sim -0.2$~dex), positive age gradients
(younger in the center) of $\sim$3-5 Gyr per decade of variation in
radius, and null or slightly positive [E/Fe] gradients (Jablonka et
al. 2007; Morelli et al. 2006). 
Moorthy \& Holtman (2006) found that  if  positive age gradients
(central regions being younger)  were found, they were invariably in barred
galaxies, although barred galaxies could also show negative gradients.
Jablonka et al. (2007), however,  found that for most galaxies of their sample,
the age gradient was positive --albeit always very small and, in many cases,
compatible with being null-- independently of them being barred-or not barred.
In fact, Jablonka et al. (2007) did not find any
difference between the stellar populations (SSP-equivalent parameters)
of barred and non-barred galaxies.  The differences between Jablonka et al. and Moorthy et al.\ studies
can be due to the
fact that Jablonka et al.\ (2007) observed  edge-on galaxies  and,
therefore, avoided the contribution of disc-like components such as bars,
rings or nuclear discs.
The results of Morelli et al. (2007) are consistent with Jablonka et al. (2007),
despite the fact that their galaxies are face-on. However, their sample is composed
of non-barred (or very weakly barred) galaxies and this limits the
number of subcomponents, such as rings and nuclear discs, present in the
sample\footnote{Although nuclear rings have been
detected  in non-barred galaxies as well, see Knapen (2010)\nocite{Kna10}.}. Furthermore, the sample
is biased towards early-type  systems.

We also found a variety of age gradients depending on the galaxy,
ranging from
flat (NGC~1365), to positive (NGC~1672), to negative (NGC~1433) and an
equivalent behaviour in the metallicity gradient.
We found that the variety of gradients in our
galaxies is due to the contrast produced by the presence of different
components,  such  as rings and nuclear discs.
As mentioned in
Sec.~\ref{sec:results:grad}, NGC~1365 presents a much lower metallicity in
the bulge region than in the rest of our galaxies (and the opposite is
true for the disc region).

We also find, in the bulge regions, a variety of [E/Fe] gradients, 
ranging from flat
(NGC~1358), to positive (NGC~1433), to negative (NGC~1672 and NGC~1365),
although the values at the very centre are always compatible with
being zero.
In particular, the decreasing [E/Fe] profile in NGC~1672 is
due to the contrast between the bulge region and the inner disc
region, where the  stellar populations  show a minimum in metallicity and
[E/Fe]. The contrast between these two regions 
is also responsible for the positive age gradient
and for the  very steep metallicity gradient in the
bulge region,  where the star
formation is proceeding  from low-metallicity  gas.
The higher values of [E/Fe] and metallicity in the central
region may be indicating that the star formation was more intense in
the central region than in the circumnuclear ring (r) .
The  negative age gradient in NGC~1365 and NGC~1433 is also due to the 
contrast between the age in the bulge and that of the nuclear ring
(r$\sim$19 arcsec for NGC~1433 and r$\sim$7 arcsec in NGC~1365).

\section{Conclusions and Discussion}

We are carrying out a project to understand the star formation history
of disc galaxies and, in particular, the influence of the bar on 
producing large scale mixing in the stellar disc.  In this paper we
have tested the techniques we are planning to use to derive star
formation histories and chemical evolution of disc galaxies using
high quality, long-slit spectroscopy, for a sample of  disc
galaxies covering a range  in  mass  and morphological type, with and without bars.

\begin{itemize}

\item To quantify the behaviour of the stellar population parameters we have derived 
the average values of age and metallicity weighting with the mass or the 
light of the stars. We discuss separately the results 
in the bulge and the disc region: 

{\bf Bulge Region:}
In the bulge region, we have found that the age gradient depends on
the presence of substructures such as circumnuclear discs or rings.  This
is in agreement with the results of Peletier et al. (2007), who
found, using 2D line-strength maps, that the young stars are found
either in rings or everywhere in the central regions. They also found
that the presence of young stars is  always associated with the presence of
dust.  This result is confirmed by studies of the morphology of star formation regions using H$\alpha$
emission, that find H$\alpha$ in circumnuclear
regions such as inner rings and that relate the presence 
of these structures with the dynamics of the host galaxy 
and its stellar bar (Knapen 2005; Knapen  et al. 2006).
We believe, in agreement with Peletier et al. (2007), 
 that the contradictory results obtained by different authors
in relation to the age of the bulges are due to differences in the sample 
selection (barred or not barred) and inclination. For example,
Jablonka et al. (2007) did not find any difference in the star
formation history of barred and non-barred galaxies because their
sample was edge-on and, therefore, they were not sensitive to the
presence of nuclear discs or rings. Morelli et al.\ (2006) found very
weak age gradients because their sample was selected to be
non-barred. On the contrary, Moorthy \& Holtman (2006) found different
age gradients in barred and non-barred galaxies. In particular, they
found that when positive gradients exist in their sample, they were
invariably in barred galaxies. 

The mass-weighted age in the bulge region is consistent with the disc
and takes values of $\sim$10  Gyr,  except in the very central region of
NGC~1672 and NGC~1433. The bulge, as the disc, is dominated by  old
stellar populations. 
Similar results were obtained by MacArthur et
al. (2009).

{\bf Inner Disc Region:}

\item The luminosity-weighted age gradient of the inner disc is always flat or slightly
negative, compatible with a moderate inside-out formation.  
This result agrees with that obtained by
MacArthur et al. (2009) for their sample of face-on galaxies or by
Mu\~noz-Mateos et al. (2007) for a sample of nearby disc galaxies
using a combination of UV and infrared colors.
However, we only reach  radii  beyond corotation for half of the sample
and, inside this radius, the stellar distribution and stellar population
gradients might be  significantly influenced by 
the bar potential.
The moderate age gradient can be 
explained either by  the disc not having
grown very much in the last $\sim$8 Gyr,
or if there has been radial mixing causing a migration of old stars to
the external parts.
   
All the discs of our sample are dominated, in mass, by an old
stellar population ($\sim$10 Gyr) and [M/H]$\sim -0.2$ dex. This is
true even at 3 scalelengths for some of our galaxies. This result
agrees with studies of resolved stellar population in nearby disc
galaxies like M81 (Williams et al.
2009\nocite{Will09a}) 
or NGC~300 (Gorgarten et al.\ 2010) where they found that, at 1.5-2
scalelengths, $\sim$70\% of the stars formed 8 Gyr ago.
However, in M33,  a similar disc in morphology and mass  seems to
contain a much larger fraction of young stars at those distances
(Williams et al. 2009). As Gogarten et al. (2010) suggest, these
differences may be due to environment, as M33 is
interacting with M31 (Braun \& Thilker 2004; Bekki 2008).  In
contrast, NGC~300 and M81 are  fairly  isolated  
systems.
  
\item Regarding the metallicity, we have found that the
luminosity-weighted and mass-weighted metallicity gradients are very
similar and always  
flat or slightly negative (NGC~1433) in the disc region.  The
mean metallicity in the disc is rather large, $-0.3<[M/H]<-0.2$ dex.
Together with the results of the item above, this indicates that most
stars at $\sim$2 scalelengths formed at z$>$1 and experienced a
rapid metal enrichment. These result coincide with those obtained
using resolved stellar populations in nearby discs -- e.g., in M81
(Williams et al. 2009) and M31 (Brown et al. 2006).

\item For two of the galaxies, NGC~1358 and NGC~1433 we have been able
to compare the metallicity and age gradient along the disc with that
along the bar, observed previously by P\'erez et al. (2009). We have
found that, in comparison with the stellar population in the disc,
bars show older ages, higher metallicities, high [E/Fe] abundances
and flatter gradients in
both parameters. This finding argues against a recent formation of
the bar from disc stars. As the stars in the bar are old, this
results is a clear indication that, at least in some galaxies -- the
two galaxies analysed here are both early-types-- this structure is
old, giving a definitive answer to a long-standing controversy
regarding the survivability of bars.   This was already  suggested in 
P\'erez et al. (2008, 2009) and
Gadotti \& Souza (2006).

\item We have found that the degeneracy between metallicity and $\sigma$ in 
methods where the kinematics is determined at the same time as  
the stellar population parameters can seriously bias the final results. 
In particular, the mass-weighted metallicity tend to be higher than 
what it should be. This is true, not only when using our code, but 
also when other methods, as {\tt starlight} is used.
Therefore, we recommend calculating the kinematics
separately by other methods and then fitting for
the stellar population parameters.
We also find that the age-metallicity degeneracy is highly reduced when 
{\tt steckmap} is used to derived these parameters as opposed to the most
classical index-index or multi-indices approaches.


\item Despite the strong degeneracy between the age and metallicity 
when these parameters are derived using line-strength indices, the 
gradients are not very different from the luminosity-weighted values 
obtained with {\tt steckmap}.

Previous studies using broadband colors (e.g. Bell \& de Jong 2000)
concluded that colour gradients were predominantly due to a gradient
in metallicity. MacArthur et al. (2004) however, found  both a 
gradient in age and metallicity and found that galaxies with stronger
metallicity gradients are those showing flatter age gradients and
vice-versa, showing the difficulty in breaking the age-metallicity-dust
degeneracy using only broad-band colours.  These different results do
not invalidate  a  specific  technique,  but show the importance of
interpreting correctly the results depending on the method used to
obtain stellar population parameters.
In a following paper we will apply the techniques presented in this paper to study the differences
in the star formation history of  barred and non-barred galaxies.
\end{itemize}

\section*{Acknowledgments}
PSB is supported by the Ministerio de Ciencia e Innovaci\'on (MICINN) 
of Spain through the Ramon y Cajal programme.
PSB also acknowledges a ERC within the 6th European Community Framework
Programme. 
PO is supported by a grant form the French Centre National de Recherche Scientifique.
BKG acknowledges
the support of UK Science \& Technology FacilitiesCouncil (STFC
Grant ST/F002432/1) and the Commonwealth Cosmology Initiative.
IP thanks the support of a postdoctoral
 fellowship from the Netherlands Organisation for Scientific Research (NWO, Veni-Grant 639.041.511) 
and she is currently supported by the Spanish Plan Nacional del Espacio del Ministerio 
de Educaci\'on y Ciencia(via grant C-CONSOLIDER AYA 2007-67625-C02-02). 
She also thanks  the Junta de Andaluc\'{\i}a for support through the  FQM-108 project.

\bibliographystyle{mn2e}
\bibliography{references}

\begin{thebibliography}{}

\bibitem[\protect\citeauthoryear{{Athanassoula}}{{Athanassoula}}{2003}]{Ath03}
{Athanassoula} E.,  2003, \mnras, 341, 1179

\bibitem[\protect\citeauthoryear{{Barbanis} \& {Woltjer}}{{Barbanis} \&
  {Woltjer}}{1967}]{BW67}
{Barbanis} B.,  {Woltjer} L.,  1967, \apj, 150, 461

\bibitem[\protect\citeauthoryear{{Barker} \& {Sarajedini}}{{Barker} \&
  {Sarajedini}}{2008}]{Bar08}
{Barker} M.~K.,  {Sarajedini} A.,  2008, \mnras, 390, 863

\bibitem[\protect\citeauthoryear{{Beauchamp} \& {Hardy}}{{Beauchamp} \&
  {Hardy}}{1997}]{BH97}
{Beauchamp} D.,  {Hardy} E.,  1997, \aj, 113, 1666

\bibitem[\protect\citeauthoryear{{Bell} \& {de Jong}}{{Bell} \& {de
  Jong}}{2000}]{BdJ00}
{Bell} E.~F.,  {de Jong} R.~S.,  2000, \mnras, 312, 497

\bibitem[\protect\citeauthoryear{{Binney} \& {Lacey}}{{Binney} \&
  {Lacey}}{1988}]{BL88}
{Binney} J.,  {Lacey} C.,  1988, \mnras, 230, 597

\bibitem[\protect\citeauthoryear{{Bruzual} \& {Charlot}}{{Bruzual} \&
  {Charlot}}{2003}]{BC03}
{Bruzual} G.,  {Charlot} S.,  2003, \mnras, 344, 1000

\bibitem[\protect\citeauthoryear{{Cappellari} \& {Emsellem}}{{Cappellari} \&
  {Emsellem}}{2004}]{CE04}
{Cappellari} M.,  {Emsellem} E.,  2004, \pasp, 116, 138

\bibitem[\protect\citeauthoryear{{Cardiel}}{{Cardiel}}{1999}]{Car99}
{Cardiel} N.,  1999, PhD thesis, , Universidad Complutense de Madrid, Spain,
  (1999)

\bibitem[\protect\citeauthoryear{{Carraro}, {Geisler}, {Villanova},
  {Frinchaboy} \& {Majewski}}{{Carraro} et~al.}{2007}]{Carr07}
{Carraro} G.,  {Geisler} D.,  {Villanova} S.,  {Frinchaboy} P.~M.,
  {Majewski} S.~R.,  2007, \aap, 476, 217

\bibitem[\protect\citeauthoryear{{Carraro}, {Ng} \& {Portinari}}{{Carraro}
  et~al.}{1998}]{Carr98}
{Carraro} G.,  {Ng} Y.~K.,    {Portinari} L.,  1998, \mnras, 296, 1045

\bibitem[\protect\citeauthoryear{{Cenarro}, {Peletier},
  {S{\'a}nchez-Bl{\'a}zquez}, {Selam}, {Toloba}, {Cardiel},
  {Falc{\'o}n-Barroso}, {Gorgas}, {Jim{\'e}nez-Vicente} \&
  {Vazdekis}}{{Cenarro} et~al.}{2007}]{Cen07}
{Cenarro} A.~J.,  {Peletier} R.~F.,  {S{\'a}nchez-Bl{\'a}zquez} P.,  {Selam}
  S.~O.,  {Toloba} E.,  {Cardiel} N.,  {Falc{\'o}n-Barroso} J.,  {Gorgas} J.,
  {Jim{\'e}nez-Vicente} J.,    {Vazdekis} A.,  2007, \mnras, 374, 664

\bibitem[\protect\citeauthoryear{{Chilingarian}, {Prugniel}, {Sil'Chenko} \&
  {Koleva}}{{Chilingarian} et~al.}{2007}]{NBURSTS}
{Chilingarian} I.,  {Prugniel} P.,  {Sil'Chenko} O.,    {Koleva} M.,  2007, in
  {Vazdekis} A.,  {Peletier} R.~F.,  eds, IAU Symposium Vol.~241 of IAU
  Symposium, {NBursts: Simultaneous Extraction of Internal Kinematics and
  Parametrized SFH from Integrated Light Spectra}.
pp 175--176

\bibitem[\protect\citeauthoryear{{Cioni}, {Irwin}, {Ferguson}, {McConnachie},
  {Conn}, {Huxor}, {Ibata}, {Lewis} \& {Tanvir}}{{Cioni} et~al.}{2009}]{Cio09}
{Cioni} M.,  {Irwin} M.,  {Ferguson} A.~M.~N.,  {McConnachie} A.,  {Conn}
  B.~C.,  {Huxor} A.,  {Ibata} R.,  {Lewis} G.,    {Tanvir} N.,  2009, \aap,
  500, 1025

\bibitem[\protect\citeauthoryear{{Coelho}, {Bruzual}, {Charlot}, {Weiss},
  {Barbuy} \& {Ferguson}}{{Coelho} et~al.}{2007}]{Coel07}
{Coelho} P.,  {Bruzual} G.,  {Charlot} S.,  {Weiss} A.,  {Barbuy} B.,
  {Ferguson} J.~W.,  2007, \mnras, 382, 498

\bibitem[\protect\citeauthoryear{{Coelho}, {Mendes de Oliveira} \& {Cid
  Fernandes}}{{Coelho} et~al.}{2009}]{Coel09}
{Coelho} P.,  {Mendes de Oliveira} C.,    {Cid Fernandes} R.,  2009, \mnras,
  396, 624

\bibitem[\protect\citeauthoryear{{Crocker}, {Baugus} \& {Buta}}{{Crocker}
  et~al.}{1996}]{Crock96}
{Crocker} D.~A.,  {Baugus} P.~D.,    {Buta} R.,  1996, \apjs, 105, 353

\bibitem[\protect\citeauthoryear{{de Jong}}{{de Jong}}{1996}]{dJ96}
{de Jong} R.~S.,  1996, \aap, 313, 377

\bibitem[\protect\citeauthoryear{{Debattista}, {Mayer}, {Carollo}, {Moore},
  {Wadsley} \& {Quinn}}{{Debattista} et~al.}{2006}]{Deb06}
{Debattista} V.~P.,  {Mayer} L.,  {Carollo} C.~M.,  {Moore} B.,  {Wadsley} J.,
    {Quinn} T.,  2006, \apj, 645, 209

\bibitem[\protect\citeauthoryear{{D'Onofrio}, {Zaggia}, {Longo}, {Caon} \&
  {Capaccioli}}{{D'Onofrio} et~al.}{1995}]{DO95}
{D'Onofrio} M.,  {Zaggia} S.~R.,  {Longo} G.,  {Caon} N.,    {Capaccioli} M.,
  1995, \aap, 296, 319

\bibitem[\protect\citeauthoryear{{Dumas}, {Emsellem} \& {Ferruit}}{{Dumas}
  et~al.}{2007}]{Dum07}
{Dumas} G.,  {Emsellem} E.,    {Ferruit} P.,  2007, in {M.~Kissler-Patig,
  J.~R.~Walsh, \& M.~M.~Roth} ed., Science Perspectives for 3D Spectroscopy {A
  3D View of the Central Kiloparsec of the Seyfert Galaxy NGC 1358}.
pp 269--+

\bibitem[\protect\citeauthoryear{{Edvardsson}, {Andersen}, {Gustafsson},
  {Lambert}, {Nissen} \& {Tomkin}}{{Edvardsson} et~al.}{1993}]{Ed93}
{Edvardsson} B.,  {Andersen} J.,  {Gustafsson} B.,  {Lambert} D.~L.,  {Nissen}
  P.~E.,    {Tomkin} J.,  1993, \aap, 275, 101

\bibitem[\protect\citeauthoryear{{Favata}, {Micela} \& {Sciortino}}{{Favata}
  et~al.}{1997}]{Fav97}
{Favata} F.,  {Micela} G.,    {Sciortino} S.,  1997, \aap, 323, 809

\bibitem[\protect\citeauthoryear{{Feltzing} \& {Gonzalez}}{{Feltzing} \&
  {Gonzalez}}{2001}]{FG01}
{Feltzing} S.,  {Gonzalez} G.,  2001, \aap, 367, 253

\bibitem[\protect\citeauthoryear{{Feltzing}, {Holmberg} \& {Hurley}}{{Feltzing}
  et~al.}{2001}]{Felt01}
{Feltzing} S.,  {Holmberg} J.,    {Hurley} J.~R.,  2001, \aap, 377, 911

\bibitem[\protect\citeauthoryear{{Flynn} \& {Morell}}{{Flynn} \&
  {Morell}}{1997}]{FM97}
{Flynn} C.,  {Morell} O.,  1997, \mnras, 286, 617

\bibitem[\protect\citeauthoryear{{Freeman} \& {Bland-Hawthorn}}{{Freeman} \&
  {Bland-Hawthorn}}{2002}]{FB02}
{Freeman} K.,  {Bland-Hawthorn} J.,  2002, \araa, 40, 487

\bibitem[\protect\citeauthoryear{{Friedli}}{{Friedli}}{1998}]{Fried98}
{Friedli} D.,  1998, in {D.~Friedli, M.~Edmunds, C.~Robert, \& L.~Drissen} ed.,
  Abundance Profiles: Diagnostic Tools for Galaxy History Vol.~147 of
  Astronomical Society of the Pacific Conference Series, {Mixing and Transfer
  of Elements by Triaxiality}.
pp 287--+

\bibitem[\protect\citeauthoryear{{Friedli}, {Benz} \& {Kennicutt}}{{Friedli}
  et~al.}{1994}]{Fried94}
{Friedli} D.,  {Benz} W.,    {Kennicutt} R.,  1994, \apjl, 430, L105

\bibitem[\protect\citeauthoryear{{Friel}, {Janes}, {Tavarez}, {Scott},
  {Katsanis}, {Lotz}, {Hong} \& {Miller}}{{Friel} et~al.}{2002}]{Friel02}
{Friel} E.~D.,  {Janes} K.~A.,  {Tavarez} M.,  {Scott} J.,  {Katsanis} R.,
  {Lotz} J.,  {Hong} L.,    {Miller} N.,  2002, \aj, 124, 2693

\bibitem[\protect\citeauthoryear{{Fuchs}}{{Fuchs}}{2001}]{Fuchs01}
{Fuchs} B.,  2001, \mnras, 325, 1637

\bibitem[\protect\citeauthoryear{{Ganda}, {Falc{\'o}n-Barroso}, {Peletier},
  {Cappellari}, {Emsellem}, {McDermid}, {de Zeeuw} \& {Carollo}}{{Ganda}
  et~al.}{2006}]{Gan06}
{Ganda} K.,  {Falc{\'o}n-Barroso} J.,  {Peletier} R.~F.,  {Cappellari} M.,
  {Emsellem} E.,  {McDermid} R.~M.,  {de Zeeuw} P.~T.,    {Carollo} C.~M.,
  2006, \mnras, 367, 46

\bibitem[\protect\citeauthoryear{{Ganda}, {Peletier}, {McDermid},
  {Falc{\'o}n-Barroso}, {de Zeeuw}, {Bacon}, {Cappellari}, {Davies},
  {Emsellem}, {Krajnovi{\'c}}, {Kuntschner}, {Sarzi} \& {van de Ven}}{{Ganda}
  et~al.}{2007}]{Gan07}
{Ganda} K.,  {Peletier} R.~F.,  {McDermid} R.~M.,  {Falc{\'o}n-Barroso} J.,
  {de Zeeuw} P.~T.,  {Bacon} R.,  {Cappellari} M.,  {Davies} R.~L.,  {Emsellem}
  E.,  {Krajnovi{\'c}} D.,  {Kuntschner} H.,  {Sarzi} M.,    {van de Ven} G.,
  2007, \mnras, 380, 506

\bibitem[\protect\citeauthoryear{{Hameed} \& {Devereux}}{{Hameed} \&
  {Devereux}}{1999}]{HD99}
{Hameed} S.,  {Devereux} N.,  1999, \aj, 118, 730

\bibitem[\protect\citeauthoryear{{Haywood}}{{Haywood}}{2001}]{Hay01}
{Haywood} M.,  2001, \mnras, 325, 1365

\bibitem[\protect\citeauthoryear{{Hook}, {J{\o}rgensen}, {Allington-Smith},
  {Davies}, {Metcalfe}, {Murowinski} \& {Crampton}}{{Hook} et~al.}{2004}]{H04}
{Hook} I.~M.,  {J{\o}rgensen} I.,  {Allington-Smith} J.~R.,  {Davies} R.~L.,
  {Metcalfe} N.,  {Murowinski} R.~G.,    {Crampton} D.,  2004, \pasp, 116, 425

\bibitem[\protect\citeauthoryear{{Houdashelt}, {Trager} \&
  {Worthey}}{{Houdashelt} et~al.}{2005}]{Houd05}
{Houdashelt} M.~L.,  {Trager} S.~C.,    {Worthey} G.,  2005, Highlights of
  Astronomy, 13, 585

\bibitem[\protect\citeauthoryear{{Knapen}}{{Knapen}}{2010}]{Kna10}
{Knapen} J.~H.,  2010, ArXiv e-prints

\bibitem[\protect\citeauthoryear{{Koleva}, {Prugniel}, {Ocvirk}, {Le Borgne} \&
  {Soubiran}}{{Koleva} et~al.}{2008}]{Kol08}
{Koleva} M.,  {Prugniel} P.,  {Ocvirk} P.,  {Le Borgne} D.,    {Soubiran} C.,
  2008, \mnras, 385, 1998

\bibitem[\protect\citeauthoryear{{Korn}, {Maraston} \& {Thomas}}{{Korn}
  et~al.}{2005}]{Korn05}
{Korn} A.~J.,  {Maraston} C.,    {Thomas} D.,  2005, \aap, 438, 685

\bibitem[\protect\citeauthoryear{{Le Borgne}, {Rocca-Volmerange}, {Prugniel},
  {Lan{\c c}on}, {Fioc} \& {Soubiran}}{{Le Borgne} et~al.}{2004}]{PEG-HR}
{Le Borgne} D.,  {Rocca-Volmerange} B.,  {Prugniel} P.,  {Lan{\c c}on} A.,
  {Fioc} M.,    {Soubiran} C.,  2004, \aap, 425, 881

\bibitem[\protect\citeauthoryear{{MacArthur}, {Courteau}, {Bell} \&
  {Holtzman}}{{MacArthur} et~al.}{2004}]{Mac04}
{MacArthur} L.~A.,  {Courteau} S.,  {Bell} E.,    {Holtzman} J.~A.,  2004,
  \apjs, 152, 175

\bibitem[\protect\citeauthoryear{{MacArthur}, {Gonz{\'a}lez} \&
  {Courteau}}{{MacArthur} et~al.}{2009}]{Mac09}
{MacArthur} L.~A.,  {Gonz{\'a}lez} J.~J.,    {Courteau} S.,  2009, \mnras, 395,
  28

\bibitem[\protect\citeauthoryear{{Maciel}, {Costa} \& {Uchida}}{{Maciel}
  et~al.}{2003}]{Mac03}
{Maciel} W.~J.,  {Costa} R.~D.~D.,    {Uchida} M.~M.~M.,  2003, \aap, 397, 667

\bibitem[\protect\citeauthoryear{{Magrini}, {Sestito}, {Randich} \&
  {Galli}}{{Magrini} et~al.}{2009}]{Mag09}
{Magrini} L.,  {Sestito} P.,  {Randich} S.,    {Galli} D.,  2009, \aap, 494, 95

\bibitem[\protect\citeauthoryear{{Martin} \& {Roy}}{{Martin} \&
  {Roy}}{1994}]{MR94}
{Martin} P.,  {Roy} J.-R.,  1994, \apj, 424, 599

\bibitem[\protect\citeauthoryear{{Moll{\'a}}, {Hardy} \&
  {Beauchamp}}{{Moll{\'a}} et~al.}{1999}]{Moll99}
{Moll{\'a}} M.,  {Hardy} E.,    {Beauchamp} D.,  1999, \apj, 513, 695

\bibitem[\protect\citeauthoryear{{Monteverde}, {Herrero}, {Lennon} \&
  {Kudritzki}}{{Monteverde} et~al.}{1997}]{Mon97}
{Monteverde} M.~I.,  {Herrero} A.,  {Lennon} D.~J.,    {Kudritzki} R.-P.,
  1997, \apjl, 474, L107+

\bibitem[\protect\citeauthoryear{{Ocvirk}}{{Ocvirk}}{2010}]{ocvirk2010}
{Ocvirk} P.,  2010, \apj, 709, 88

\bibitem[\protect\citeauthoryear{{Ocvirk}, {Pichon}, {Lan{\c c}on} \&
  {Thi{\'e}baut}}{{Ocvirk} et~al.}{2006a}]{Ocv06b}
{Ocvirk} P.,  {Pichon} C.,  {Lan{\c c}on} A.,    {Thi{\'e}baut} E.,  2006a,
  \mnras, 365, 74

\bibitem[\protect\citeauthoryear{{Ocvirk}, {Pichon}, {Lan{\c c}on} \&
  {Thi{\'e}baut}}{{Ocvirk} et~al.}{2006b}]{Ocv06a}
{Ocvirk} P.,  {Pichon} C.,  {Lan{\c c}on} A.,    {Thi{\'e}baut} E.,  2006b,
  \mnras, 365, 46

\bibitem[\protect\citeauthoryear{{Pagel} \& {Edmunds}}{{Pagel} \&
  {Edmunds}}{1981}]{PE81}
{Pagel} B.~E.~J.,  {Edmunds} M.~G.,  1981, \araa, 19, 77

\bibitem[\protect\citeauthoryear{{Pappalardo}, {Lancon}, {Vollmer}, {Ocvirk},
  {Boissier} \& {Boselli}}{{Pappalardo} et~al.}{2010}]{pappalardo2010}
{Pappalardo} C.,  {Lancon} A.,  {Vollmer} B.,  {Ocvirk} P.,  {Boissier} S.,
  {Boselli} A.,  2010, ArXiv e-prints

\bibitem[\protect\citeauthoryear{{Peletier} \& {Balcells}}{{Peletier} \&
  {Balcells}}{1996}]{PB96}
{Peletier} R.~F.,  {Balcells} M.,  1996, \aj, 111, 2238

\bibitem[\protect\citeauthoryear{{Peletier}, {Falc{\'o}n-Barroso}, {Bacon},
  {Cappellari}, {Davies}, {de Zeeuw}, {Emsellem}, {Ganda}, {Krajnovi{\'c}},
  {Kuntschner}, {McDermid}, {Sarzi} \& {van de Ven}}{{Peletier}
  et~al.}{2007}]{Pel07}
{Peletier} R.~F.,  {Falc{\'o}n-Barroso} J.,  {Bacon} R.,  {Cappellari} M.,
  {Davies} R.~L.,  {de Zeeuw} P.~T.,  {Emsellem} E.,  {Ganda} K.,
  {Krajnovi{\'c}} D.,  {Kuntschner} H.,  {McDermid} R.~M.,  {Sarzi} M.,    {van
  de Ven} G.,  2007, \mnras, 379, 445

\bibitem[\protect\citeauthoryear{{P{\'e}rez}, {S{\'a}nchez-Bl{\'a}zquez} \&
  {Zurita}}{{P{\'e}rez} et~al.}{2007}]{PSZ07}
{P{\'e}rez} I.,  {S{\'a}nchez-Bl{\'a}zquez} P.,    {Zurita} A.,  2007, \aap,
  465, L9

\bibitem[\protect\citeauthoryear{{P{\'e}rez}, {S{\'a}nchez-Bl{\'a}zquez} \&
  {Zurita}}{{P{\'e}rez} et~al.}{2009}]{PSZ09}
{P{\'e}rez} I.,  {S{\'a}nchez-Bl{\'a}zquez} P.,    {Zurita} A.,  2009, \aap,
  495, 775

\bibitem[\protect\citeauthoryear{{Pfenniger} \& {Friedli}}{{Pfenniger} \&
  {Friedli}}{1991}]{PF91}
{Pfenniger} D.,  {Friedli} D.,  1991, \aap, 252, 75

\bibitem[\protect\citeauthoryear{{Proctor}, {Forbes} \& {Beasley}}{{Proctor}
  et~al.}{2004}]{Pro04b}
{Proctor} R.~N.,  {Forbes} D.~A.,    {Beasley} M.~A.,  2004, \mnras, 355, 1327

\bibitem[\protect\citeauthoryear{{Proctor}, {Forbes}, {Hau}, {Beasley}, {De
  Silva}, {Contreras} \& {Terlevich}}{{Proctor} et~al.}{2004}]{Pro04a}
{Proctor} R.~N.,  {Forbes} D.~A.,  {Hau} G.~K.~T.,  {Beasley} M.~A.,  {De
  Silva} G.~M.,  {Contreras} R.,    {Terlevich} A.~I.,  2004, \mnras, 349, 1381

\bibitem[\protect\citeauthoryear{{Rocha-Pinto} \& {Maciel}}{{Rocha-Pinto} \&
  {Maciel}}{1996}]{RPM96}
{Rocha-Pinto} H.~J.,  {Maciel} W.~J.,  1996, \mnras, 279, 447

\bibitem[\protect\citeauthoryear{{Rocha-Pinto}, {Maciel}, {Scalo} \&
  {Flynn}}{{Rocha-Pinto} et~al.}{2000}]{RP00}
{Rocha-Pinto} H.~J.,  {Maciel} W.~J.,  {Scalo} J.,    {Flynn} C.,  2000, \aap,
  358, 850

\bibitem[\protect\citeauthoryear{{Ro{\v s}kar}, {Debattista}, {Quinn},
  {Stinson} \& {Wadsley}}{{Ro{\v s}kar} et~al.}{2008}]{Rosk08b}
{Ro{\v s}kar} R.,  {Debattista} V.~P.,  {Quinn} T.~R.,  {Stinson} G.~S.,
  {Wadsley} J.,  2008, \apjl, 684, L79

\bibitem[\protect\citeauthoryear{{Ryder}, {Fenner} \& {Gibson}}{{Ryder}
  et~al.}{2005}]{Ryd05}
{Ryder} S.~D.,  {Fenner} Y.,    {Gibson} B.~K.,  2005, \mnras, 358, 1337

\bibitem[\protect\citeauthoryear{{S{\'a}nchez-Bl{\'a}zquez}, {Courty}, {Gibson}
  \& {Brook}}{{S{\'a}nchez-Bl{\'a}zquez} et~al.}{2009}]{SB09a}
{S{\'a}nchez-Bl{\'a}zquez} P.,  {Courty} S.,  {Gibson} B.~K.,    {Brook} C.~B.,
   2009, \mnras, 398, 591

\bibitem[\protect\citeauthoryear{{S{\'a}nchez-Bl{\'a}zquez}, {Gorgas},
  {Cardiel} \& {Gonz{\'a}lez}}{{S{\'a}nchez-Bl{\'a}zquez}
  et~al.}{2006a}]{SB06a}
{S{\'a}nchez-Bl{\'a}zquez} P.,  {Gorgas} J.,  {Cardiel} N.,    {Gonz{\'a}lez}
  J.~J.,  2006a, \aap, 457, 787

\bibitem[\protect\citeauthoryear{{S{\'a}nchez-Bl{\'a}zquez}, {Gorgas},
  {Cardiel} \& {Gonz{\'a}lez}}{{S{\'a}nchez-Bl{\'a}zquez}
  et~al.}{2006b}]{SB06b}
{S{\'a}nchez-Bl{\'a}zquez} P.,  {Gorgas} J.,  {Cardiel} N.,    {Gonz{\'a}lez}
  J.~J.,  2006b, \aap, 457, 809

\bibitem[\protect\citeauthoryear{{S{\'a}nchez-Bl{\'a}zquez}, {Peletier},
  {Jim{\'e}nez-Vicente}, {Cardiel}, {Cenarro}, {Falc{\'o}n-Barroso}, {Gorgas},
  {Selam} \& {Vazdekis}}{{S{\'a}nchez-Bl{\'a}zquez} et~al.}{2006}]{SB06}
{S{\'a}nchez-Bl{\'a}zquez} P.,  {Peletier} R.~F.,  {Jim{\'e}nez-Vicente} J.,
  {Cardiel} N.,  {Cenarro} A.~J.,  {Falc{\'o}n-Barroso} J.,  {Gorgas} J.,
  {Selam} S.,    {Vazdekis} A.,  2006, \mnras, 371, 703

\bibitem[\protect\citeauthoryear{{Sarzi}, {Falc{\'o}n-Barroso}, {Davies},
  {Bacon}, {Bureau}, {Cappellari}, {de Zeeuw}, {Emsellem}, {Fathi},
  {Krajnovi{\'c}}, {Kuntschner}, {McDermid} \& {Peletier}}{{Sarzi}
  et~al.}{2006}]{Sar06}
{Sarzi} M.,  {Falc{\'o}n-Barroso} J.,  {Davies} R.~L.,  {Bacon} R.,  {Bureau}
  M.,  {Cappellari} M.,  {de Zeeuw} P.~T.,  {Emsellem} E.,  {Fathi} K.,
  {Krajnovi{\'c}} D.,  {Kuntschner} H.,  {McDermid} R.~M.,    {Peletier} R.~F.,
   2006, \mnras, 366, 1151

\bibitem[\protect\citeauthoryear{{Schiavon}}{{Schiavon}}{2007}]{Schi07}
{Schiavon} R.~P.,  2007, \apjs, 171, 146

\bibitem[\protect\citeauthoryear{{Sellwood}}{{Sellwood}}{1981}]{Sell81}
{Sellwood} J.~A.,  1981, \aap, 99, 362

\bibitem[\protect\citeauthoryear{{Sellwood} \& {Binney}}{{Sellwood} \&
  {Binney}}{2002}]{SB02}
{Sellwood} J.~A.,  {Binney} J.~J.,  2002, \mnras, 336, 785

\bibitem[\protect\citeauthoryear{{Sellwood} \& {Wilkinson}}{{Sellwood} \&
  {Wilkinson}}{1993}]{SW93}
{Sellwood} J.~A.,  {Wilkinson} A.,  1993, Reports on Progress in Physics, 56,
  173

\bibitem[\protect\citeauthoryear{{Serra} \& {Trager}}{{Serra} \&
  {Trager}}{2007}]{ST07}
{Serra} P.,  {Trager} S.~C.,  2007, \mnras, 374, 769

\bibitem[\protect\citeauthoryear{{S{\'e}rsic} \& {Pastoriza}}{{S{\'e}rsic} \&
  {Pastoriza}}{1965}]{SP65}
{S{\'e}rsic} J.~L.,  {Pastoriza} M.,  1965, \pasp, 77, 287

\bibitem[\protect\citeauthoryear{{Sharina} \& {Davoust}}{{Sharina} \&
  {Davoust}}{2009}]{sharina09}
{Sharina} M.,  {Davoust} E.,  2009, \aap, 497, 65

\bibitem[\protect\citeauthoryear{{Spitzer} Jr. \& {Schwarzschild}}{{Spitzer} \&
  {Schwarzschild}}{1953}]{SS53}
{Spitzer} Jr. L.,  {Schwarzschild} M.,  1953, \apj, 118, 106

\bibitem[\protect\citeauthoryear{{Stanghellini}, {Guerrero}, {Cunha},
  {Manchado} \& {Villaver}}{{Stanghellini} et~al.}{2006}]{Stan06}
{Stanghellini} L.,  {Guerrero} M.~A.,  {Cunha} K.,  {Manchado} A.,
  {Villaver} E.,  2006, \apj, 651, 898

\bibitem[\protect\citeauthoryear{{Tantalo}, {Chiosi} \& {Bressan}}{{Tantalo}
  et~al.}{1998}]{Tan98}
{Tantalo} R.,  {Chiosi} C.,    {Bressan} A.,  1998, \aap, 333, 419

\bibitem[\protect\citeauthoryear{{Thomas}, {Maraston} \& {Bender}}{{Thomas}
  et~al.}{2003}]{TMB03}
{Thomas} D.,  {Maraston} C.,    {Bender} R.,  2003, \mnras, 339, 897

\bibitem[\protect\citeauthoryear{{Toloba}, {Boselli}, {Cenarro}, {Peletier},
  {Gorgas}, {Gil de Paz} \& {Mu{\~n}oz-Mateos}}{{Toloba} et~al.}{2011}]{T11}
{Toloba} E.,  {Boselli} A.,  {Cenarro} A.~J.,  {Peletier} R.~F.,  {Gorgas} J.,
  {Gil de Paz} A.,    {Mu{\~n}oz-Mateos} J.~C.,  2011, \aap, 526, A114+

\bibitem[\protect\citeauthoryear{{Trager}, {Dalcanton} \& {Weiner}}{{Trager}
  et~al.}{1999}]{TDW99}
{Trager} S.~C.,  {Dalcanton} J.~J.,    {Weiner} B.~J.,  1999, in
  {C.~M.~Carollo, H.~C.~Ferguson, \& R.~F.~G.~Wyse} ed., The Formation of
  Galactic Bulges {Integrated Stellar Populations of Bulges: First Results}.
pp 42--+

\bibitem[\protect\citeauthoryear{{Trager}, {Faber}, {Worthey} \&
  {Gonz{\'a}lez}}{{Trager} et~al.}{2000}]{T00a}
{Trager} S.~C.,  {Faber} S.~M.,  {Worthey} G.,    {Gonz{\'a}lez} J.~J.,  2000,
  \aj, 119, 1645

\bibitem[\protect\citeauthoryear{{Treuthardt}, {Salo}, {Rautiainen} \&
  {Buta}}{{Treuthardt} et~al.}{2008}]{Treut08}
{Treuthardt} P.,  {Salo} H.,  {Rautiainen} P.,    {Buta} R.,  2008, \aj, 136,
  300

\bibitem[\protect\citeauthoryear{{Tripicco} \& {Bell}}{{Tripicco} \&
  {Bell}}{1995}]{TB95}
{Tripicco} M.~J.,  {Bell} R.~A.,  1995, \aj, 110, 3035

\bibitem[\protect\citeauthoryear{{Twarog}}{{Twarog}}{1980a}]{Twa80a}
{Twarog} B.~A.,  1980a, \apjs, 44, 1

\bibitem[\protect\citeauthoryear{{Twarog}}{{Twarog}}{1980b}]{Twa80b}
{Twarog} B.~A.,  1980b, \apj, 242, 242

\bibitem[\protect\citeauthoryear{{Vazdekis}}{{Vazdekis}}{1999}]{Vaz99}
{Vazdekis} A.,  1999, \apj, 513, 224

\bibitem[\protect\citeauthoryear{{Veron-Cetty} \& {Veron}}{{Veron-Cetty} \&
  {Veron}}{1986}]{VV86}
{Veron-Cetty} M.,  {Veron} P.,  1986, \aaps, 66, 335

\bibitem[\protect\citeauthoryear{{Vila-Costas} \& {Edmunds}}{{Vila-Costas} \&
  {Edmunds}}{1992}]{VCE92}
{Vila-Costas} M.~B.,  {Edmunds} M.~G.,  1992, \mnras, 259, 121

\bibitem[\protect\citeauthoryear{{Williams}, {Dalcanton}, {Dolphin}, {Holtzman}
  \& {Sarajedini}}{{Williams} et~al.}{2009}]{Will09b}
{Williams} B.~F.,  {Dalcanton} J.~J.,  {Dolphin} A.~E.,  {Holtzman} J.,
  {Sarajedini} A.,  2009, \apjl, 695, L15

\bibitem[\protect\citeauthoryear{{Williams}, {Dalcanton}, {Seth}, {Weisz},
  {Dolphin}, {Skillman}, {Harris}, {Holtzman}, {Girardi}, {de Jong}, {Olsen},
  {Cole}, {Gallart}, {Gogarten}, {Hidalgo}, {Mateo}, {Rosema}, {Stetson} \&
  {Quinn}}{{Williams} et~al.}{2009}]{Will09a}
{Williams} B.~F.,  {Dalcanton} J.~J.,  {Seth} A.~C.,  {Weisz} D.,  {Dolphin}
  A.,  {Skillman} E.,  {Harris} J.,  {Holtzman} J.,  {Girardi} L.,  {de Jong}
  R.~S.,  {Olsen} K.,  {Cole} A.,  {Gallart} C.,  {Gogarten} S.~M.,  {Hidalgo}
  S.~L.,  {Mateo} M.,  {Rosema} K.,  {Stetson} P.~B.,    {Quinn} T.,  2009,
  \aj, 137, 419

\bibitem[\protect\citeauthoryear{{Wyse} \& {Gilmore}}{{Wyse} \&
  {Gilmore}}{1995}]{WG95}
{Wyse} R.~F.~G.,  {Gilmore} G.,  1995, \aj, 110, 2771

\bibitem[\protect\citeauthoryear{{Yoachim} \& {Dalcanton}}{{Yoachim} \&
  {Dalcanton}}{2008}]{YD08}
{Yoachim} P.,  {Dalcanton} J.~J.,  2008, \apj, 683, 707

\bibitem[\protect\citeauthoryear{{Yong}, {Carney}, {Teixera de Almeida} \&
  {Pohl}}{{Yong} et~al.}{2006}]{Yong06}
{Yong} D.,  {Carney} B.~W.,  {Teixera de Almeida} M.~L.,    {Pohl} B.~L.,
  2006, \aj, 131, 2256

\end{thebibliography}
\label{lastpage}

\appendix
\section{Comparison between MILES and BC03 stellar population models}
\label{a:comparisonmodels}
\begin{figure*}
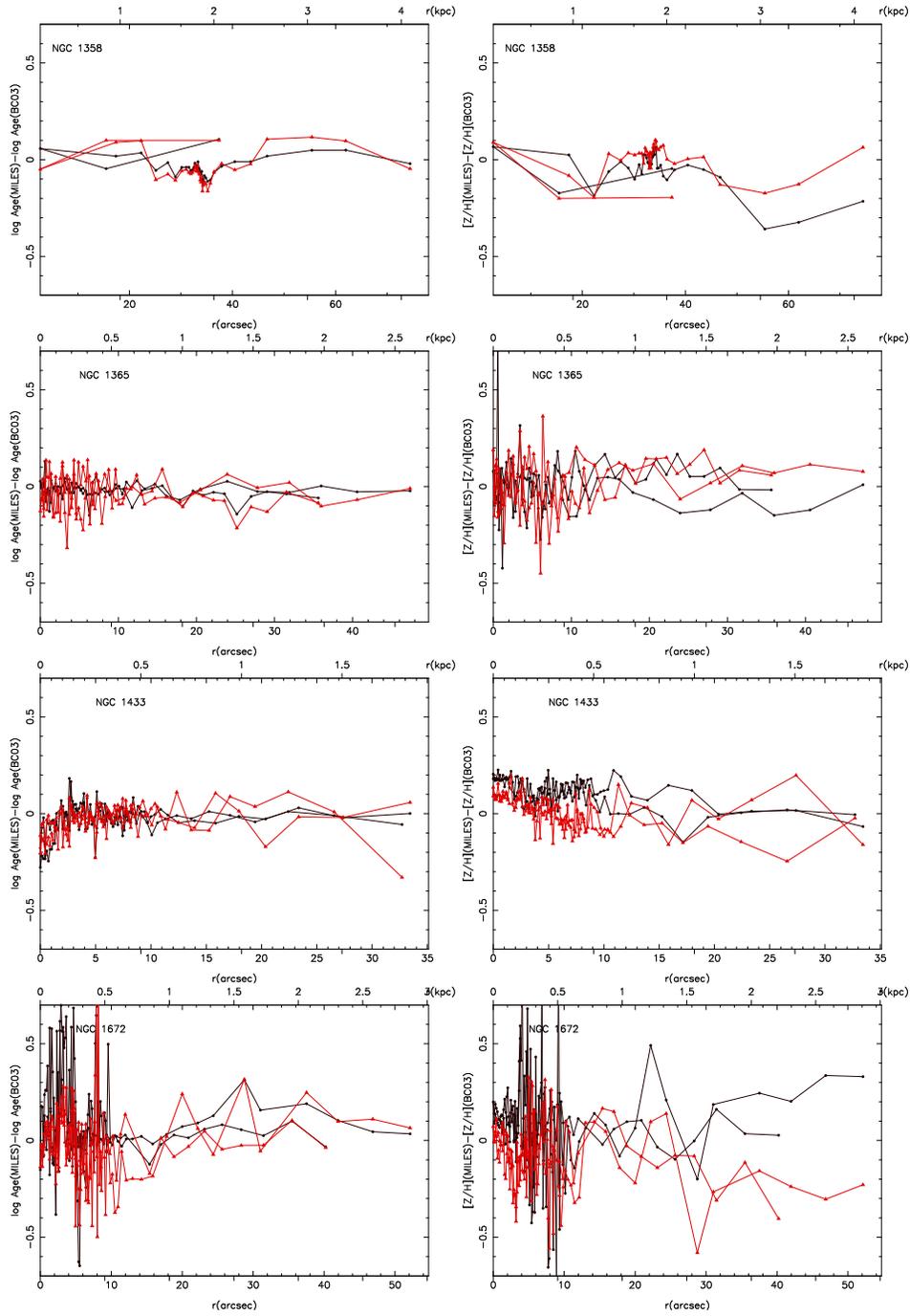

\resizebox{0.35\textwidth}{!}{\includegraphics[angle=-90]{n1358.difage.new.ps}}
\resizebox{0.35\textwidth}{!}{\includegraphics[angle=-90]{n1358.difz.new.ps}}
\resizebox{0.35\textwidth}{!}{\includegraphics[angle=-90]{n1365.difa.new.ps}}
\resizebox{0.35\textwidth}{!}{\includegraphics[angle=-90]{n1365.difz.new.ps}}
\resizebox{0.35\textwidth}{!}{\includegraphics[angle=-90]{n1433.difage.new.ps}}
\resizebox{0.35\textwidth}{!}{\includegraphics[angle=-90]{n1433.difz.new.ps}}
\resizebox{0.35\textwidth}{!}{\includegraphics[angle=-90]{n1672.difa.new.ps}}
\resizebox{0.35\textwidth}{!}{\includegraphics[angle=-90]{n1672.difz.new.ps}}
\caption{Differences in the luminosity-weighted (red) and mass weighted (black)
mean age and metallicity obtained with the MILES and the BC03 stellar
population models as a function of radius.\label{comparison_models}}
\end{figure*}

Figure~\ref{comparison_models} shows the difference between 
the luminosity- and mass-weighted
ages and metallicities obtained with two different stellar population 
models, MILES (Vazdekis et al. 2010)
and BC03 (Bruzual \& Charlot 2003) as a function of radius.
It can be seen that, despite for some radial bins the differences can be 
very large, in general they scatter around zero, and the gradients of 
the differences are compatible with being null. The largest differences
are obtained for populations with ages younger than $\sim$3 Gyr (ie, the bulge 
region of NGC~1433 and the ring region in NGC~1672). This is, in principle, 
in agreement with the results of Koleva et al. (2008) although we can not explore
the parameter space in a systematic way as it was done in that work.
However, we do not find large differences for populations with subsolar 
metallicity, like for example in NGC~1365.
\label{compa_models} 

\section{Degeneracy Between $\sigma$ and Metallicity }
\label{a:sigma_z}
It has been known for a long time that the determined velocity dispersion
of a galaxy spectrum  depends on the spectral 
type and metallicity of the stellar template (ie., Laird \& Levison 1985; Bender 1990;
Koleva et al. 2007). A deeper absorption feature 
can be obtained by either increasing the 
metallicity or decreasing the broadening. This  leads  to degeneracies when 
trying to derive both parameters at the same time.
Koleva et al. (2007) quantify this degeneracy as:

\begin{equation}
\delta(\sigma)/\sigma \sim 0.4 \times \delta([Fe/H])
\end{equation}
\noindent
with $\sigma$ in km/s and $[$Fe/H$]$ in dex - a 0.1 dex mismatch in 
metallicity results in a 4\% error in the velocity dispersion. 

\begin{figure*}
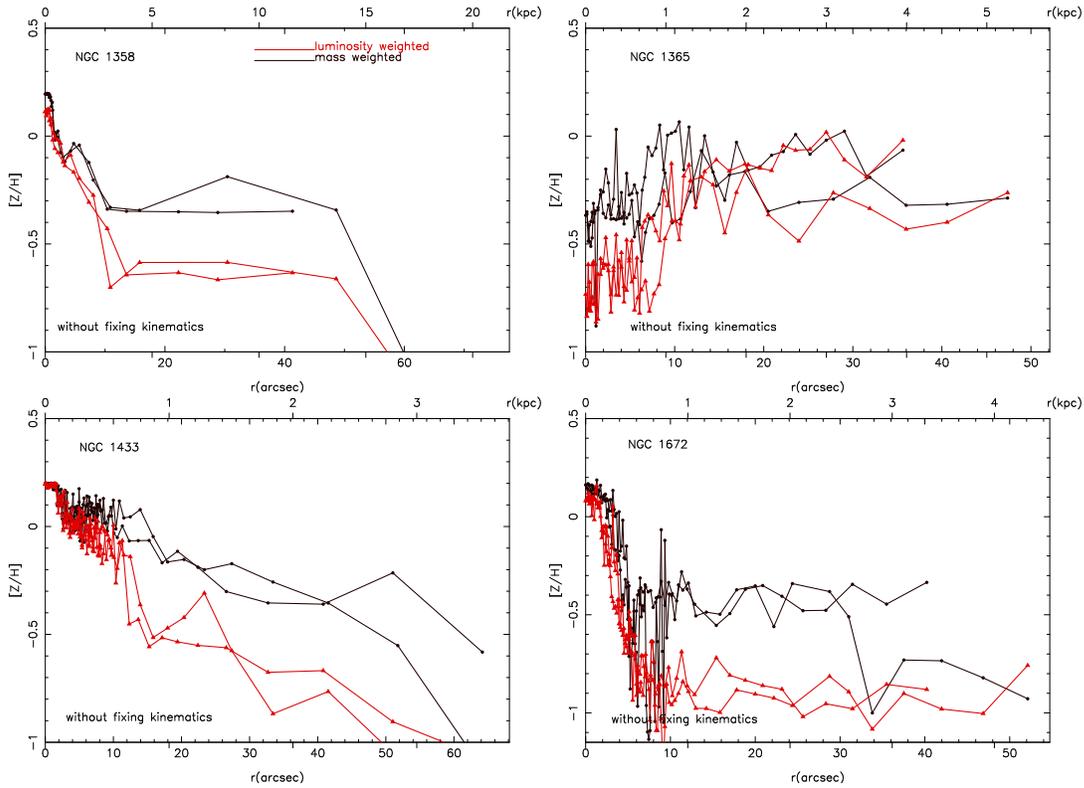

\resizebox{0.4\textwidth}{!}{\includegraphics[angle=-90]{n1358.z.sfr.nofixing.ps}}
\resizebox{0.4\textwidth}{!}{\includegraphics[angle=-90]{n1365.z.sfr.nofixing.ps}}
\resizebox{0.4\textwidth}{!}{\includegraphics[angle=-90]{n1433.z.sfr.nofixing.extra.ps}}
\resizebox{0.4\textwidth}{!}{\includegraphics[angle=-90]{n1672.z.sfr.nofixing.ps}}
\caption{Mass- and luminosity-weighted  metallicity gradients derived from the 
recovered star formation history, when the kinematics are not fixed\label{age_z.weighted.nofixing}}
\end{figure*}

In order to quantify the differences in the obtained parameters with and without fixing the 
kinematics, we have perform the calculations using both methods. Figure~\ref{age_z.weighted.nofixing}
shows the results of the mass- and luminosity-weighted  mean metallicity profiles when the 
kinematics are not fixed, derived using {\tt steckmap}.
As can be seen, in all cases, the luminosity-weighted values are lower than the mass-weighted ones.
This is very puzzling, in principle, because it is contrary to what is  expected in  
a normal chemical evolution scenario.

\begin{figure*}
\resizebox{0.4\textwidth}{!}{\includegraphics[angle=-90]{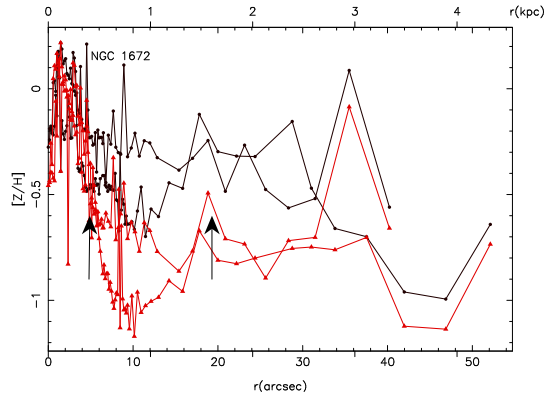}}
\caption{Mass- (black line) and luminosity-weighted (red line) age and [Z/H] gradients derived from the 
recovered star formation history using {\tt STARLIGHT}.\label{grad_starlight}}
\end{figure*}

To check if this behaviour  is a systematic error or an artifact of  the method we are using, we repeated
the experiment using the code {\tt STARLIGHT}
(Cid-Fernandes et al.\ 2005) for one of our galaxies, NGC~1672.
Although {\tt STARLIGHT} uses the  continuum  shape of the spectrum to perform the
analysis, we have removed it from both the templates and the
galaxies, using spline fits to avoid the
uncertainties due to dust extintion.  Figure~\ref{grad_starlight}
shows the results.  As can be seen, although the absolute values in
some regions differ depending on the code that we use, the general
shape of the gradients are very similar, independent of the method
used to derive star formation histories. The way {\tt STARLIGHT} and
{\tt STECKMAP} find the optimal solution is quite different and the
similarity of the results lends support to our final conclusions.
As can be seen in the figure, {\tt STARLIGHT} also gives lower luminosity-weighted
than mass-weighted metallicities, contrary to what would be expected in a normal 
chemical evolution scenario.

\begin{figure*}
 \resizebox{0.8\textwidth}{!}{\includegraphics[angle=-90]{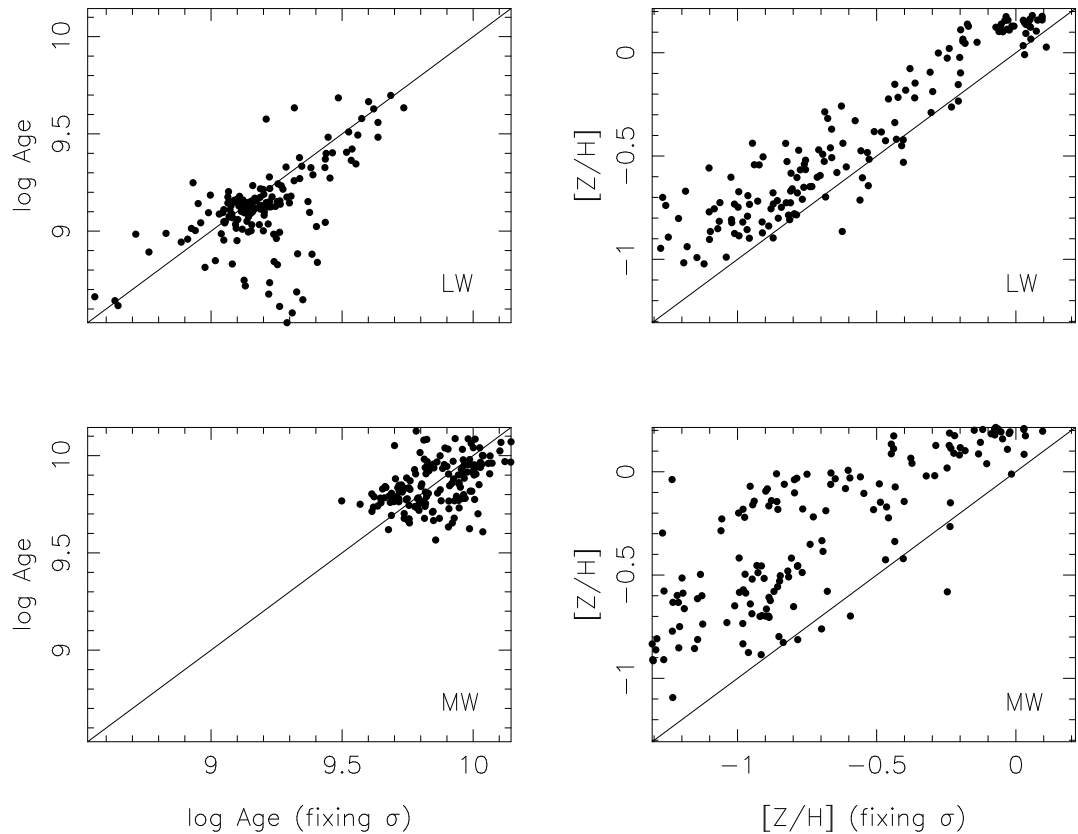}}
 \caption{Comparison of the stellar parameters obtained when the $\sigma$ is 
fixed 
 and calculated using {\tt Steckmap}. \label{compara}}
 \end{figure*}

To understand this effect better, we compare, in Fig.~\ref{compara}, the mean 
stellar population parameters obtained fixing and without fixing the kinematics when 
running {\tt Steckmap}. 
As can be seen, while the mean ages are very similar, the 
metallicities obtained in both cases show 
clear differences. The effect is larger for luminosity-weighted values than for the mass-weighted ones.
The net result is that the luminosity-weighted metallicities appear to be lower than 
the mass-weighted ones. 

The question then arises,  
how do we know which value is correct? 
We have mentioned previsouly that a scenario
where the stellar metallicity decrease with time is, in principle, 
contrary to what 
would be expected for a standard chemical evolution scenario. 
However, this by itself  is 
not an argument to discard the result.
For example, it has been claimed
that in our own Milky Way there has been 
chemical dilution (see, e.g., Tsujimoto et al. 2010).
These authors present a compilation of 
metallicities measured in Cepheids (aged of about 10 Myr) and open clusters older
than 1 Gyr, finding that the metallicity of the latter is 
higher than that of the Cepheids in the inner disc (up to 
more than four scalelenghts), although the trend seems to 
reverse at larger distances.  
They are able to reproduce this evolution in
metallicity  with a model where the inner
disc stars form from a metal enriched wind originating from the bulge. This wind is  metal rich due to early efficient star formation. Later, accretion of low-metallicity gas from
the halo dilutes the metals of the disk, producing a decreasing metallicity with time.

To explore the systematics  associated when the kinematics are fixed, as
opposed to 
when it is not,  under different evolutionary scenarios, we have  performed a series of 
simulations.
For these simulations we used an exponentially declining star formation history 
$\exp(-t/\tau)$ with different values of $\tau$. 
The SSP models were truncated at ages$=[0.3-17]$ Gyr and the metallicities were chosen so
as to remain in the ${\rm [Z/H]}>-1$ regime, to ensure that we work in a 
 parameter space where the models are most reliable, and also to ensure
that we bracket the parameter
  space covered by the galaxies we analyse.
 We have tested three different chemical enrichment scenarios. In the following equations 
 $t$ can be regarded as the age of the corresponding stellar component or the lookback time in Gyr.
\begin{enumerate}
\item{{\em Standard chemical enrichment}: 
\begin{equation}
{\rm [Z/H]}= +0.2-0.51 \times (\log(t)+0.5) \, .
\end{equation}
In this model the youngest stars have ${\rm [Z/H]}=+0.2$, and the oldest stars have
 ${\rm [Z/H]}=-0.7$.}\\
\item{{\em no chemical evolution}: ${\rm [Z/H]}=0$ (i.e. solar metallicity) for all the
 generations of stars.}\\
\item{{\em Reverse chemical evolution} or chemical dilution: 
\begin{equation}
{\rm [Z/H]}= -0.7+0.51 \times (\log(t)+0.5) \, .
\end{equation}
In this model the youngest stars have ${\rm [Z/H]}=-0.7$ while the oldest stars have ${\rm [Z/H]}=+0.2$.}
\end{enumerate}

\begin{figure*}
\resizebox{0.9\textwidth}{!}{\includegraphics[angle=90]{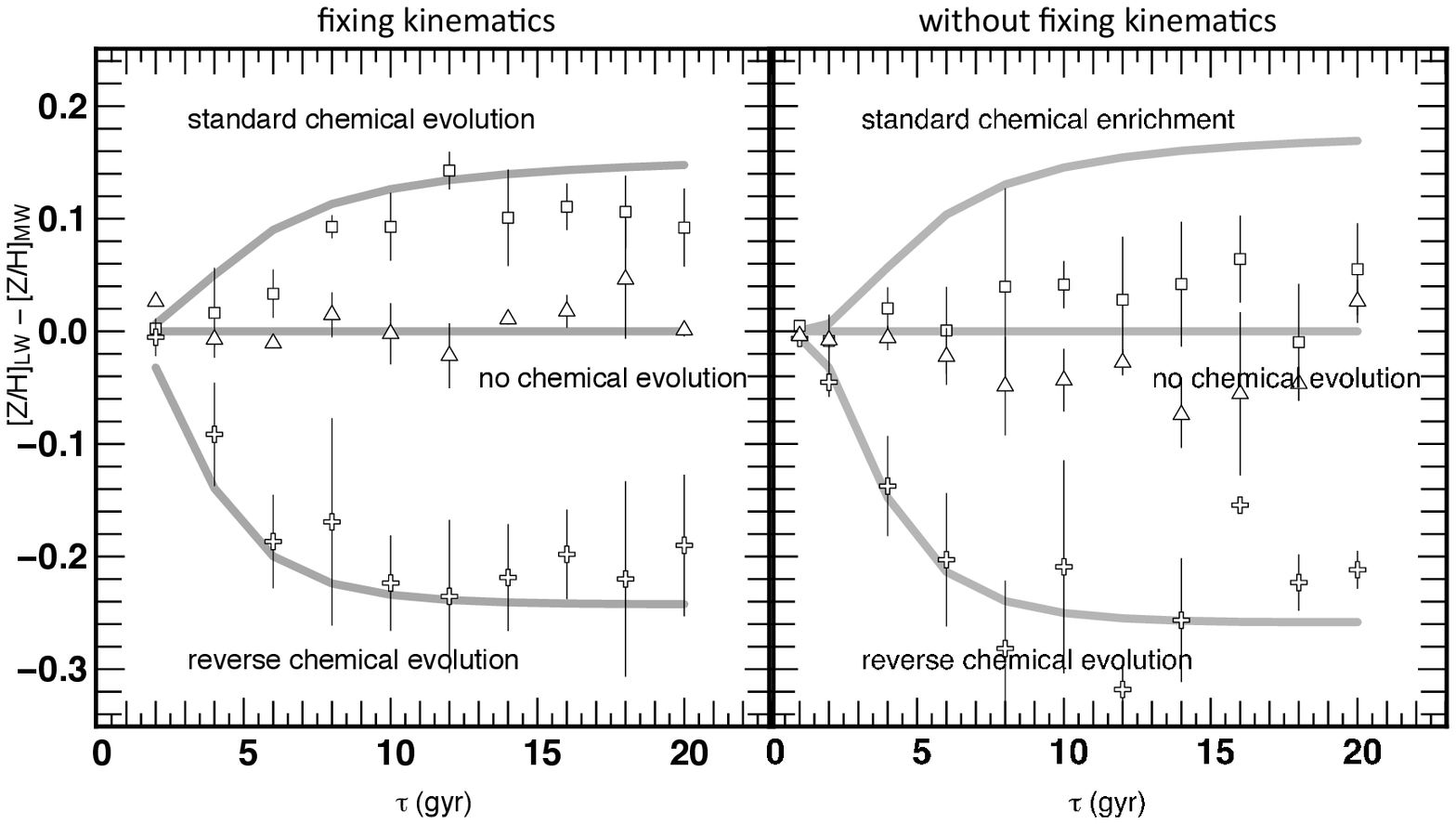}}
\caption{Differences between the mass and luminosity-weighted metallicities
for exponentially declining star formation histories with various timescales $\tau$ (x-axis) generated
as explained in Sec.~\ref{sec:test2} for three different chemical evolution 
scenarios. The thick grey lines are the values obtained from the models, and the symbols show 
the values recovered by {\tt STECKMAP} fixing the kinematics (left panel) and 
without fixing it (right panels). The error bars are obtained via 
Monte Carlo simulations.
\label{diferencias_mass_lum}}
\end{figure*}

We have used, for this experiment, the 
models by V10, broadened to 200 km/s. In order to mimic the data, we have 
added noise to the spectra to reach a signal-to-noise per Angstrom of 50.
Figure~\ref{diferencias_mass_lum} 
shows the difference between the luminosity- and mass-weighted
metallicities as a function of $\tau$ for the three chemical evolution 
scenarios in both cases, when the kinematics is fixed and when it is not. 
The thick grey lines show the model for the 3 different scenarios, and
 the symbols are the values as computed from the {\tt STECKMAP} inversions of the mock spectra.
The error bars in the figure are computed as the RMS dispersion of the values obtained 
in  50 MonteCarlo simulations in which each pixel of the synthetic spectrum 
was modified randomly following a Gaussian distribution with a width given by the noise spectrum.

As can be seen, when the kinematics are fixed the difference between the luminosity- and 
the mass-weighted metallicity are much better reproduced than when the velocity dispersion is calculated
at the same time as the stellar population parameters. As the effect is stronger  for  young ages and 
at larger metallicities, the "standard chemical evolution" scenario where the younger populations 
are more metal rich is the one that is more affected. This is the reason why 
when the $\sigma$ is not fixed, we tend to get lower luminosity-weighted than mass-weighted metallicities.

\subsection{Testing the velocity dispersion}
\begin{figure}
\resizebox{0.4\textwidth}{!}{\includegraphics[angle=-90]{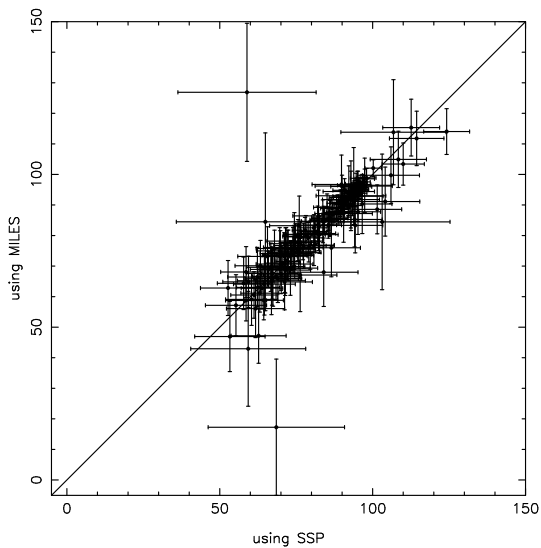}}
\caption{Comparison of the velocity dispersions using the SSP-models by V10 as templates
and using a subset of 250 stars from the MILES library.\label{compara_sigmas}}
\end{figure}

\begin{figure}
\resizebox{0.4\textwidth}{!}{\includegraphics[angle=-90]{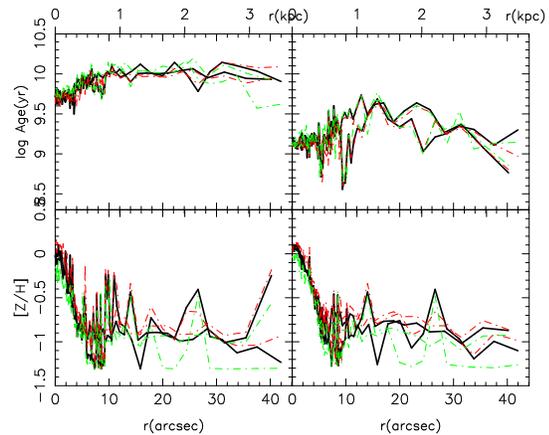}}
\caption{Mass (left panels)- and luminosity- (right panels) weighted ages and metallicities
for the galaxy NGC1672 fixing the kinematics to the observed values (solid black line) and 
fixing the kinematics to   $\sigma$+error in $\sigma$ (red dashed line and
$\sigma$-error in $\sigma$ (green dashed line).\label{compara_sigmas2}}
\end{figure}

There was a concern, raised by the referee, that the kinematics we are measuring could be biased as we have used
the same SSP-models than those used in {\tt steckmap}, suggesting than the template mixing 
process behind the {\tt pPXF} kinematics fit may well suffer from the same degeneracy that we
are trying to avoid.
To deal with this possibility we have perform two different tests:
\begin{enumerate}
\item We have   re-calculated the kinematics using a subset of the MILES stellar library, instead of 
the stellar population models. We have selected 250 stars coveraing a wide range in the atmospheric
parameters and run {\tt pPXF} using these as a templates. 
The results can be seen in Fig.~\ref{compara_sigmas}. The offset between the velocity dispersions 
measured with the models and the stars is $-0.36$kms$^{-1}$ with an RMS dispersion of 12.6 kms$^{-1}$.
That is, there is not a significant difference between the two values. 
\item We have studied the impact  that the errors in the measured $\sigma$ have in the {\tt steckmap} results.
To do this we have run {\tt steckmap} fixing the kinematics to the values of $\sigma+e\sigma$ and 
$\sigma-e\sigma$ for one of the galaxies, NGC~1672, where $e\sigma$ represent 1.5 of the formal errors.
The results are displayed in Fig.~\ref{compara_sigmas2}. In the figure we show the mean stellar population 
parameters  obtained in each case. It can be seen that the errors in the derived  $\sigma$ does not affect
significatively the results of this work.
\end{enumerate}
\subsection{Dependence of the Results on the Initial Guess and Best Initial Parameters}
\begin{figure*}
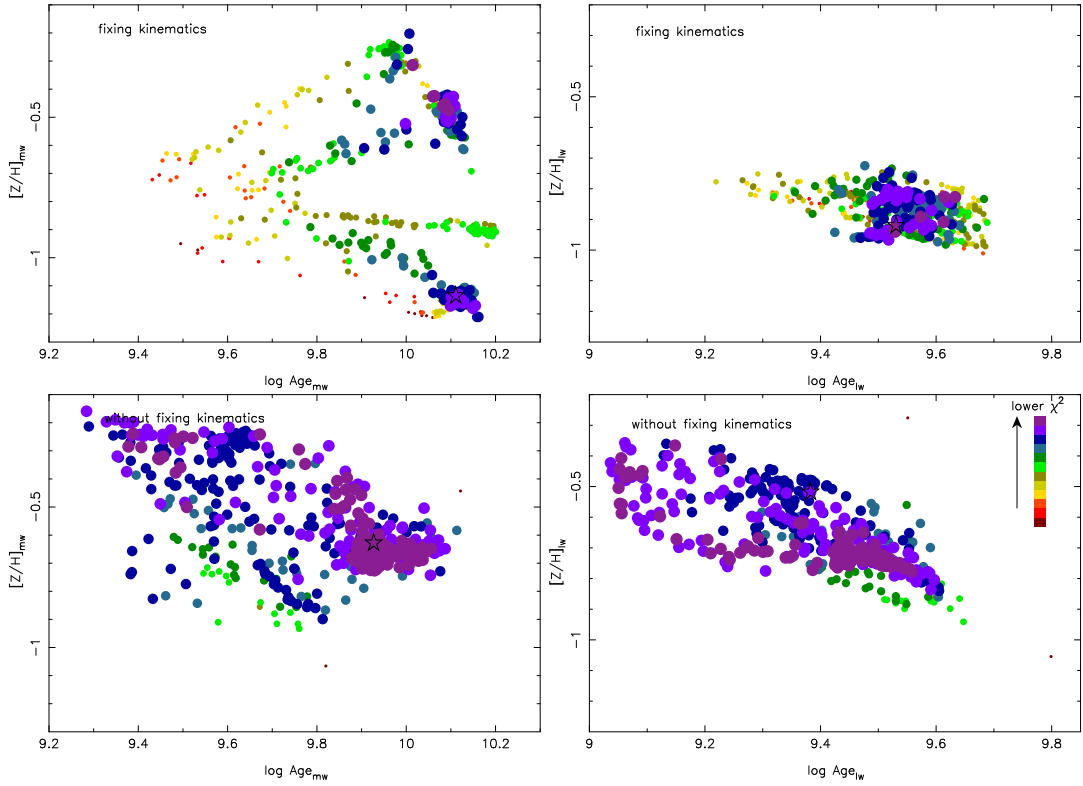

\resizebox{0.4\textwidth}{!}{\includegraphics[angle=-90]{age_z.mw.chi.ps}}
\resizebox{0.4\textwidth}{!}{\includegraphics[angle=-90]{age_z.lw.chi.ps}}
\resizebox{0.4\textwidth}{!}{\includegraphics[angle=-90]{age_z.mw.nofixing.ps}}
\resizebox{0.4\textwidth}{!}{\includegraphics[angle=-90]{age_z.lw.nofixing.ps}}
\caption{Luminosity- and mass-weighted metallicity vs. age for different Monte Carlo realisations,
each with a different initial guess for {\tt steckmap}. 
The initial guesses that we have used here 
consist of random bursts of star formation with random age and metallicity. The empty star
shows the location of the solution found with the {\bf flat initial guess}. This experiment was performed for 
galaxy NGC~1672 at
position R=10.2 arcsec\label{fig:chi}}
\end{figure*}

All iterative optimisation methods require a starting point, which is usually referred 
to as the `initial guess'.
Ideally, the method is insensitive to the choice of this 
guess, which is rather arbitrary. We have used for our real data an initial 
guess
consisting of flat  distributions for all fields. 
The reason for this choice is that
it always converges  to a good solution (in the $\chi^2$ and regularization sense), which 
is not true with all the first guesses. However, a bad choice for 
an initial guess can 
lead  to 
bad fits  or to secondary  minima. To be sure that we reach a global minimum and not a secondary
minimum, we have run {\tt Steckmap} over several spectra using a family of initial guesses. 
Here we show the experiments  with first guesses consisting  of  random bursts happening at 
random times and  with random widths. The age-metallicity relation of the model is flat with a random metallicity
within the range of the models and the broadening function is a Gaussian centred at zero. 
We have performed 500 simulations.  Figure~\ref{fig:chi} shows the mean  age-metallicity plane 
using the two weights, luminosity and mass, for the different realisations. We have colour-coded the 
symbols depending on the $\chi^2$ of the fit. The $\chi^2$ of the fit is also inverserly proportional 
to the size of the symbol. We show as well, for comparison, the position 
of the value calculated with an initial guess of a flat distribution.
We have performed  the experiment for the two cases 
of fixing and not fixing the kinematics.
As can be seen, the results may depend considerably on the choice of fitting 
or not fitting the broadening function. In the first case, we usually 
obtain two local minima, although one of them  is clearer than 
the other.  Furthermore, this minimum  always coincides
with the values that we obtain using a flat distribution as the intial guess.
In the case where we try to calculate the kinematics, there is a clear 
age-metallicity degeneracy, especially 
for the luminosity weighted values.  

These tests validate the 
flat initial guess as a good choice, and therefore this 
approach was adopted for this work. It allows users to analyse
real data 
without having to experiment with a family of initial 
guesses, to find the best minimum of $\chi^2$.

\end{document}